\documentclass[onecolumn,nofootinbib]{revtex4}
\usepackage{setspace}
\usepackage{amsmath,amssymb}
\usepackage{graphicx}
\usepackage{dcolumn}
\usepackage{bm,color}
\usepackage{hyperref}
\usepackage{accents}
\usepackage{amssymb,float}
\usepackage{amsmath}
\usepackage{multirow}
\usepackage{tikz}
\usepackage{subcaption}
\usepackage[rightcaption]{sidecap}
\usepackage{float}
\sidecaptionvpos{figure}{t}
\usepackage{enumerate}
\usepackage{cancel}

\newcolumntype{P}[1]{>{\centering\arraybackslash}p{#1}}
\newcolumntype{M}[1]{>{\centering\arraybackslash}m{#1}}
\hypersetup{
    colorlinks=true,
    linkcolor=red,
    filecolor=magenta,      
    citecolor=blue
}
\usepackage{rotating}

\newcommand{\udt}[3]{#1^{#2}_{\phantom{#2}#3}}
\newcommand{\udut}[4]{#1^{#2\phantom{#3}#4}_{\phantom{#2}#3\phantom{#4}}}

\newcommand{\dut}[3]{#1_{#2}^{\phantom{#2}#3}}
\newcommand{\dudt}[4]{#1_{#2\phantom{#3}#4}^{\phantom{#2}#3}}
\newcommand{\dd}{\mathrm{d}}
\newcommand{\lc}[1]{\accentset{\circ}{#1}}

\begin{document}

\title{Gauge invariant perturbations in teleparallel Horndeski gravity}

\author{Bobomurat Ahmedov}
\email{ahmedov@astrin.uz}
\affiliation{New Uzbekistan University, Movarounnahr str. 1, Tashkent 100000, Uzbekistan}
\affiliation{Institute of Fundamental and Applied Research,
National Research University TIIAME, Kori Niyoziy 39, Tashkent 100000, Uzbekistan}

\author{Maria Caruana}
\email{maria.caruana.16@um.edu.mt}
\affiliation{Institute of Space Sciences and Astronomy, University of Malta, Malta, MSD 2080}
\affiliation{Department of Physics, University of Malta, Malta}

\author{Konstantinos F. Dialektopoulos}
\email{kdialekt@gmail.com}
\affiliation{Department of Mathematics and Computer Science, Transilvania University of Brasov, 500091, Brasov, Romania}

\author{Jackson Levi Said}
\email{jackson.said@um.edu.mt}
\affiliation{Institute of Space Sciences and Astronomy, University of Malta, Malta, MSD 2080}
\affiliation{Department of Physics, University of Malta, Malta}

\author{Abdurakhmon Nosirov}
\email{abdurahmonnosirov000203@gmail.com}
\affiliation{Center for Astronomy and Astrophysics, Center for Field Theory and Particle Physics and Department of Physics, Fudan University, 200438 Shanghai, China}

\author{Zinovia Oikonomopoulou}
\email{zhnobia.oikonomopoulou.21@um.edu.mt}
\affiliation{Institute of Space Sciences and Astronomy, University of Malta, Malta, MSD 2080}

\author{Odil Yunusov}
\email{odilbekhamroev@gmail.com}
\affiliation{National Research University TIIAME, Kori Niyoziy 39, Tashkent 100000, Uzbekistan}
\affiliation{Shahrisabz State Pedagogical Institute, Shahrisabz Str. 10, Shahrisabz 181301, Uzbekistan}

\date{\today}

\begin{abstract}
We present in the form of a catalogue the cosmological perturbations within the Bahamonde-Dialektopoulos-Levi Said (BDLS) theory, which serves as the teleparallel counterpart of Horndeski gravity. To understand structure formation in cosmological models, it is essential to study both the background and perturbative aspects of their cosmology. While extensive analysis of both Horndeski gravity and its teleparallel analog exists in the literature, a quantitative understanding requires a detailed examination of their cosmological perturbations. We review here all the different gauges for the scalar, vector and tensor perturbations of a cosmological background up to second order and we hope this will help people who work with observations, to incorporate it in existing codes.
\end{abstract}

\maketitle

\section{Introduction}

For several deacdes, the $\Lambda$CDM model understood to be the best way to unify cosmological observations at different scales as well as to accurately describe astrophysical phenomena~\cite{Peebles:2002gy,Copeland:2006wr}. For this setting, the cosmological constant ($\Lambda$) drives a late-time accelerated expansion~\cite{Riess:1998cb,Perlmutter:1998np} while gravitational interactions are expressed through general relativity (GR) whereas cold dark matter (CDM) plays a key role in the primordial formation of the seeds of large-scale structure formation in the early Universe and as a stabilizing agent in galactic structures in the late-Universe~\cite{Baudis:2016qwx,Bertone:2004pz}. The $\Lambda$CDM model poses many foundational questions such as the fine-tuning of the cosmological constant~\cite{Weinberg:1988cp}, but many others. Putting aside these important concerns, the concordance model description of the Universe is increasingly coming into tension when observational surveys from the early and late Universe are contrasted~\cite{DiValentino:2020vhf,DiValentino:2020zio,DiValentino:2020vvd,Staicova:2021ajb}. Simultaneously, the prospect of direct observations of CDM appear to be less probably as further measurements continue not report any direct detections~\cite{LUX:2016ggv,Gaitskell:2004gd}.

A large spectrum of competing modifications to the $\Lambda$CDM model have been proposed in the literature together with the search for different physical phenomena that may resolve the most pressing open challenges to the concordance model. The vast battery of alterations of the concordance model include new proposals for the behaviour of CDM~\cite{Feng:2010gw,Dodelson:1993je,Joyce:2014kja,Abazajian:2012ys}, adding dynamical features to dark energy~\cite{Copeland:2006wr,Benisty:2021gde,Benisty:2020otr,Bamba:2012cp}, as well as extensions to the GR description of gravity~\cite{Clifton:2011jh,CANTATA:2021ktz,Bahamonde:2021gfp,AlvesBatista:2021gzc,Addazi:2021xuf,Capozziello:2011et}, among many others. By and large, these models consist of additional layers of complexity on top of the baseline $\Lambda$CDM model. Another approach is to reconsider the foundations of $\Lambda$CDM. Teleparallel gravity (TG) is grounded in the exchange of the curvature associated with the Levi-Civita connection with geometric torsion as the description of gravitational interactions \cite{Bahamonde:2021gfp,Krssak:2018ywd,Cai:2015emx}. This has some advantages in foundational aspects of the theory such as having a well defined gravitational energy momentum tensor \cite{Aldrovandi:2013wha}, among others, and may also offer a vehicle to meet the growing body of observational challenges. 

The novel architecture of teleparallel gravity is curvature-free, satisfies metricity, and features a formulation which is dynamically equivalent to GR called the teleparallel equivalent of general relativity (TEGR) \cite{Maluf:2013gaa,aldrovandi1995introduction}. TEGR differs from GR by a boundary term which plays an important role in modifications of the model, and may also produce new IR limit possible realizations. As in regular curvature-based formulations of gravity, TEGR can be modified in different ways to expose various new physics. The most direct modification of TEGR is known as $f(T)$ gravity \cite{Ferraro:2006jd,Ferraro:2008ey,Bengochea:2008gz,Linder:2010py,Chen:2010va,Bahamonde:2019zea,Paliathanasis:2017htk,Farrugia:2020fcu,Bahamonde:2021srr,Bahamonde:2020bbc,Bahamonde:2022ohm,Bahamonde:2020lsm}, and is a direct generalization of the so-called torsion scalar $T$ to an arbitrary functional of this variable. However, unlike most generic modifications of the gravitational sector, $f(T)$ gravity produces generally second order field equations, and agrees with a widening spectrum of observational phenomena. Building on these toy models, there has also been growing interest in exploring possible scalar-tensor generalizations of the underlying theory in an analogous fashion to Horndeski gravity \cite{Horndeski:1974wa}.

Horndeski gravity encompasses the largest class of second order scalar-tensor models in which only one scalar field is adopted and geometric curvature is retained. However, the constraints brought about by the observation of the gravitational wave event GW170817 \cite{TheLIGOScientific:2017qsa} and its electromagnetic counterpart GRB170817A \cite{Goldstein:2017mmi} has placed severely limiting constraints its most exotic branches of models which was a focal point within the literature \cite{Kobayashi:2019hrl}. Building on this background, a teleparallel gravity formulation of the scalar-tensor Horndeski gravity was proposed in Ref.~\cite{Bahamonde:2019shr}. The speed of gravitational wave constraint was revisited in Ref.~\cite{Bahamonde:2019ipm} where it was found to be circumvented as a limiting factor in the construction of models within the class of theories, where as in Ref.~\cite{Bahamonde:2021dqn} the gravitational wave polarization modes were determined for various subclasses of models. The post-Newtonian parametrization framework was investigated in Ref.~\cite{Bahamonde:2020cfv} where the standard tests were found to be observed for most for the majority of functional models of the theory. While efforts to build models motivated by Noether symmetries were explored in Ref.~\cite{Dialektopoulos:2021ryi} which also contains the full classification of these models. Other features have been probed such as the well-temperating of the class of theories~\cite{Bernardo:2021bsg,Bernardo:2021izq} as well as conditions on the stability of the system~\cite{Capozziello:2023foy} and reconstructed classes of models \cite{Bernardo:2021qhu}.

For models within the class of theories contained in the teleparallel analogue of Horndeski gravity to more robustly be investigated against the latest observational measurements, its cosmological perturbations must be determined. This is also important to fully understand the behavior of these models both in the early Universe as well as through its evolution into the late Universe. This was first investigated in Ref.~\cite{Ahmedov:2023num} where the full calculation is undertaken for a particular gauge choice. Here, the primordial power spectrum and the alpha parametrization of the perturbations is described in detail. In the present work, we explore the gauge-invariant formulation of the cosmological perturbations which is important for exploring different phenomena which have their own natural gauge choices in further works. To facilitate these expressions of the different perturbation equations of motion, we also show the different gauge scenarios for the gauge-invariant cosmological perturbation equations. In Sec.~\ref{sec:Intro_BDLS}, the BDLS technical details are discussed while the cosmological perturbations are presented in Sec.~\ref{sec:cosmological_perturbations}. The expressions are analysed for potential ghost and stability conditions in Sec.~\ref{sec:stability}. The main results are summarized and discussed in Sec.~\ref{sec:conclusion}. Units in which $c = 1$ are assumed unless otherwise stated.

\section{BDLS Gravity: The Teleparallel Analogue of Horndeski Theory} \label{sec:Intro_BDLS}

To begin, we will provide an overview of teleparallel gravity, including its foundational principles and the dynamics that govern its behavior in a cosmological background. This introduction will offer insight into the core concepts of teleparallel gravity and set the stage for understanding its role in cosmological evolution. We will also discuss the basic equations and key assumptions underlying the cosmological dynamics within this framework, establishing a basis for exploring perturbations and structure formation in teleparallel theories.

\subsection{Teleparallel Gravity} \label{tg_foundations}

Gravity theories based on curvature, like General Relativity (GR), are built upon a geometric framework where the Levi-Civitat connection, $\udt{\lc{\Gamma}}{\sigma}{\mu\nu}$ (with over-circles indicating quantities derived from this connection), serves as the foundation for the theory's core geometric structures, such as the Riemann tensor. Consequently, curvature-based gravity models rely heavily on the Levi-Civita connection, as seen in the formulation of the Einstein-Hilbert action via the Ricci scalar. Teleparallel Gravity (TG), however, presents an alternative framework, where the curvature-dependent connection is replaced by the torsion-based connection, $\udt{\Gamma}{\sigma}{\mu\nu}$ \cite{Aldrovandi:2013wha,Bahamonde:2021gfp,Cai:2015emx,Krssak:2018ywd}.

Practically speaking, curvature-based and torsion-based theories of gravity differ significantly in their mathematical frameworks. Curvature-based theories like GR utilize the metric tensor $g_{\mu\nu}$ and its derivatives, whereas TG is formulated using the tetrad $\udt{e}{A}{\mu}$, which defines the system's gravitational variables, along with a flat spin connection $\udt{\omega}{B}{C\nu}$. Here, greek indices represent coordinates on the general manifold, while Latin ones refer to the local Minkowski spacetime. Although, the tetrad and spin connection also appear in GR, their roles are more complex and less practical; in TG, the spin connection serves as an inertial object. The tetrad directly related to the metric tensor by
\begin{align}
    g_{\mu\nu} = \udt{e}{A}{\mu}\udt{e}{B}{\nu} \eta_{AB}\quad {\rm and}\quad \eta_{AB} = \dut{E}{A}{\mu}\dut{E}{B}{\nu} g_{\mu\nu}\,,\label{eq:metr_trans}
\end{align}
where $\dut{E}{A}{\mu}$ represents the inverse tetrad. This relationship shows the flexibility in choosing tetrad components, with the spin connection preserving diffeomorphism invariance across these choices. 

The tetrad-spin connection pair defines the potential components for a given spacetime in TG, enabling the teleparallel connection to be expressed as \cite{Cai:2015emx,Krssak:2018ywd}
\begin{equation}
    \Gamma^{\lambda}{}_{\nu\mu}=\dut{E}{A}{\lambda}\partial_{\mu}\udt{e}{A}{\nu}+\dut{E}{A}{\lambda}\udt{\omega}{A}{B\mu}\udt{e}{B}{\nu}\,,
\end{equation}
where the flatness of the spin connection is ensured by the condition \cite{Bahamonde:2021gfp}
\begin{equation}
    \partial_{[\mu}\udt{\omega}{A}{|B|\nu]} + \udt{\omega}{A}{C[\mu}\udt{\omega}{C}{|B|\nu]} \equiv 0\,.
\end{equation}
There are also specific frames for any spacetime where all spin connection terms vanish for certain tetrad choices. This configuration is known as the Weitzenb\"{o}ck gauge \cite{Weitzenbock:1923efa} and is consistently applied in cases where the spin connection field equations vanish identically for these particular tetrad selections.

In TG, gravitational scalars are constructed by substituting the Levi-Civita connection with its teleparallel counterpart. As a result, the Riemann tensor vanishes identically, $\udt{R}{\alpha}{\beta\gamma\epsilon}(\udt{\Gamma}{\sigma}{\mu\nu}) \equiv 0$, even though the standard Riemann tensor based on the Levi-Civita connection remains nonzero, $\udt{\lc{R}}{\alpha}{\beta\gamma\epsilon}(\udt{\lc{\Gamma}}{\sigma}{\mu\nu}) \neq 0$. To proceed, we introduce a torsion tensor that relies only on the teleparallel connection, defined as \cite{Aldrovandi:2013wha,Ortin:2004ms}
\begin{equation}
    \udt{T}{A}{\mu\nu} := 2\udt{\Gamma}{A}{[\nu\mu]}\,,
\end{equation}
where the square brackets indicate antisymmetrization. This torsion tensor, which serves as the field strength in the theory \cite{Bahamonde:2021gfp} is invariant under both local Lorentz transformations and diffeomorphisms \cite{Krssak:2015oua}.

The torsion tensor in TG can be decomposed into three fundamental irreducible components \cite{Hayashi:1979qx,Bahamonde:2017wwk,Bahamonde:2024zkb}
\begin{align}
    a_{\mu} & :=\frac{1}{6}\epsilon_{\mu\nu\lambda\rho}T^{\nu\lambda\rho}\,,\\[4pt]
    v_{\mu} & :=\udt{T}{\lambda}{\lambda\mu}\,,\\[4pt]
    t_{\lambda\mu\nu} & :=\frac{1}{2}\left(T_{\lambda\mu\nu}+T_{\mu\lambda\nu}\right)+\frac{1}{6}\left(g_{\nu\lambda}v_{\mu}+g_{\nu\mu}v_{\lambda}\right)-\frac{1}{3}g_{\lambda\mu}v_{\nu}\,,
\end{align}
representing the axial, vector and tensor parts respectively. Here, $\epsilon _{\mu\nu\lambda\rho}$ is the totally antisymmetric Levi-Civita tensor in four dimensions. Using this decomposition, distinct gravitational scalar invariants can be defined as follows \cite{Bahamonde:2015zma}
\begin{align}
    T_{\text{ax}} & := a_{\mu}a^{\mu} = -\frac{1}{18}\left(T_{\lambda\mu\nu}T^{\lambda\mu\nu}-2T_{\lambda\mu\nu}T^{\mu\lambda\nu}\right)\,,\\[4pt]
    T_{\text{vec}} & :=v_{\mu}v^{\mu}=\udt{T}{\lambda}{\lambda\mu}\dut{T}{\rho}{\rho\mu}\,,\\[4pt]
    T{_{\text{ten}}} & :=t_{\lambda\mu\nu}t^{\lambda\mu\nu}=\frac{1}{2}\left(T_{\lambda\mu\nu}T^{\lambda\mu\nu}+T_{\lambda\mu\nu}T^{\mu\lambda\nu}\right)-\frac{1}{2}\udt{T}{\lambda}{\lambda\mu}\dut{T}{\rho}{\rho\mu}\,.
\end{align}
These scalars encompass all general non-parity-violating scalar invariants involving up to quadratic contractions of the torsion tensor. 

A particular linear combination of the axial, vector, and purely tensorial scalar invariants yields the torsion scalar, defined as \cite{Bahamonde:2021gfp}
\begin{equation}
    T:=\frac{3}{2}T_{\text{ax}}+\frac{2}{3}T_{\text{ten}}-\frac{2}{3}T{_{\text{vec}}}=\frac{1}{2}\left(E_{A}{}^{\lambda}g^{\rho\mu}E_{B}{}^{\nu}+2E_{B}{}^{\rho}g^{\lambda\mu}E_{A}{}^{\nu}+\frac{1}{2}\eta_{AB}g^{\mu\rho}g^{\nu\lambda}\right)T^{A}{}_{\mu\nu}T^{B}{}_{\rho\lambda}\,.
\end{equation}
The torsion scalar is essential, as it equates to the Ricci scalar up to a total derivative term \cite{Bahamonde:2015zma}
\begin{equation}
    R=\lc{R}+T-\frac{2}{e}\partial_{\mu}\left(e\udut{T}{\lambda}{\lambda}{\mu}\right)=0\,,
\end{equation}
where $R$ represents the Ricci scalar computed with the teleparallel connection, which vanishes as mentioned earlier, and $e=\text{det}\left(\udt{e}{A}{\mu}\right)=\sqrt{-g}$ is the determinant of the tetrad. The conventional Ricci scalar computed from the Levi-Civita connection can then be written as
\begin{equation}
    \lc{R}=-T+\frac{2}{e}\partial_{\mu}\left(e\udut{T}{\lambda}{\lambda}{\mu}\right) := -T+B\,,
\end{equation}
where $B$ represents a total divergence term.

An action based on the linear form of the torsion scalar leads to the Teleparallel Equivalent of GR (TEGR), which is dynamically equivalent to GR \cite{Hehl:1994ue,Aldrovandi:2013wha}. By analogy to gravity theories based on curvature, the TEGR action can be generalized to an $f(T)$ gravity framework \cite{Ferraro:2006jd,Ferraro:2008ey,Bengochea:2008gz,Paliathanasis:2017htk,Linder:2010py,Chen:2010va,Bahamonde:2019zea,Cai:2015emx,Farrugia:2016qqe,Iorio:2012cm,Deng:2018ncg} where the Lagrangian is promoted from $T$ to an arbitrary function of it, $f(T)$. A notable advantage of $f(T)$ gravity is that, unlike its curvature-analog, $f(R)$, the resulting field equations remain second order in derivatives of the tetrad, simplifying thus the equations of motion.

In this work, we explore the teleparallel version of Horndeski gravity, which studies the interactions of the metric with a scalar field. In TG, scalar fields couple to matter in the same manner as in GR through a minimal coupling approach, where the partial derivatives are elevated to Levi-Civita covariant derivatives, i.e. \cite{Aldrovandi:2013wha,BeltranJimenez:2020sih}
\begin{equation}
    \partial_{\mu} \rightarrow \mathring{\nabla}_{\mu}\,,
\end{equation}
which applies only to the matter sector. Building on the minimal coupling prescription, gravitational objects such as the torsion tensor are associated with the teleparallel connection, while scalar and other matter fields are analogously connected with the regular Levi-Civita connection. Through this scheme, teleparallel theories can be built. Based on that, we can examine the recently introduced teleparallel analog of Horndeski gravity\cite{Bahamonde:2019shr,Bahamonde:2019ipm,Bahamonde:2020cfv}, also known as the Bahamonde-Dialektopoulos-Levi Said (BDLS) theory. The formulation of this theory is based on three fundamental conditions: (i) the field equations must maintain a maximum order of second derivatives of the tetrads; (ii) the scalar invariants involved should not violate parity; and (iii) the number of contractions with the torsion tensor must be limited to a maximum of quadratic order. Although, higher-order contractions of the torsion tensor can yield second-order field equations, BDLS theory was intentionally crafted under these criteria, which we adhere in our analysis. 

Due to the second-order nature of various extensions of TG, the resulting action serves as an extension of the conventional Horndeski gravity. Consequently, the conditions outlined above yield both the standard terms of Horndeski gravity and additional terms that are linear in contractions with the torsion tensor \cite{Bahamonde:2019shr}
\begin{equation}
    I_2 = v^{\mu} \phi_{;\mu}\,,
\end{equation}
where $\phi$ denotes the scalar field. Furthermore, we obtain terms that are quadratic in this context, including
\begin{align}
    J_{1} & =a^{\mu}a^{\nu}\phi_{;\mu}\phi_{;\nu}\,,\\[4pt]
    J_{3} & =v_{\sigma}t^{\sigma\mu\nu}\phi_{;\mu}\phi_{;\nu}\,,\\[4pt]
    J_{5} & =t^{\sigma\mu\nu}\dudt{t}{\sigma}{\alpha}{\nu}\phi_{;\mu}\phi_{;\alpha}\,,\\[4pt]
    J_{6} & =t^{\sigma\mu\nu}\dut{t}{\sigma}{\alpha\beta}\phi_{;\mu}\phi_{;\nu}\phi_{;\alpha}\phi_{;\beta}\,,\\[4pt]
    J_{8} & =t^{\sigma\mu\nu}\dut{t}{\sigma\mu}{\alpha}\phi_{;\nu}\phi_{;\alpha}\,,\\[4pt]
    J_{10} & =\udt{\epsilon}{\mu}{\nu\sigma\rho}a^{\nu}t^{\alpha\rho\sigma}\phi_{;\mu}\phi_{;\alpha}\,,
\end{align}
where the semicolons denote covariant derivatives associated with the Levi-Civita connection.

Therefore, we can express the teleparallel analogue of Horndeski gravity in the form of an action, given by
\begin{equation}\label{action}
    \mathcal{S}_{\text{BDLS}} = \int d^4 x\, e\mathcal{L}_{\text{Tele}} + \sum_{i=2}^{5} \int d^4 x\, e\mathcal{L}_i+ \int d^4x \, e\mathcal{L}_{\rm m}\,,
\end{equation}
where the contributions from standard Horndeski gravity remain present as seen below \cite{Horndeski:1974wa}
\begin{align}
    \mathcal{L}_{2} & :=G_{2}(\phi,X)\,,\label{eq:LagrHorn1}\\[4pt]
    \mathcal{L}_{3} & :=-G_{3}(\phi,X)\mathring{\Box}\phi\,,\\[4pt]
    \mathcal{L}_{4} & :=G_{4}(\phi,X)\left(-T+B\right)+G_{4,X}(\phi,X)\left[\left(\mathring{\Box}\phi\right)^{2}-\phi_{;\mu\nu}\phi^{;\mu\nu}\right]\,,\\[4pt]
    \mathcal{L}_{5} & :=G_{5}(\phi,X)\mathring{G}_{\mu\nu}\phi^{;\mu\nu}-\frac{1}{6}G_{5,X}(\phi,X)\left[\left(\mathring{\Box}\phi\right)^{3}+2\dut{\phi}{;\mu}{\nu}\dut{\phi}{;\nu}{\alpha}\dut{\phi}{;\alpha}{\mu}-3\phi_{;\mu\nu}\phi^{;\mu\nu}\,\mathring{\Box}\phi\right]\,.\label{eq:LagrHorn5}
\end{align}
These terms are analogous to their standard Horndeski counterparts but are computed using the teleparallel quantities instead of the metric. Nevertheless, they yield the same contributions to the equations of motion for specific systems. Here we define
\begin{equation}
\label{eq:LTele}
    \mathcal{L}_{\text{Tele}}:= G_{\text{Tele}}\left(\phi,X,T,T_{\text{ax}},T_{\text{vec}},I_2,J_1,J_3,J_5,J_6,J_8,J_{10}\right)\,,
\end{equation}
where the kinetic term is expressed as $X:=-\frac{1}{2}\partial^{\mu}\phi\partial_{\mu}\phi$, $\mathcal{L}_{\rm m}$ represents the matter Lagrangian in the Jordan conformal frame and $\lc{G}_{\mu\nu}$ denoted the regular Einstein tensor. In this notation,  commas signify standard partial derivatives. Notably, when $G_{\text{Tele}} = 0$, we retrieve the original form of Horndeski gravity.

\subsection{Background Cosmology} \label{sec:TG_FLRW}

By varying the action with respect to the tetrad, spin connection, and scalar field, we can derive the field equations as detailed in Ref.~\cite{Bahamonde:2020cfv}. The broader scope of this theory results in significantly more complex field equations. Therefore, we will focus exclusively on the equations of motion relevant to a flat Friedmann-Lema\^{i}tre-Robertson-Walker (FLRW) cosmology, described by the metric
\begin{equation}
    \dd s^2 = -N(t)^2 \dd t^2 + a(t)^2(\dd x^2 + \dd y^2 + \dd z^2)\,,
\end{equation}
where $N(t)$ denotes the lapse function (which we can set to one after obtaining the equations of motion), and $a(t)$ representes the scale factor. To derive the modified equations of motion, we adopt the tetrad choice $\udt{e}{a}{\mu} = \textrm{diag}(N(t),a(t),a(t),a(t))$ which aligns with the Weitzenb\"{o}ck gauge \cite{Krssak:2018ywd,Bahamonde:2021gfp}.

By varying the point-like Lagrangian with respect to the dynamical variables $N(t)$, $a(t)$, and $\phi(t)$, we derive the equations of motion of the system for a flat homogeneous and isotropic background. This leads us to the Friedmann equation
\begin{equation}
    \mathcal{E}_{\rm Tele} + \sum_{i=2}^5 \mathcal{E}_i = 0\,,
\end{equation}
where
\begin{align}
    \mathcal{E}_{\rm Tele} &= 6 H\dot{\phi}\tilde{G}_{6,I_2}+12 H^2 \tilde{G}_{6,T}+2X \tilde{G}_{6,X}-\tilde{G}_{6}\,,\\
    \mathcal{E}_2 &= 2XG_{2,X}-G_2\,,\\
    \mathcal{E}_3 &= 6X\dot\phi HG_{3,X}-2XG_{3,\phi}\,,\\
    \mathcal{E}_4 &= -6H^2G_4+24H^2X(G_{4,X}+XG_{4,XX}) - 12HX\dot\phi G_{4,\phi X}-6H\dot\phi G_{4,\phi }\,,\\
    \mathcal{E}_5 &= 2H^3X\dot\phi\left(5G_{5,X}+2XG_{5,XX}\right) - 6H^2X\left(3G_{5,\phi}+2XG_{5,\phi X}\right)\,,
\end{align}
and
\begin{equation}
    \mathcal{L}_{\rm Tele}=\tilde{G}_6(\phi,X,T,I_{2})\,,
\end{equation}
which encompasses all the nonzero scalars associated with $G_{\rm{Tele}}$. The Hubble parameter is defined as $H = \dot{a}/a$, and dots indicate derivatives with respect to cosmic time. The torsion scalar is expressed as $T = 6H^2$, while $I_2 = 3H\dot{\phi}$ and $X = \frac{1}{2} \dot{\phi}^2$, with commas indicating partial derivatives. Next, varying with respect to the scalar factor leads us to the second Friedmann equation
\begin{equation}
    \mathcal{P}_{\rm Tele}+\sum_{i=2}^5 \mathcal{P}_i=0\,,
\end{equation}
where
\begin{align}
    \mathcal{P}_{\rm Tele}&=-3 H\dot{\phi}\tilde{G}_{6,I_2}-12 H^2\tilde{G}_{6,T}-\frac{d}{dt}\Big(4H \tilde{G}_{6,T}+\dot{\phi}\,\tilde{G}_{6,I_2}\Big)+\tilde{G}_6\,,\\
    \mathcal{P}_2&=G_2\,,\\
    \mathcal{P}_3&=-2X\left(G_{3,\phi}+\ddot\phi G_{3,X} \right) \,,\\
    \mathcal{P}_4&=2\left(3H^2+2\dot H\right) G_4 - 12 H^2 XG_{4,X}-4H\dot X G_{4,X} - 8\dot HXG_{4,X}\nonumber\\
    & \phantom{gggg}-8HX\dot X G_{4,XX} +2\left(\ddot\phi+2H\dot\phi\right) G_{4,\phi} + 4XG_{4,\phi\phi} + 4X\left(\ddot\phi-2H\dot\phi\right) G_{4,\phi X}\,,\\
    \mathcal{P}_5&=-2X\left(2H^3\dot\phi+2H\dot H\dot\phi+3H^2\ddot\phi\right)G_{5,X} - 4H^2X^2\ddot\phi G_{5,XX}\nonumber\\
    & \phantom{gggg} +4HX\left(\dot X-HX\right)G_{5,\phi X} + 2\left[2\frac{d}{dt}\left(HX\right)+3H^2X\right]G_{5,\phi} + 4HX\dot\phi G_{5,\phi\phi}\,.
\end{align}
Finally, the modified Klein-Gordon equation can be obtained by varying with respect to the scalar field, resulting in 
\begin{equation}
    \frac{1}{a^3}\frac{\dd}{\dd t}\Big[a^3 (J+J_{\rm Tele})\Big]=P_{\phi}+P_{\rm Tele}\,,
\end{equation}
where the standard Horndeski terms appear as $J$ and $P_{\phi}$, arising from the Lagrangian terms $\mathcal{L}_i$, with $i=2,..,5$ as shown in \cite{Kobayashi:2011nu}
\begin{align}
    J &= \dot\phi G_{2,X} +6HXG_{3,X}-2\dot\phi G_{3,\phi} + 6H^2\dot\phi\left(G_{4,X}+2XG_{4,XX}\right)-12HXG_{4,\phi X}\nonumber\\
    & \phantom{gggggggg} + 2H^3 X\left(3G_{5,X}+2XG_{5,XX}\right) - 6H^2\dot\phi\left(G_{5,\phi}+XG_{5,\phi X}\right)\,,\\
    P_{\phi} &= G_{2,\phi} -2X\left(G_{3,\phi\phi}+\ddot\phi G_{3,\phi X}\right) + 6\left(2H^2+\dot H\right)G_{4,\phi} \nonumber\\
    & \phantom{gggggii} + 6H\left(\dot X+2HX\right)G_{4,\phi X} -6H^2XG_{5,\phi\phi}+2H^3X\dot\phi G_{5,\phi X}\,,
\end{align}
where $J_{\rm Tele}$ and $P_{\rm Tele}$ denote additional terms associated with the teleparallel Horndeski, given by
\begin{align}
    J_{\rm Tele} &= \dot{\phi}\tilde{G}_{6,X}\,,\\
    P_{\rm Tele} &= -9 H^2\tilde{G}_{6,I_2}+\tilde{G}_{6,\phi}-3  \frac{\dd}{\dd t}\left(H\tilde{G}_{6,I_2}\right)\,.
\end{align}
Notably, the parameters of \(\tilde{G}_6\) are independent of \(T_{\rm ax}\) and \(T_{\rm ten}\), as these values are zero for the flat FLRW metric. Consequently, we can express the contributions solely in terms of \(T_{\rm vec}\), given that \(T = -(2/3)T_{\rm vec} = 6H^2\) in this scenario. It is important to recognize that this relationship may not hold true at the perturbative level \cite{Bahamonde:2019ipm}. In this study, we investigate the cosmological perturbations within the teleparallel analogue of Horndeski gravity. This approach will facilitate more thorough explorations of specific models within the framework, potentially enhancing our understanding of which models best align with observations regarding the cosmic evolution of the Universe.

\section{Cosmological Perturbations}\label{sec:cosmological_perturbations}

We will now calculate the perturbations around a FLRW spacetime up to second order for the BDLS theory, which is given by the Lagrangian~\eqref{action}. Since the dynamical variable in our formulation of the theory is the tetrad, we will perturb it as
\begin{equation}
    e^A{}_\mu=\bar{e}^A{}_\mu+\delta e^A{}_\mu\,,\label{per}
\end{equation}
where $\bar{e}^A{}_\mu$ is the background FLRW diagonal tetrad, i.e. 
\begin{equation}
    \bar{e}^A{}_\mu = {\rm diag}[N(t),a(t),a(t),a(t)]\,,
\end{equation}
and $\delta e^A{}_\mu$ is the tetrad perturbation. We decompose it according to the group of spatial rotations and thus it is described by
\begin{align} \label{eq:tetrad_pert}
    \udt{\delta e}{A}{\mu} = \begin{pmatrix}
        \Phi & a \left(\partial_{i}\beta + u_{i}\right)\\
        \delta^{A}_{i} \left(\partial^{i} B + v^i \right) & a \delta^{A j} \left(\delta_{i j} \psi + \partial_{i}\partial_{j}E + 2 \partial_{(i} w_{j)} +\frac{1}{2} h_{ij} + \epsilon_{ijk} ( \partial^{k}\sigma + V^k )\right)
    \end{pmatrix}
\end{align}
The 16 components of this field are split into one traceless and transverse tensor $h_{ij}$, three traceless vectors $u_i,v^i$ and $w_i$, one traceless pseudovector $V^k$ and five scalars $\Phi, \psi, B, \beta$ and $E$ as well as one pseudoscalar $\sigma$. One can easily find that at the linearized level, the (pseudo)scalar, (pseudo)vector, and tensor perturbations decouple. In the following subsections, we will treat these modes separately, and study the degrees of freedom (DoFs) of the theory, but before that, let's see how the perturbative variables transform under diffeomorphisms.

\subsection{Gauge symmetry}
The BDLS action \eqref{action} is invariant under infinitesimal coordinate transformations \begin{equation}\label{eq:gauge-transformation}
x^\mu \rightarrow x'^{\mu} = x^\mu + \xi ^\mu (x).    
\end{equation} 
Under these transformations the tetrad becomes
$$e^A{}_\mu \rightarrow e'^A{}_\mu = e^A{}_\nu \frac{\partial x^\nu}{\partial x'^\mu}.$$ We decompose the transformation vector as $$\xi^{\mu} = \left[\xi^{0}, \frac{1}{a} \big(\xi^{i} + \delta^{ij} \partial_{j}\xi\big) \right],$$ with $\xi ^0$ and $\xi$ being scalars and $\xi ^i$ a divergenceless vector. The perturbative variables on FLRW transform as
\begin{subequations}
\label{eq:gauge_invariant_calculations}
\begin{align}
    \tilde{\sigma} &= \sigma\,, &&&\\
    \tilde{\delta\phi} &= \delta\phi + \dot{\phi} \xi^{0}\,, &&&  \\
    \tilde{\Phi} &= \Phi + \dot{\xi}^{0}\,, &&&\\
    \tilde{\psi} &= \psi + H \xi^{0}\,, &\tilde{V}^{k} &= V^{k} + \frac{1}{2a} \epsilon^{ijk} \partial_{i}\xi_{j}\,, &&\\
    \tilde{\beta} &= \beta + \frac{1}{a} \xi^{0}\,, & \tilde{u}_{i} &= u_{i}\,, &&\\
     \tilde{B} &= B - H \xi + \dot{\xi}\,, & \tilde{v}_{i} &= v_{i} - H \xi_{i} + \dot{\xi}_{i}\,, &&\\
    \tilde{E} &= E + \frac{1}{a} \xi\,, & \tilde{w}_{i} &= w_{i} + \frac{1}{2a} \xi_{i}\,, & \tilde{h}_{ij} &= h_{ij}\,,
\end{align}
\end{subequations}
In what follows, we present the perturbative analysis of the different modes separately.

\subsection{Scalars}

Regarding the scalar perturbations, apart from the five scalar modes discussed above, we have an additional one coming from the perturbation of the scalar field. Specifically, we have $$\phi = \phi_0 + \delta \phi .$$ The BDLS action~\eqref{action} up to second order terms becomes
\begin{align} \label{eq:scalar-quadratic-gauge-dependent-action}
    \mathcal{S}_{\text{S}} = \int \text{d}t \text{d}^3 x a^3 \Bigg[&
    \mathcal{A}_{1} \left(\frac{\nabla^{2}(B - a \dot{E})}{a^2}\right)^2 + \frac{\left( B - a \dot{E}\right)}{a^2}\left(\mathcal{A}_{2} \nabla^2 \Phi + \mathcal{A}_{3} \nabla^2 \dot{\psi} + \mathcal{A}_{4} \nabla^2 \delta\phi + \mathcal{A}_{5} \nabla^2 \dot{\delta\phi}\right) 
    \nonumber \\ & + \frac{\beta}{a^2} \left( \mathcal{A}_{6} \nabla^2 \Phi + \mathcal{A}_{8} \nabla^2 \delta\phi + \mathcal{A}_{9} \nabla^2 \dot{\delta\phi} + \mathcal{A}_{10} \nabla^2 \psi + \mathcal{A}_{12} \nabla^2 \dot{\psi} \right) + \frac{\dot{\beta}}{a^2} \left(\mathcal{A}_{7} \nabla^2 \Phi + \mathcal{A}_{11} \nabla^{2}\psi \right)  
    \nonumber \\ & + \mathcal{A}_{13} \left(\frac{\nabla \dot{\beta}}{a}\right)^2 + \Phi \left(\mathcal{A}_{14} \frac{\nabla^2 \psi}{a^2} + \mathcal{A}_{15} \dot{\psi} + \mathcal{A}_{16} \delta\phi + \mathcal{A}_{17} \frac{\nabla^2 \delta\phi}{a^2} + \mathcal{A}_{18} \dot{\delta\phi}\right) + \mathcal{A}_{19} \left( \frac{\nabla \Phi}{a}\right)^2
    \nonumber \\ & 
    + \mathcal{A}_{20}  \Phi^2 + \mathcal{A}_{21} \left( \frac{\nabla \psi}{a}\right)^2 + \mathcal{A}_{22} \dot{\psi}^2 + \mathcal{A}_{23} \frac{\nabla \psi \nabla \delta\phi}{a^2} + \mathcal{A}_{24} \delta\phi \dot{\psi} + \mathcal{A}_{25} \dot{\delta\phi} \dot{\psi} + \mathcal{A}_{26} \left( \frac{\nabla \delta\phi}{a}\right)^2 
    \nonumber \\ & 
    + \mathcal{A}_{27} \delta\phi^2 + \mathcal{A}_{28} \dot{\delta\phi}^2
    \Bigg] \,,
\end{align}
where all the coefficients $\mathcal{A}_1$ to $\mathcal{A}_{28}$ are presented in the appendix \ref{appendix_coefficients}. Note that further integration by parts would reduce the number of terms present in the action, but for simplicity this is omitted. Coefficients $\{\mathcal{A}_{1}, \mathcal{A}_{7}, \mathcal{A}_{10}, \mathcal{A}_{11}, \mathcal{A}_{13}, \mathcal{A}_{19}\}$ are purely teleparallel and would completely vanish in the standard Horndeski limit. 

As already mentioned above, we have the gauge freedom to choose one element out of each set $\{\delta\phi, \Phi,\beta,\psi\}$ and $\{B,E\}$ to be set to zero. From the above action, it is easy to obtain the results expressed in different gauge choices. In this section we will work with gauge invariant variables, but in the Appendix \ref{appendix_gauge_choices} we will present the same results in flat, unitary, Newtonian and synchronous gauges. In that way, people who want to emphasize on different things can choose any gauge they want to work with.

Since $\{\delta\phi, \Phi,\beta,\psi,B,E\}$ are gauge dependent under \eqref{eq:gauge-transformation}, we define
\begin{subequations}\label{eq:gauge_invariant_quantities}
    \begin{align}
    \mathcal{X}_{1} &= \delta\phi - a \dot{\phi_0}\beta \,, \\
    \mathcal{X}_{2} &= \Phi - a (H \beta + \dot{\beta}) \,, \\
    \mathcal{X}_{3} &= \psi - a H \beta\,, \\
    \mathcal{X}_{4} &= -B + a \dot{E}\,, 
\end{align}
\end{subequations}
which are gauge invariant. Substituting these in \eqref{eq:scalar-quadratic-gauge-dependent-action} we get
\begin{align}\label{eq:scalar_gauge_invariant}
    \mathcal{S}_{\text{S}} = \int \text{d}t \, \text{d}^3x \, a^3 \Bigg[& 
    \tilde{\mathcal{A}}_{1} \, \mathcal{X}_{2}^2 + \tilde{\mathcal{A}}_{2} \, \left(\frac{\nabla \mathcal{X}_{2}}{a}\right)^2 + \tilde{\mathcal{A}}_{3} \, \left(\frac{\nabla \mathcal{X}_{3}}{a}\right)^2 + \tilde{\mathcal{A}}_{4} \dot{\mathcal{X}}_{3}^2 \, + \tilde{\mathcal{A}}_{5} \, \mathcal{X}_{1}^2 + \tilde{\mathcal{A}}_{6} \, \dot{\mathcal{X}}_{1}^{2} + \tilde{\mathcal{A}}_{7} \, \left(\frac{\nabla \mathcal{X}_{1}}{a}\right)^2  
    \nonumber \\ & 
    + \tilde{\mathcal{A}}_{8} \, \left(\frac{\nabla^2 \mathcal{X}_{4}}{a^2}\right)^2 + \tilde{\mathcal{A}}_{9} \mathcal{X}_{2}\dot{\mathcal{X}}_{3} + \tilde{\mathcal{A}}_{10} \, \frac{\nabla \mathcal{X}_{2} \nabla \mathcal{X}_{3}}{a^2} + \tilde{\mathcal{A}}_{11} \, \mathcal{X}_{2} \mathcal{X}_{1} + \tilde{\mathcal{A}}_{12} \, \mathcal{X}_{2} \dot{\mathcal{X}}_{1} + \tilde{\mathcal{A}}_{13} \, \frac{\nabla \mathcal{X}_{2} \nabla \mathcal{X}_{1}}{a^2} 
    \nonumber \\ &
    + \tilde{\mathcal{A}}_{15} \, \dot{\mathcal{X}}_{3} \mathcal{X}_{1} + \tilde{\mathcal{A}}_{16} \, \dot{\mathcal{X}}_{3} \dot{\mathcal{X}}_{1} + \tilde{\mathcal{A}}_{17} \, \frac{\nabla \mathcal{X}_{3} \nabla\mathcal{X}_{1}}{a^2} + \frac{\nabla^2\mathcal{X}_{4}}{a^2} (\tilde{\mathcal{A}}_{14} \, \mathcal{X}_{2} + \tilde{\mathcal{A}}_{18} \, \dot{\mathcal{X}}_{3} + \tilde{\mathcal{A}}_{19} \, \mathcal{X}_{1} + \tilde{\mathcal{A}}_{20} \, \dot{\mathcal{X}}_{1})
    \Bigg]\,,
\end{align}
where again, the coefficients $\tilde{\mathcal{A}}_1 - \tilde{\mathcal{A}}_{20}$ are presented in the appendix~\ref{appendix_coefficients}. We notice that not all the scalar modes are propagating and specifically $\mathcal{X}_2$ and $\mathcal{X}_4$ are auxiliary. In order to find the action with the propagating modes only, we need to vary the above action with respect to the auxiliary fields, find the constraint equations and substitute them back to the action. This will be done in the next section \ref{sec:stability}.

\subsection{Pseudoscalar perturbations}
Moving on to the pseudoscalar mode $\sigma$ and perturbing the BDLS action up to second order, we get
\begin{equation}\label{eq:pseudoscalar_action}
    \mathcal{S}_{\rm PS} = \int \dd t \dd ^3 x \frac{a^3}{9}\left[ \mathcal{B}_{1}(\partial _i \dot{\sigma})^2 + \mathcal{B}_{2} (\Delta \sigma )^2\right]\, \,
\end{equation}
where
\begin{subequations}
\begin{align}
    \mathcal{B}_{1} &= 4 G_{{\rm Tele},T_{\rm ax}} + 9 X G_{{\rm Tele},{J_5}} - 12 X G_{{\rm Tele},J_{10}}\,,\\
    \mathcal{B}_{2} &= \frac{4}{a^2} \left( 2 X G_{{\rm Tele},J_1} - G_{{\rm Tele},T_{\rm ax}}\right)\,.
\end{align}
\end{subequations}
The pseudoscalar in the tetrad, introduced to account for changes in sign under parity transformations, only contributes to the antisymmetric part of the tetrad. It can be seen that its contribution vanishes when calculating the symmetric metric, which is further highlighted in the action~\eqref{eq:pseudoscalar_action}, where all coefficients are from the teleparallel sector.

In some classes of teleparallel theories, like Type I, II and V New General Relativity (NGR) the pseudoscalar mode propagates \cite{Bahamonde:2024zkb}, while in others, like $f(T)$ it does not because of the presence of remnant symmetries \cite{Izumi:2012qj,Ferraro:2014owa}. Since both NGR and $f(T)$ theories are subclasses of BDLS we cannot extract a unique result for all subclasses regarding the propagation of the pseudoscalar mode. However, in case they're propagating, in order for them to be ghost-free, the condition
\begin{equation}\label{eq:pseudoscalar-stability-condition}
    \mathcal{B}_1 > 0\,,
\end{equation}
has to be satisfied.

\subsection{Vectors}

Let us now study the vector modes. In the curvature-based theory, the vector sector decays immediately. Here, where the fundamental dynamical object is the tetrad, the same perturbation procedure is applied to the vector sector, yielding

\begin{align}\label{eq:vector_action}
    \mathcal{S}_{\text{V}} = \int \text{d}t\, \text{d}^3x\, a^{3} \bigg[&
    \mathcal{C}_{1} \frac{(\nabla \textbf{\textit{v}})^2}{a^2} + \mathcal{C}_{2} \frac{(\Delta \textbf{w})^2}{a^2} + \mathcal{C}_{3} \, (\nabla\dot{\textbf{w}})^2 + \mathcal{C}_{2} \frac{(\nabla \textbf{V})^2}{a^2} + \mathcal{C}_{4} \, \dot{\textbf{V}}^2 + \mathcal{C}_{5} \, \dot{\textbf{u}}^2+ \mathcal{C}_{6} \frac{(\nabla \textbf{u})^2}{a^2} - \mathcal{C}_{3}  \frac{(\nabla \textbf{\textit{v}})(\nabla\dot{\textbf{w}})}{a}
    \nonumber \\ & 
     + \mathcal{C}_{7} \frac{(\nabla \textbf{\textit{v}})(\nabla \textbf{u})}{a^2} + \mathcal{C}_{8}  \frac{(\nabla \textbf{u})(\nabla \textbf{w})}{a} + \mathcal{C}_{9}  \frac{(\nabla \textbf{u})(\nabla \dot{\textbf{w}})}{a}+ \mathcal{C}_{10}  \frac{(\nabla\dot{\textbf{u}})(\nabla \textbf{w})}{a}  +2 \mathcal{C}_{2} \, \frac{(\nabla\times \textbf{V})\,\Delta \textbf{w}}{a^2}
    \nonumber \\ &
    + \frac{1}{a} \bigg( \mathcal{C}_{4} \, \dot{\textbf{V}} \, (\nabla\times \textbf{\textit{v}}) + \mathcal{C}_{8}\, \textbf{V}\, (\nabla\times \textbf{u})+ \mathcal{C}_{10}\, \textbf{V} \, (\nabla\times\dot{\textbf{u}})  + \mathcal{C}_{11} \dot{\textbf{V}} \, (\nabla\times \textbf{u}) \bigg)\bigg]\,,
\end{align}
where as previously, the coefficients $\mathcal{C}_1 - \mathcal{C}_11$ are shown in the appendix. 

Note that the pseudovector modes couple with the vector ones already at linear order (in constrast with the rest of the modes which couple at second order and beyond) and thus cannot be treated separately\footnote{In principle, one could set the pseudovectors to zero as a gauge choice, since one of the $\{v_i,u_i, V_i\}$ can vanish. This will be studied in the next section; here we want to study gauge invariant variables.} . While non-teleparallel coefficients appear in the action, further integration by parts would show that any remaining contributions correspond to a non-dynamical mode, which in this case is $v_{i}$. By fixing $v_{i}$, $w_{i}$ or $V_{i}$ to zero we can obtain a gauge fixed result (see Appendix~\ref{appendix_gauge_choices} for further details). Nonetheless, the gauge invariant route is opted. 

We define the following gauge invariant variables \begin{subequations}\label{eq:gauge_invariant_vectors}
    \begin{align}
    \mathcal{Y}_{i} &= v_{i} - 2 a \dot{w}_{i} \,, \\
    \mathcal{Z}_{i} &= V_{i} + \epsilon_{ijk} \partial^{j}w^k\,.
\end{align}
\end{subequations} 
and the action \eqref{eq:vector_action} takes the form
\begin{align}    \label{eq:vector action2}
\mathcal{S}_{\text{V}} = \int \text{d}t\, \text{d}^3x\, a^{3} \bigg[& \tilde{\mathcal{C}}_1 \, \dot{\textbf{u}}^2 \,  + \tilde{\mathcal{C}}_2 \frac{(\nabla \textbf{u})^2}{a^2} + \tilde{\mathcal{C}}_3 \frac{\mathcal{\textbf{Z}} \, (\nabla \times\textbf{u})}{a} +\tilde{\mathcal{C}}_4 \frac{(\nabla \textbf{u}) \, (\nabla\mathcal{\textbf{Y}})}{a^2} - \tilde{\mathcal{C}}_5 \frac{(\nabla\times\dot{\textbf{u}}) \, \mathcal{\textbf{Z}}}{a}
\nonumber \\ & 
+ \tilde{\mathcal{C}}_6 \frac{\dot{\mathcal{\textbf{Z}}}\, (\nabla\times \mathcal{\textbf{Y})}}{a} + \tilde{\mathcal{C}}_7\,\dot{\mathcal{\textbf{Z}}}^2 + \tilde{\mathcal{C}}_8 \frac{(\nabla\mathcal{\textbf{Z}})^2}{a^2} + \tilde{\mathcal{C}}_9 \frac{(\nabla\mathcal{\textbf{Y}})^2}{a^2} \bigg]
\end{align}
where again the coefficients are shown in the appendix \ref{appendix_coefficients}. The non-teleparallel terms only appear in $\{\tilde{\mathcal{C}}_{2}, \tilde{\mathcal{C}}_{4}, \tilde{\mathcal{C}}_{9}\}$ such that if the Horndeski limit is applied to the system, all modes would be non-dynamical and give a trivial solution, as expected. However, as can be seen from the action \eqref{eq:vector action2}, not all modes are propagating; this will be studied in the next section \ref{sec:stability}.

\subsection{Tensor perturbations}

The quadratic action of the tensor modes reads
\begin{equation} 
    \mathcal{S}_{\rm T}= \int \dd t \dd ^3 x \, \,\frac{a^3}{4} \left[\,\mathcal{D}_1 \dot{h}_{ij}^2 - \frac{\mathcal{D}_2}{a^2}(\bm{\nabla} h_{ij})^2\,\right]\,,
    \label{eq:ten_pert_action}
\end{equation}
where
\begin{subequations}
\begin{align}
    \mathcal{D}_{1} &= 2 \left(G_4-2 X G_{4,X}+ X G_{5,\phi_0}-H X \dot{\phi_0} G_{5,X} + 2 X G_{{\rm Tele}, J_8}+\frac{X}{2} G_{{\rm Tele}, J_5} -  G_{{\rm Tele},T}\right)\,,
    \label{scalarcoef1} \\
    \mathcal{D}_{2} &= 2 \left(G_4-X G_{5,\phi_0}-X \ddot{\phi_0} G_{5,X}-G_{{\rm Tele},T} \right)\,,
    \label{scalarcoef2}
\end{align}
\end{subequations}
where $G_{i,A}$ denotes the derivative of $G_i$ with respect to $A$. As expected, when $G_{\rm Tele} \rightarrow 0$, Eqs.~\eqref{scalarcoef1} and \eqref{scalarcoef2} take the standard Horndeski form for the perturbed action, as presented in~\cite{Kobayashi:2011nu}.

Variation of the action~\eqref{eq:ten_pert_action} yields the propagation equation of gravitational waves, which confirms the result in Ref.~\cite{Bahamonde:2019ipm}. The standard $\alpha_i$ parametrization of cosmological perturbations is presented in Ref.~\cite{Ahmedov:2023num}. The squared speed of the tensor modes reads
\begin{equation}
    c_{\rm T} ^2 = \frac{\mathcal{D}_2}{\mathcal{D}_1}\,,
\end{equation}
which in general, is not equal to unity. Furthermore, as will be discussed in Sec.~\ref{sec:stability}, we can deduce from the action \eqref{eq:ten_pert_action} that the conditions
\begin{equation}
    \mathcal{D}_1 > 0 \quad \text{and} \quad \mathcal{D}_2 > 0 \, ,
    \label{tensorstability}
\end{equation}
must hold in order to avoid ghost and gradient instabilities.

\section{Ghost and Laplacian Stability Conditions}
\label{sec:stability}

In the previous section we presented the quadratic actions for all the modes of the perturbations, tensors, pseudoscalar, scalars and vectors. The tensor modes are two, since $h_{ij}$ is transverse and traceless, and in order for them to propagate healthily, conditions \eqref{tensorstability} must be satisfied. Similarly for the pseudoscalar mode condition \eqref{eq:pseudoscalar-stability-condition} should hold. In this section, we study first of all which are the scalar and vector modes which propagate and second which are the stability conditions for them to be healthy. 

Ghosts represent extra degrees of freedom that may propagate due to a negative kinetic term, these instabilities can have a detrimental effect on the health of the underlying theory. In order to identify the possible appearance of ghosts, the action is expanded up to second order about the FLRW background. In Fourier space, the auxiliary modes are determined and eliminated by appropriate variations of the action with respect to those non-dynamical modes. In practice the auxiliary field variations are substituted back into the original system and thus removed. In this way, only dynamical modes are retained in the resulting action. This is performed separately for the scalar, vector, and tensor sectors. 

Firstly, a gauge invariant action is formulated by taking its variations with respect to spatial and temporal parts of an arbitrary vector field transformation, and imposing that this vanishes. In turn, a diagonalized kinetic matrix $\mathbf{K}$ can be produced wherein a constraint can be associated with every entry. Due to the time dependence of the background spacetime, these constraints may also express some time-dependence. At this point the gradient or Laplacian instability can be defined as the imposition that the speed of propagation of these modes is positive definite. In this way, both the ghost and gradient instabilities are confirmed to be stable. Due to the high energy regime being the most likely region where ghosts appear, a high $k$ limit is assumed throughout the procedure.

The general approach rests on an action perturbed up to second order where non-dynamical modes are identified. Once these modes are nullified and then substituted back into the original action, it will take on a form
\begin{align}\label{eq:stability_action_form}
    \mathcal{S} =  \frac{1}{2\kappa^2} \int \text{d}t \frac{\text{d}^{3}k}{(2 \pi)^{3}} \, a^3 \left[ \dot{\vec{\chi}}^{\intercal} \, \mathbf{K}(t,k) \, \dot{\vec{\chi}}  + \vec{\chi}^{\intercal} \, \left(\frac{k^{2}}{a^2} \mathbf{G}(t,k) + \mathbf{M}(t,k) \right) \vec{\chi} + \vec{\chi}^{\intercal} \mathbf{Q}(t,k) \dot{\vec{\chi}} \right]\,,
\end{align}
where $\vec{\chi}$ is a vector of modes, $\mathbf{G}$ is the gradient matrix, $\mathbf{M}$ is the matrix corresponding to non-dynamical portions of the modes, and $\mathbf{Q}$ is an imaginary matrix. For the high-$k$ limit, ghost modes correspond to entries of $\mathbf{K}$, while positive definite propagation speeds correspond to gradient instability corrections.

\subsection{Scalars}

The scalar and pseudoscalar modes do not suffer any mixing and so can be treated separately. This approach is identical to the Arnowitt-Deser-Misner (ADM) decomposition used in other torsion based treatments \cite{Capozziello:2023foy,Gonzalez-Espinoza:2021mwr}, and similarly leads to the same metric as curvature-based formulations. Taking the perturbation up to the second order followed by integration by parts and the insertion of the background equations of motion, the gauge invariant action \eqref{eq:scalar_gauge_invariant} produces the Fourier transformed form given by
\begin{align}\label{eq:scalar_gauge_invariant_fourier}
    \mathcal{S}_{\text{S}} = \int \text{d}t \, \frac{\text{d}^3k}{(2\pi)^{\frac{3}{2}}} \, a^3 \Bigg[& 
    \tilde{\mathcal{A}}_{6} \, \dot{\mathcal{X}}_{1}^{2} + \tilde{\mathcal{A}}_{4} \, \dot{\mathcal{X}}_{3}^2
    + \frac{k^2}{a^2} \bigg(\frac{k^2}{a^2} \tilde{\mathcal{A}}_{8} \, \mathcal{X}_{4}^2 + \tilde{\mathcal{A}}_{2} \, \mathcal{X}_{2}^2 + \tilde{\mathcal{A}}_{3} \, \mathcal{X}_{3}^2 + \tilde{\mathcal{A}}_{7} \, \mathcal{X}_{1}^2 + \tilde{\mathcal{A}}_{10} \, \mathcal{X}_{2} \mathcal{X}_{3} + \tilde{\mathcal{A}}_{13} \mathcal{X}_{2} \mathcal{X}_{1}  
    \nonumber \\ & \quad 
    + \tilde{\mathcal{A}}_{17} \, \mathcal{X}_{3} \mathcal{X}_{1} - \mathcal{X}_{4} (\tilde{\mathcal{A}}_{14} \, \mathcal{X}_{2} + \tilde{\mathcal{A}}_{18} \, \dot{\mathcal{X}}_{3} + \tilde{\mathcal{A}}_{19} \, \mathcal{X}_{1} + \tilde{\mathcal{A}}_{20} \, \dot{\mathcal{X}}_{1}) \bigg)
    + \tilde{\mathcal{A}}_{9} \mathcal{X}_{2}\dot{\mathcal{X}}_{3} + \tilde{\mathcal{A}}_{12} \, \mathcal{X}_{2} \dot{\mathcal{X}}_{1} 
    \nonumber \\ & \quad + \tilde{\mathcal{A}}_{15} \, \dot{\mathcal{X}}_{3} \mathcal{X}_{1} + \tilde{\mathcal{A}}_{16} \, \dot{\mathcal{X}}_{3} \dot{\mathcal{X}}_{1}
    + \tilde{\mathcal{A}}_{1} \, \mathcal{X}_{2}^2 + \tilde{\mathcal{A}}_{5} \, \mathcal{X}_{1}^2 + \tilde{\mathcal{A}}_{11} \, \mathcal{X}_{2} \mathcal{X}_{1} \Bigg]\,,
\end{align}
where $\mathcal{X}_{2}$ and $\mathcal{X}_{4}$ are non-dynamical modes. Taking variation with respect these two quantities results in the respective relationships
\begin{align}
    0 &= \frac{k^2}{a^2}\left(2 \tilde{\mathcal{A}}_{2} \mathcal{X}_{2} + \tilde{\mathcal{A}}_{10} \mathcal{X}_{3} + \tilde{\mathcal{A}}_{13} \mathcal{X}_{1} - \tilde{\mathcal{A}}_{14}\mathcal{X}_{4} \right) + \tilde{\mathcal{A}}_{9} \dot{\mathcal{X}}_{3} + \tilde{\mathcal{A}}_{12} \dot{\mathcal{X}}_{1} + 2 \tilde{\mathcal{A}}_{1} \mathcal{X}_{2} + \tilde{\mathcal{A}}_{11} \mathcal{X}_{1}\,, \\
    0 &= 2 \frac{k^2}{a^2} \tilde{\mathcal{A}}_{8} \mathcal{X}_{4} - \tilde{\mathcal{A}}_{19} \mathcal{X}_{1} - \tilde{\mathcal{A}}_{14} \mathcal{X}_{2} - \tilde{\mathcal{A}}_{20} \dot{\mathcal{X}}_{1} - \tilde{\mathcal{A}}_{18} \dot{\mathcal{X}}_{3}\,\,,
\end{align}
which when substituted back in the action~\eqref{eq:scalar_gauge_invariant_fourier} gives
\begin{align}\label{eq:scalar_action_minimised}
     \mathcal{S}_{\text{S}} &= \frac{1}{2\kappa^2}\int \text{d}t \, \frac{\text{d}^3k}{(2\pi)^{\frac{3}{2}}} \, a^3 
    \left[
    \hat{\mathcal{A}}_{1} \mathcal{X}_{1}^2 + \hat{\mathcal{A}}_{2} \mathcal{X}_{1} \dot{\mathcal{X}}_{1} + \hat{\mathcal{A}}_{3} \dot{\mathcal{X}}_{1}^2 + \hat{\mathcal{A}}_{4} \mathcal{X}_{3}^2 + \hat{\mathcal{A}}_{5} \mathcal{X}_{3} \dot{\mathcal{X}}_{3} + \hat{\mathcal{A}}_{6} \dot{\mathcal{X}}_{3}^2 \right.
    \nonumber \\ & \qquad\qquad\qquad\qquad\qquad \left.
    + \hat{\mathcal{A}}_{7} \mathcal{X}_{1} \mathcal{X}_{3} + \hat{\mathcal{A}}_{8} \dot{\mathcal{X}}_{1} \mathcal{X}_{3} + \hat{\mathcal{A}}_{9} \mathcal{X}_{1} \dot{\mathcal{X}}_{3} + \hat{\mathcal{A}}_{10} \dot{\mathcal{X}}_{1} \dot{\mathcal{X}}_{3} 
    \right] \,,
\end{align}
where
\begin{subequations}
\allowdisplaybreaks
\begin{align}
    \tilde{a}\hat{\mathcal{A}}_{1} &= \tilde{\mathcal{A}}_{11} \tilde{\mathcal{A}}_{14}\tilde{\mathcal{A}}_{19}  - \tilde{\mathcal{A}}_{1} \tilde{\mathcal{A}}_{19}^2 - \tilde{\mathcal{A}}_{5}\tilde{\mathcal{A}}_{14}^2 - \tilde{\mathcal{A}}_{8} \left(\tilde{\mathcal{A}}_{11}^2 - 4 \tilde{\mathcal{A}}_{1} \tilde{\mathcal{A}}_{5}\right) + \frac{k^2}{a^2} \left(\tilde{\mathcal{A}}_{13} \tilde{\mathcal{A}}_{14}\tilde{\mathcal{A}}_{19} - \tilde{\mathcal{A}}_{2} \tilde{\mathcal{A}}_{19}^2  \right.
    \nonumber \\ & \quad \left.
    - \tilde{\mathcal{A}}_{7} \tilde{\mathcal{A}}_{14}^2 + 2 \tilde{\mathcal{A}}_{8} \left(-\tilde{\mathcal{A}}_{11} \tilde{\mathcal{A}}_{13} + 2 \tilde{\mathcal{A}}_{2} \tilde{\mathcal{A}}_{5} + 2 \tilde{\mathcal{A}}_{1} \tilde{\mathcal{A}}_{7}\right) \right) - \frac{k^4}{a^4} \tilde{\mathcal{A}}_{8}\left(\tilde{\mathcal{A}}_{13}^2 - 4 \tilde{\mathcal{A}}_{2} \tilde{\mathcal{A}}_{7}\right) \,, \\
    \tilde{a}\hat{\mathcal{A}}_{2} &= -\tilde{\mathcal{A}}_{20} \left(-\tilde{\mathcal{A}}_{11} \tilde{\mathcal{A}}_{14} + 2 \tilde{\mathcal{A}}_{1} \tilde{\mathcal{A}}_{19}\right) + \tilde{\mathcal{A}}_{12} \left(\tilde{\mathcal{A}}_{14} \tilde{\mathcal{A}}_{19} - 2 \tilde{\mathcal{A}}_{8} \tilde{\mathcal{A}}_{11}\right) - \frac{k^2}{a^2} \left(2 \tilde{\mathcal{A}}_{2} \tilde{\mathcal{A}}_{19} \tilde{\mathcal{A}}_{20} \right.
    \nonumber \\ & \quad \left.
    + \tilde{\mathcal{A}}_{13} \left(-\tilde{\mathcal{A}}_{14}\tilde{\mathcal{A}}_{20} + 2 \tilde{\mathcal{A}}_{8}\tilde{\mathcal{A}}_{12}\right)\right)\,, \\ 
    \tilde{a}\hat{\mathcal{A}}_{3} &= \tilde{\mathcal{A}}_{12} \tilde{\mathcal{A}}_{14} \tilde{\mathcal{A}}_{20} - \tilde{\mathcal{A}}_{1} \tilde{\mathcal{A}}_{20}^2 - \tilde{\mathcal{A}}_{6} \tilde{\mathcal{A}}_{14}^2 - \tilde{\mathcal{A}}_{8}\left(\tilde{\mathcal{A}}_{12}^2 - 4 \tilde{\mathcal{A}}_{1} \tilde{\mathcal{A}}_{6}\right) - \frac{k^2}{a^2}  \tilde{\mathcal{A}}_{2} \left(\tilde{\mathcal{A}}_{20}^2 - 4 \tilde{\mathcal{A}}_{6} \tilde{\mathcal{A}}_{8}\right)\,, \\ 
    \tilde{a}\hat{\mathcal{A}}_{4} &= -\frac{k^2}{a^2} \tilde{\mathcal{A}}_{3}\left(\tilde{\mathcal{A}}_{14}^2 - 4 \tilde{\mathcal{A}}_{1}\tilde{\mathcal{A}}_{8}\right) - \frac{k^4}{a^4} \tilde{\mathcal{A}}_{8} \left(\tilde{\mathcal{A}}_{10}^2 - 4 \tilde{\mathcal{A}}_{2}\tilde{\mathcal{A}}_{3}\right) \,, \\ 
    \tilde{a}\hat{\mathcal{A}}_{5} &= \frac{k^2}{a^2} \tilde{\mathcal{A}}_{10} \left(\tilde{\mathcal{A}}_{14}\tilde{\mathcal{A}}_{18} - 2 \tilde{\mathcal{A}}_{8} \tilde{\mathcal{A}}_{9}\right)\,, \\ 
    \tilde{a}\hat{\mathcal{A}}_{6} &= -\tilde{\mathcal{A}}_{4} \tilde{\mathcal{A}}_{14}^2 - \tilde{\mathcal{A}}_{1} \left(\tilde{\mathcal{A}}_{18}^2 - 4 \tilde{\mathcal{A}}_{4} \tilde{\mathcal{A}}_{8}\right) + \tilde{\mathcal{A}}_{9} \left(\tilde{\mathcal{A}}_{14} \tilde{\mathcal{A}}_{18} - \tilde{\mathcal{A}}_{8} \tilde{\mathcal{A}}_{9}\right) - \frac{k^2}{a^2} \tilde{\mathcal{A}}_{2} \left(\tilde{\mathcal{A}}_{18}^2 - 4 \tilde{\mathcal{A}}_{4}\tilde{\mathcal{A}}_{8}\right)\,, \\
    \tilde{a}\hat{\mathcal{A}}_{7} &= \frac{k^2}{a^2}\left[-\tilde{\mathcal{A}}_{14}^2 \tilde{\mathcal{A}}_{17} + \tilde{\mathcal{A}}_{10} \tilde{\mathcal{A}}_{14} \tilde{\mathcal{A}}_{19} - 2 \tilde{\mathcal{A}}_{8} \tilde{\mathcal{A}}_{10} \tilde{\mathcal{A}}_{11} + 4 \tilde{\mathcal{A}}_{1} \tilde{\mathcal{A}}_{8} \tilde{\mathcal{A}}_{17}\right] - 2 \frac{k^4}{a^4} \tilde{\mathcal{A}}_{8} \left(\tilde{\mathcal{A}}_{10}\tilde{\mathcal{A}}_{13} - 2 \tilde{\mathcal{A}}_{2}\tilde{\mathcal{A}}_{17}\right)\,,\\
    \tilde{a}\hat{\mathcal{A}}_{8} &= -\frac{k^2}{a^2} \tilde{\mathcal{A}}_{10} \left(-\tilde{\mathcal{A}}_{14} \tilde{\mathcal{A}}_{20} + 2 \tilde{\mathcal{A}}_{8} \tilde{\mathcal{A}}_{12}\right) \,, \\
    \tilde{a}\hat{\mathcal{A}}_{9} &= -\tilde{\mathcal{A}}_{14}^2 \tilde{\mathcal{A}}_{15} + \tilde{\mathcal{A}}_{11} \tilde{\mathcal{A}}_{14} \tilde{\mathcal{A}}_{18} - 2 \tilde{\mathcal{A}}_{1} \tilde{\mathcal{A}}_{18} \tilde{\mathcal{A}}_{19} + 4 \tilde{\mathcal{A}}_{1} \tilde{\mathcal{A}}_{8} \tilde{\mathcal{A}}_{15} + \tilde{\mathcal{A}}_{9} \tilde{\mathcal{A}}_{14} \tilde{\mathcal{A}}_{19} - 2 \tilde{\mathcal{A}}_{8} \tilde{\mathcal{A}}_{9} \tilde{\mathcal{A}}_{11}
    \nonumber \\ & \quad 
    + \frac{k^2}{a^2} \left(\tilde{\mathcal{A}}_{13} \tilde{\mathcal{A}}_{14} \tilde{\mathcal{A}}_{18} - 2 \tilde{\mathcal{A}}_{2} \tilde{\mathcal{A}}_{18} \tilde{\mathcal{A}}_{19} + 4 \tilde{\mathcal{A}}_{2} \tilde{\mathcal{A}}_{8} \tilde{\mathcal{A}}_{15} - 2 \tilde{\mathcal{A}}_{8} \tilde{\mathcal{A}}_{9} \tilde{\mathcal{A}}_{13}\right)\,, \\
    \tilde{a}\hat{\mathcal{A}}_{10} &= -\tilde{\mathcal{A}}_{14}^2 \tilde{\mathcal{A}}_{16} + \tilde{\mathcal{A}}_{12} \tilde{\mathcal{A}}_{14} \tilde{\mathcal{A}}_{18} - 2 \tilde{\mathcal{A}}_{1} \tilde{\mathcal{A}}_{18} \tilde{\mathcal{A}}_{20} + 4 \tilde{\mathcal{A}}_{1} \tilde{\mathcal{A}}_{8} \tilde{\mathcal{A}}_{16} - \tilde{\mathcal{A}}_{9}\left(-\tilde{\mathcal{A}}_{14}\tilde{\mathcal{A}}_{20} + 2 \tilde{\mathcal{A}}_{8}\tilde{\mathcal{A}}_{12}\right)\nonumber \\ & \quad 
    + 2 \frac{k^2}{a^2}  \tilde{\mathcal{A}}_{2} \left(-\tilde{\mathcal{A}}_{18} \tilde{\mathcal{A}}_{20} + 2 \tilde{\mathcal{A}}_{8} \tilde{\mathcal{A}}_{16}\right)\,,
\end{align}
\end{subequations}
and $\tilde{a} = -\tilde{\mathcal{A}}_{14}^2 + 4 \tilde{\mathcal{A}}_{8} \left(\tilde{\mathcal{A}}_{1} + \frac{k^2}{a^2} \tilde{\mathcal{A}}_{2}\right)$. Provided that $\tilde{a} \neq 0$ the following analysis holds, otherwise the analysis would need to be altered at the \eqref{eq:scalar_gauge_invariant_fourier} stage of the calculation. Additionally, emphasizes have been made to keep the $k$-dependencies. At a later stage this becomes crucial to obtain high-$k$ limit. Through integration by parts Eq.~\eqref{eq:scalar_action_minimised} changes to
\begin{align}\label{eq:scalar_action_before_diagionalisation}
    \mathcal{S}_{\text{S}} &= \frac{1}{2\kappa^2}\int \text{d}t \, \frac{\text{d}^3k}{(2\pi)^{\frac{3}{2}}} \, a^3 
    \left[
    \left(\hat{\mathcal{A}}_{1} - \frac{1}{2a^3} \frac{\text{d}}{\text{d}t}\left( a^3 \hat{\mathcal{A}}_{2}\right) \right) \mathcal{X}_{1}^2 + \hat{\mathcal{A}}_{3} \dot{\mathcal{X}}_{1}^2 + \left(\hat{\mathcal{A}}_{4} - \frac{1}{2a^3} \frac{\text{d}}{\text{d}t}\left( a^3 \hat{\mathcal{A}}_{5}\right)\right) \mathcal{X}_{3}^2  \right.
    \nonumber \\ & \quad \left.
    +  \hat{\mathcal{A}}_{6} \dot{\mathcal{X}}_{3}^2 + \left(\hat{\mathcal{A}}_{7} - \frac{1}{a^3} \frac{\text{d}}{\text{d}t} \left(a^3 \hat{\mathcal{A}}_{8}\right)\right) \mathcal{X}_{1} \mathcal{X}_{3} + \left(\hat{\mathcal{A}}_{9} - \hat{\mathcal{A}}_{8} \right)\mathcal{X}_{1} \dot{\mathcal{X}}_{3} + \hat{\mathcal{A}}_{10} \dot{\mathcal{X}}_{1} \dot{\mathcal{X}}_{3} 
    \right] \,.
\end{align} 
When looking at the kinetic contributions, it was noted that there is a mixing of modes of $\dot{\mathcal{X}}_{1} \dot{\mathcal{X}}_{3}$ corresponding to the coefficient $\hat{\mathcal{A}}_{10}$. To obtain a diagonal kinetic matrix, as required in Eq.~\eqref{eq:stability_action_form}, the gauge invariant quantities are redefined as
\begin{align}\label{eq:Psi1_Psi2_definition}
    \mathcal{X}_{1} := \Psi_{1}\,, \qquad 
    \mathcal{X}_{3} := \Psi_{2} - \frac{1}{2} \left( \frac{\hat{\mathcal{A}}_{10}}{\hat{\mathcal{A}}_{6}} \right) \Psi_{1}\,, 
\end{align} 
where $\Psi_{1}$ and $\Psi_{2}$ are still gauge invariant as they are a linear combination of the previous dynamical modes. Due to the complexity of the action upon substituting these new fields, the generalised result has been omitted. By considering the high-$k$ limit, the result simplifies to
\begin{align}\label{eq:scalar_action_diagionalisation}
    \mathcal{S}_{\text{S}} &= \frac{1}{2\kappa^2}\int \text{d}t \, \frac{\text{d}^3k}{(2\pi)^{\frac{3}{2}}} \, a^3 
    \left[
    \breve{\mathcal{A}}_{1} \dot{\Psi}_{1}^2 + \breve{\mathcal{A}}_{2} \dot{\Psi}_{2}^2 + \frac{k^2}{a^2} \left(\breve{\mathcal{A}}_{3} \Psi_{1}^2 + \breve{\mathcal{A}}_{4} \Psi_{2}^2 + \breve{\mathcal{A}}_{5} \Psi_{1} \Psi_{2}\right) \right.
    \nonumber \\ & \qquad\qquad\qquad\qquad\qquad\qquad\qquad\qquad \left.
    + \breve{\mathcal{A}}_{6} \Psi_{1}^2 + \breve{\mathcal{A}}_{7} \Psi_{2}^2 + \breve{\mathcal{A}}_{8} \Psi_{1} \Psi_{2} + \breve{\mathcal{A}}_{9} \Psi_{1} \dot{\Psi}_{2}
    \right]\,.
\end{align}
Before listing the coefficients of $\breve{A}_{i}$ for $i \in [1, 9]$, we consider the temporal Fourier transformation $\dot{\Psi} \rightarrow i \omega \Psi$ such that the Lagrangian portion of Eq~\eqref{eq:scalar_action_diagionalisation} can be expressed as follows:
\begin{align}
    \mathcal{L}_{\text{S}} &= 
    \Big( \Psi_{1} \quad \Psi_{2} \Big)
    \begin{pmatrix}
        -\omega^2 \breve{\mathcal{A}}_{1} + \frac{k^2}{a^2} \breve{\mathcal{A}}_{3} + \breve{\mathcal{A}}_{6} & \frac{1}{2} \left(\frac{k^2}{a^2} \breve{\mathcal{A}}_{5} + \breve{\mathcal{A}}_{8} + i \omega \breve{\mathcal{A}}_{9}\right) \\
        \frac{1}{2} \left(\frac{k^2}{a^2} \breve{\mathcal{A}}_{5} + \breve{\mathcal{A}}_{8} + i \omega \breve{\mathcal{A}}_{9}\right) & -\omega^2 \breve{\mathcal{A}}_{2} + \frac{k^2}{a^2} \breve{\mathcal{A}}_{4} + \breve{\mathcal{A}}_{7}
    \end{pmatrix}
    \begin{pmatrix}
        \Psi_{1} \\
        \Psi_{2}
    \end{pmatrix}\,.
\end{align}
The dispersion relation given by $\omega = c_{\text{S}} \frac{k}{a}$ shows that $\omega$ is of the same order of $k$. If we consider only the leading order of $k$, coefficients $\{ \breve{\mathcal{A}}_{6}\,, \breve{\mathcal{A}}_{7}\,, \breve{\mathcal{A}}_{8}\,, \breve{\mathcal{A}}_{9}\}$ do not contribute, thus
\begin{align}\label{eq:scalar_action_highk_limit}
    \mathcal{S}_{\text{S}} &= \frac{1}{2\kappa^2}\int \text{d}t \, \frac{\text{d}^3k}{(2\pi)^{\frac{3}{2}}} \, a^3 
    \left[
    \breve{\mathcal{A}}_{1} \dot{\Psi}_{1}^2 + \breve{\mathcal{A}}_{2} \dot{\Psi}_{2}^2 + \frac{k^2}{a^2} \left(\breve{\mathcal{A}}_{3} \Psi_{1}^2 + \breve{\mathcal{A}}_{4} \Psi_{2}^2 + \breve{\mathcal{A}}_{5} \Psi_{1} \Psi_{2}\right) \right]\,,
\end{align}
where
\begin{subequations}
\allowdisplaybreaks
\begin{align}
    \breve{\mathcal{A}}_{1} &=  \tilde{\mathcal{A}}_{6} +  \frac{-\tilde{\mathcal{A}}_{16}\tilde{\mathcal{A}}_{18}\tilde{\mathcal{A}}_{20}+\tilde{\mathcal{A}}_{4}\tilde{\mathcal{A}}_{20}^2+\tilde{\mathcal{A}}_{8}\tilde{\mathcal{A}}_{16}^2}{\tilde{\mathcal{A}}_{18}^2-
    4\tilde{\mathcal{A}}_{4}\tilde{\mathcal{A}}_{8}}\,, \\
    \breve{\mathcal{A}}_{2} &= \tilde{\mathcal{A}}_{4} - \frac{\tilde{\mathcal{A}}_{18}^2}{4 \tilde{\mathcal{A}}_{8}}\,, \\
    \breve{\mathcal{A}}_{3} &= 
    \tilde{\mathcal{A}}_{7} - \frac{\tilde{\mathcal{A}}_{13}^2}{4 \tilde{\mathcal{A}}_{2}} 
    +
    \frac{(\tilde{\mathcal{A}}_{10} \tilde{\mathcal{A}}_{13} -2 \tilde{\mathcal{A}}_{2} \tilde{\mathcal{A}}_{17}) (\tilde{\mathcal{A}}_{18} \tilde{\mathcal{A}}_{20} - 2 \tilde{\mathcal{A}}_{16} \tilde{\mathcal{A}}_{8})}{2 \tilde{\mathcal{A}}_{2} (\tilde{\mathcal{A}}_{18}^2 - 4 \tilde{\mathcal{A}}_{4} \tilde{\mathcal{A}}_{8})}
   \nonumber \\ & \quad
    - 
    \frac{(\tilde{\mathcal{A}}_{10}^2 - 4 \tilde{\mathcal{A}}_{2} \tilde{\mathcal{A}}_{3}) (\tilde{\mathcal{A}}_{18} \tilde{\mathcal{A}}_{20} - 2 \tilde{\mathcal{A}}_{8} \tilde{\mathcal{A}}_{16})^{2}}{4 \tilde{\mathcal{A}}_{2} (\tilde{\mathcal{A}}_{18}^2 - 4 \tilde{\mathcal{A}}_{4} \tilde{\mathcal{A}}_{8})^2}\,, \\
    \breve{\mathcal{A}}_{4} &= \tilde{\mathcal{A}}_{3} - \frac{\tilde{\mathcal{A}}_{10}^2}{4\tilde{\mathcal{A}}_{2}}\,, \\
    \breve{\mathcal{A}}_{5} &= \tilde{\mathcal{A}}_{17} - \frac{\tilde{\mathcal{A}}_{10}\tilde{\mathcal{A}}_{13}}{2\tilde{\mathcal{A}}_{2}} - \frac{(\tilde{\mathcal{A}}_{10}^2 - 4\tilde{\mathcal{A}}_{2}\tilde{\mathcal{A}}_{3})(-\tilde{\mathcal{A}}_{18}\tilde{\mathcal{A}}_{20}+2\tilde{\mathcal{A}}_{8}\tilde{\mathcal{A}}_{16})}{2\tilde{\mathcal{A}}_{2}(\tilde{\mathcal{A}}_{18}^2 - 4\tilde{\mathcal{A}}_{4}\tilde{\mathcal{A}}_{8})}\,.
\end{align}
\end{subequations}
These results hold given $\tilde{\mathcal{A}}_{18}^2 - 4 \tilde{\mathcal{A}}_{4} \tilde{\mathcal{A}}_{8} \neq 0$, $\tilde{\mathcal{A}}_{2} \neq 0$ and $\tilde{\mathcal{A}}_{8} \neq 0$. Next, taking variations with respect to the dynamical modes $\Psi_{1}$ and $\Psi_{2}$ results in the field equations
\begin{align}
    0 &\approx \ddot{\Psi}_{1} + \left(\frac{\breve{\mathcal{A}}_{1}'}{\breve{\mathcal{A}}_{1}}\right) \dot{\Psi}_{1} - \frac{k^2}{a^2}\left(\frac{\breve{\mathcal{A}}_{3}}{\breve{\mathcal{A}}_{1}}\right)\Psi_{1} - \frac{k^2}{a^2}\left(\frac{\breve{\mathcal{A}}_{5}}{2\breve{\mathcal{A}}_{1}}\right)\Psi_{2}  \,, \\
    0 &\approx \ddot{\Psi}_{2} + \left(\frac{\breve{\mathcal{A}}_{2}'}{\breve{\mathcal{A}}_{2}}\right) \dot{\Psi}_{2} - \frac{k^2}{a^2}\left(\frac{\breve{\mathcal{A}}_{4}}{\breve{\mathcal{A}}_{2}}\right)\Psi_{2} - \frac{k^2}{a^2}\left(\frac{\breve{\mathcal{A}}_{5}}{2\breve{\mathcal{A}}_{2}}\right)\Psi_{1} \,.
\end{align}
respectively, indicating that a mixing of modes occurs due to $\breve{\mathcal{A}}_{5}$ contribution. These two systems decouple for the condition
\begin{align}\label{eq:scalar_breveA5_condition}
    \breve{\mathcal{A}}_{5} = 0\,.
\end{align}
Under the restriction of Eq.~\eqref{eq:scalar_breveA5_condition}, the propagating speed for each mode is given by the coefficient of the gradient term such that
\begin{align}\label{eq:scalar_speed}
    c_{\Psi_{1}}^2 := -\frac{\breve{\mathcal{A}}_{3}}{\breve{\mathcal{A}}_{1}} > 0\,, \qquad \text{and} \qquad
    c_{\Psi_{2}}^2 := -\frac{\breve{\mathcal{A}}_{4}}{\breve{\mathcal{A}}_{2}} > 0\,,
\end{align}
and ghost stability can be attained provided that
\begin{align}\label{eq:scalar_ghost}
    \mathcal{M}_{\Psi_{1}} := \breve{\mathcal{A}}_{1} > 0\,, \qquad \text{and} \qquad \mathcal{M}_{\Psi_{2}} := \breve{\mathcal{A}}_{2} > 0\,,
\end{align}
and Laplacian/gradient stability for
\begin{align}\label{eq:scalar_laplacian}
    \mathcal{N}_{\Psi_{1}} := -\breve{\mathcal{A}}_{3} > 0\,, \qquad \text{and} \qquad \mathcal{N}_{\Psi_{2}} := -\breve{\mathcal{A}}_{4} > 0\,.
\end{align}
By satisfying these conditions, ghost and gradient instabilities can be removed from the resulting models.


\subsection{Vectors}

We next consider the vector perturbations keeping with the gauge invariant approach. For this sector, the second-order perturbation where integration by parts and the background equations of motion are resubstituted produces
\begin{align} \label{eq:vector_action_fourier}   
    \mathcal{S}_{\text{V}} = \int \text{d}t\, \frac{\text{d}^3k}{(2\pi)^{\frac{3}{2}}}\, a^{3} \bigg[& \tilde{\mathcal{C}}_1 \, \dot{u}_{i} \, \dot{u}^{i} + \tilde{\mathcal{C}}_7\,\dot{\mathcal{Z}}_{i}\,\dot{\mathcal{Z}}^{i} + \frac{k^2}{a^2} \left( \tilde{\mathcal{C}}_2 \, u_i \, u^{i} +\tilde{\mathcal{C}}_4 \, u_i \, \mathcal{Y}^i + \tilde{\mathcal{C}}_8 \, \mathcal{Z}_i \, \mathcal{Z}^i + \tilde{\mathcal{C}}_9 \,\mathcal{Y}_i \, \mathcal{Y}^i \right) 
    \nonumber \\ & 
    - \frac{i \, \epsilon^{ijk} \, k_{j}}{a} \left(\tilde{\mathcal{C}}_3  \, \mathcal{Z}_i \, u_k  - \tilde{\mathcal{C}}_5 \, \dot{u}_k \, \mathcal{Z}_i + \tilde{\mathcal{C}}_6 \, \dot{\mathcal{Z}}_{i}\, \mathcal{Y}_k \right) \bigg]\,.
\end{align}
Now, taking the variation with respect to the non-dynamical mode $\mathcal{Y}_{k}$ yields the equation
\begin{align}
    0 &= \frac{k^2}{a^2} \left(\tilde{\mathcal{C}}_{4} u_{k} + 2 \tilde{\mathcal{C}}_{9} \mathcal{Y}_{k}\right) - \frac{i \epsilon^{ijk} k_{j}}{a} \tilde{\mathcal{C}}_{6} \dot{\mathcal{Z}}_{i}\,,
\end{align}
which when solved for $\mathcal{Y}_{k}$ and substituted back in action~\eqref{eq:vector_action_fourier} yields
\begin{align}
    \mathcal{S}_{\text{V}} = \int \text{d}t\, \frac{\text{d}^3k}{(2\pi)^{\frac{3}{2}}}\, a^{3} \Bigg[& 
    \tilde{\mathcal{C}}_{1} \, \dot{u}_{i} \, \dot{u}^{i}+ \left(\tilde{\mathcal{C}}_{7} + \frac{\tilde{\mathcal{C}}_{6}^2}{4 \tilde{\mathcal{C}}_{9}}\right) \dot{\mathcal{Z}}_{i} \, \dot{\mathcal{Z}}^{i} +\frac{k^2}{a^2} \left(\tilde{\mathcal{C}}_{8} \, \mathcal{Z}_{i} \, \mathcal{Z}^{i} - \left(\tilde{\mathcal{C}}_{2} - \frac{\tilde{\mathcal{C}}_{4}^2}{4 \tilde{\mathcal{C}}_{9}}\right) u_{i} \, u^{i}\right)
    \nonumber \\ & \quad 
    - \frac{i \, \epsilon^{ijk} \, k_{j}}{a} \left( \left(\tilde{\mathcal{C}}_{3} - H \tilde{\mathcal{C}}_{5} + \dot{\tilde{\mathcal{C}}}_{5}\right) \mathcal{Z}_{i} \, u_{k} + \left( \tilde{\mathcal{C}}_{5} - \frac{\tilde{\mathcal{C}}_{4}\tilde{\mathcal{C}}_{6}}{2 \tilde{\mathcal{C}}_{9}} \right) \dot{\mathcal{Z}}_{i} \, u_{k}\right)
    \Bigg]\,.
\end{align}
Both the kinetic and the gradients portions of the action are a diagonal matrix when cast in the form given by Eq.~\eqref{eq:stability_action_form}. The term corresponding to $\mathcal{Z}_{i} u_{k}$ can be omitted within the high-$k$ limit as it is of a lower order of $k$-dependency in comparison with the rest of the terms. In contrast with the scalar sector, the term $\dot{\mathcal{Z}}_{i} u_{k}$ appears to have the same order of $k$ upon substituting the temporal Fourier transformation and $\omega^2 = c_{\text{V}}^2 \frac{k^2}{a^2}$. To further analyse this scenario, we take the variation with respect to the both $u_{i}$ and $\mathcal{Z}_{i}$, respectively 
\begin{align}
    0 &\approx \ddot{u}_{k} + \left(3H + \frac{\dot{\tilde{\mathcal{C}}}_{1}}{\tilde{\mathcal{C}}_{1}}\right) \dot{u}_{k} - \frac{k^2}{a^2} \left(\frac{\tilde{\mathcal{C}}_{4}^2 - 4 \tilde{\mathcal{C}}_{9}\tilde{\mathcal{C}}_{2}}{4 \tilde{\mathcal{C}}_{9}\tilde{\mathcal{C}}_{1}}\right) u_{k} + \frac{i \, \epsilon_{ijk} \, k^{j}}{a} \left(\frac{ 2 \tilde{\mathcal{C}}_{9}\tilde{\mathcal{C}}_{5} - \tilde{\mathcal{C}}_{4}\tilde{\mathcal{C}}_{6} }{2\tilde{\mathcal{C}}_{9}\tilde{\mathcal{C}}_{1}}\right) \dot{\mathcal{Z}}^{i} \,, \\
    0 &\approx \ddot{\mathcal{Z}}_{k} + \left(3H + \frac{1}{a^3} \frac{\frac{\text{d}}{\text{d}t}\left(a^3 \left(\tilde{\mathcal{C}}_{7} + \frac{\tilde{\mathcal{C}}_{6}^2}{4 \tilde{\mathcal{C}}_{9}}\right)\right)}{\left(\tilde{\mathcal{C}}_{7} + \frac{\tilde{\mathcal{C}}_{6}^2}{4 \tilde{\mathcal{C}}_{9}}\right)}\right) \dot{\mathcal{Z}}_{k} - \frac{k^2}{a^2}\left( \frac{4 \tilde{\mathcal{C}}_{9}\tilde{\mathcal{C}}_{8}}{\left(4 \tilde{\mathcal{C}}_{9}\tilde{\mathcal{C}}_{7} + \tilde{\mathcal{C}}_{6}^2 \right)}\right) \mathcal{Z}_{k} + \frac{i \, \epsilon_{ijk} \, k^{j}}{a} \left(\frac{2 \tilde{\mathcal{C}}_{9}\tilde{\mathcal{C}}_{5} - \tilde{\mathcal{C}}_{4}\tilde{\mathcal{C}}_{6}}{4 \tilde{\mathcal{C}}_{9}\tilde{\mathcal{C}}_{7} + \tilde{\mathcal{C}}_{6}^2}\right) \dot{u}^{i}\,,
\end{align}
for which the two modes are said to decouple provided that 
\begin{align} 
2 \tilde{\mathcal{C}}_{9}\tilde{\mathcal{C}}_{5} - \tilde{\mathcal{C}}_{4}\tilde{\mathcal{C}}_{6} = 0\,.
\end{align}
When this condition applies, the propagating speed for both modes can be determined:
\begin{align} \label{eq:vector_grad_inst}
    c_{u}^2 := \frac{ 4 \tilde{\mathcal{C}}_{9}\tilde{\mathcal{C}}_{2} -\tilde{\mathcal{C}}_{4}^2 }{4 \tilde{\mathcal{C}}_{9}\tilde{\mathcal{C}}_{1}} > 0\,, \qquad \text{and} \qquad c_{\mathcal{Z}}^2 := -\left( \frac{4 \tilde{\mathcal{C}}_{9}\tilde{\mathcal{C}}_{8}}{\left(4 \tilde{\mathcal{C}}_{9}\tilde{\mathcal{C}}_{7} + \tilde{\mathcal{C}}_{6}^2 \right)}\right) > 0\,.
\end{align}
Thus, ghost stability can be obtained when
\begin{align} \label{eq:vector_ghost_inst}
    \mathcal{M}_{u} := \tilde{\mathcal{C}}_{1} > 0\,, \qquad \text{and} \qquad \mathcal{M}_{\mathcal{Z}} := \tilde{\mathcal{C}}_{7} + \frac{\tilde{\mathcal{C}}_{6}^2}{4 \tilde{\mathcal{C}}_{9}} > 0\,,
\end{align}
and Laplacian stability satisfied provided that
\begin{align} \label{eq:vector_laplacian_inst}
    \mathcal{N}_{u} := \tilde{\mathcal{C}}_{2} - \frac{\tilde{\mathcal{C}}_{4}^2}{4 \tilde{\mathcal{C}}_{9}} > 0\,, \qquad \text{and} \qquad \mathcal{N}_{\mathcal{Z}} := -\tilde{\mathcal{C}}_{8} > 0\,.
\end{align}
which both depend on a combination of first order derivatives of the action functionals.



\section{Conclusion} \label{sec:conclusion}

The teleparallel analogue of Horndeski gravity expressed through the BDLS action in Eq.~\eqref{action} offers an interesting way to circumvent the restrictions placed on regular Horndeski gravity by the multimessenger events associated with GW170817 and GRB170817A \cite{Ezquiaga:2017ekz}. In the present work, we have explored the gauge-invariant cosmological perturbations of this framework theory together with some initial stability and ghost conditions imposed on these classes of models. This builds on previous work \cite{Ahmedov:2023num} where the cosmological perturbations were determined for a specific gauge. The current work broadens the applicability of this analysis, and provides a practical way forward on both the gauge invariant form of these perturbations, as well as example cases of the most popular gauge choices, which appear in App.~\ref{appendix_gauge_choices}.

To perform this analysis in a gauge invariant form, the tetrad inherits the perturbative form from the metric through the choice in Eq.~\eqref{eq:tetrad_pert}, which importantly remains within the Weitzenb\"{o}ck gauge in terms of the form of the spin connection components. The general cosmological perturbations are then probed for their scalar, pseudoscalar, vector and tensor modes in Sec.~\ref{sec:cosmological_perturbations}. The gauge invariant nature of these perturbations, coupled with the complexity of the underlying model framework results in highly involved expressions for the perturbation modes. For this reason, we provide example cases of the most popular gauges in the literature in App.~\ref{appendix_gauge_choices} for the scalar and vector modes.

Already, at this level of analysis an assessment can be made on the propagation of modes and whether they contain ghosts or other instabilities, which are then probed in Sec.~\ref{sec:stability}. The generality of this class of models means that it must be imposed that the scalar field kinetic term be positive, while the speed of propagation of propagating modes should also be positive, which will respectively alleviate ghost and Laplacian instabilities immediately. The gauge-invariant approach results in fairly convoluted conditions on the underlying Lagrangian terms, which for the scalar sector give Eqs.~(\ref{eq:scalar_ghost}--\ref{eq:scalar_laplacian}), while for the vector section Eqs.~(\ref{eq:vector_grad_inst}--\ref{eq:vector_laplacian_inst}), and tensor sector Eqs.~\eqref{tensorstability}. By exploring models that satisfy this combination of conditions, one can produce cosmologies that do not suffer from perturbative stability pathologies.

\begin{table}[h!]
    \centering
    \begin{tabular}{|c|c||c|c|c|}
    \hline
         \textbf{Theory} & \textbf{PDoFs} & \textbf{(Pseudo)Scalars} & \textbf{Vectors} & \textbf{Tensors}   \\
         \hline
         BDLS & 9 & $\Psi_{1}, \Psi_{2}, \sigma$ & $u_{i}, \mathcal{Z}_{i}$ & $h_{ij}$ \\
         GR & 2 & - & - & $h_{ij}$\\
         $f(T)$ & 2 & - & - & $h_{ij}$ \\
         $f(\phi, X, T)$ & 3 & $\Psi_{2}$ & - & $h_{ij}$ \\
         Generalized Teleparallel Dark Energy & 3 & $\mathcal{X}_{1}$ & - & $h_{ij}$ \\
         Generalized Scalar Tensor & 3 & $\Psi_{2}$ & - & $h_{ij}$ \\
         NGR & 8 & $\mathcal{X}_{3}, \sigma$ & $u_{i}, \mathcal{Z}_{i}$ & $h_{ij}$\\
         Horndeski & 3 & $\Psi_{2}$ & - & $h_{ij}$\\ \hline
    \end{tabular}
    \caption{List of propagating degrees of freedom for GR, $f(T)$, $f(\phi,X,T)$, NGR and Horndeski gravity. In the case of $f(\phi,X,T)$ and NGR, diagionalisation of the kinetic matrix is required thus scalar modes are expressed in terms of $\{\Psi_1, \Psi_2\}$. The scalar sector of the rest of the theories can be expressed in terms of $\{\mathcal{X}_1, \mathcal{X}_3\}$.}
    \label{tab:comparison_literature}
\end{table}

The general BDLS framework can be reduced for specific subclasses which represent a smaller number of propagating DoF. For the full treatment, it is found that there are 9 propagating DoFs which include 2 scalars, 1 pseudoscalar, 2 pairs of vectors modes, and the regular pair of tensor modes. This confirms the indications made in Ref.~\cite{Bahamonde:2021dqn} where the DoFs were obtained through the respective dispersion relations of the different perturbative sectors. In Table~\ref{tab:comparison_literature} the number of propagating DoFs is listed for several leading subclasses of the BDLS frame. The action of the generalized teleparallel dark energy theory is given by $S \sim \int ( -A(\phi) T - 1/2 (\partial \phi)^2 - V(\phi))$, for the generalized teleparallel scalar tensor it is $ S \sim \int (F(\phi) T + G_2 (\phi, X) - G_3(\phi,X) \square \phi)$, while NGR is the linear combination of all the three parity preserving, quadratic torsion contractions. By substituting these subclasses into our results, we confirm the expected 2 DoFs for GR and $f(T)$ gravity, as well as an extra scalar mode for $f(\phi, X, T)$, regular Horndeski gravity, and other forms. It is only NGR that also contains a large number of propagating DoFs which in that case results in 8 DoFs.

In the case of $f(T)$ gravity, the number of propagating DoFs falls to 2 due to the case when the background is Minkowski wherein the extra scalar DoF is rendered non-dynamical~\cite{Bahamonde:2021dqn}. By exploring the longitudinal gauge with the added assumption that $\beta=0$ gives identical results as in Ref.~\cite{Izumi:2012qj}. However, this may yield over-fixing since the gauge choice only imposes that $\beta = B$. As for the NGR case, the non-trivial linearised field equations for the Minkowski limit imply that $G_{\text{Tele},T_{\text{ax}}} = 0$ which forces the psuedoscalar contribution to be non-dynamical. The other DoFs propagate in agreement with Ref.~\cite{Bahamonde:2024zkb}. This does not occur in the general BDLS framework since the field equations also depend on the invariants $T_{\text{ax}}, J_{1}, J_{5}, J_{10}$ which incur contributions from both $\mathcal{B}_{1}$ and $\mathcal{B}_{2}$. 

The central motivation of this work was to determine the gauge-invariant cosmological perturbations of the teleparallel analogue of Horndeski gravity as expressed through the BDLS formalism. It would now be interesting to investigate the numerical strength of the propagating DoFs and their impact on changes to the standard cosmological paradigm. To do this, further physically motivated models are needed, which can be analysed against a spectrum of observational data connected to the power spectra of these perturbative modes. The cosmological perturbations can also be further investigated in order to assess condition on producing healthy theories.

\begin{acknowledgements}
This article is also based upon work from COST Action CA21136 Addressing observational tensions in cosmology with systematics and fundamental physics (CosmoVerse) supported by COST (European Cooperation in Science and Technology). K.F.D. was supported by the PNRR-III-C9-2022–I9 call, with project number 760016/27.01.2023. JLS would also like to acknowledge funding from ``Xjenza Malta'' as part of the ``FUSION R\&I: Research Excellence Programme'' REP-2023-019 (CosmoLearn) Project, and the ``Net4Tensions'' as part of the ``Research Networking Scheme''.
\end{acknowledgements}

\appendix

\section{Coefficients of the quadratic actions}  \label{appendix_coefficients}

The scalar quadratic action of BDLS theory in terms of the gauge-dependent variables is given by Eq.~\eqref{eq:scalar-quadratic-gauge-dependent-action}. The coefficients $\mathcal{A}_1 - \mathcal{A}_{28}$ are shown here 
\begin{subequations}
\allowdisplaybreaks
\begin{align}
    \mathcal{A}_{1} &= 
    \frac{a^2}{6} \Big[-6 G_{\text{Tele},T_\text{vec}} + 12 H^2 (9 G_{\text{Tele},T_\text{vec} T_{\text{vec}}} - 12 G_{\text{Tele},TT_\text{vec}} + 4 G_{\text{Tele},TT}) + 12 H \dot{\phi_0} (- 3 G_{\text{Tele},T_\text{vec} I_{2}} 
    \nonumber \\ & \quad
    + 2 G_{\text{Tele},T I_{2}}) + \dot{\phi_0}^2 (4 G_{\text{Tele},J_{8}} + G_{\text{Tele},J_5} + 3 G_{\text{Tele},I_{2} I_{2}})\Big]\,, \\
    \mathcal{A}_{2} &= - a \Big[
    -12 H^3 (9 G_{\text{Tele},T_\text{vec} T_\text{vec}} + 4 (-3 G_{\text{Tele},T T_\text{vec}} + G_{\text{Tele},TT}) ) + \dot{\phi_0} (2 G_{4,\phi_0} - G_{\text{Tele},I_{2}} 
    \nonumber \\ & \quad 
    - 2 X (G_{3,X} - 2 G_{4,\phi_0 X} + G_{\text{Tele},X I_{2}})) - H^2 \dot{\phi_0} (-54 G_{\text{Tele},T_\text{vec} I_2} + 36 G_{\text{Tele},T I_{2}} + 10 X G_{5,X} + 4 X^2 G_{5,XX}) 
    \nonumber \\ & \quad 
    + 2 H \big[
    2 G_{4} + 3 G_{\text{Tele},T_\text{vec}} - 2 G_{\text{Tele},T} + 2 X (-4 G_{4,X} - 2 G_{\text{Tele},XT} + 3(G_{5,\phi_0} - G_{\text{Tele},I_{2}I_{2}} + G_{\text{Tele},X T_\text{vec}})) 
    \nonumber \\ & \quad
    + 4 X^2 (-2G_{4,XX} + G_{5,\phi_0 X})\big] \Big] \,, \\
    \mathcal{A}_{3} &= a \Big[4 G_{4} + 6 G_{\text{Tele},T_\text{vec}} - 4 G_{\text{Tele},T} - 12 H^2 (9 G_{\text{Tele},T_\text{vec} T_\text{vec}} - 12 G_{\text{Tele},T\,T_\text{vec}} + 4 G_{\text{Tele},TT}) 
    \nonumber \\ & \quad 
    + 2 X (-4 G_{4,X} + 2 G_{5,\phi_0} - 3 G_{\text{Tele},I_2 I_2}) - 2 H \dot{\phi_0} (-18 G_{\text{Tele},T_\text{vec}\,I_{2}} + 12 G_{\text{Tele},T\,I_{2}} + 2 X G_{5,X}) \Big]\,, \\
    \mathcal{A}_{4} &= a \Big[ H (-2 G_{4,\phi_0} + 6 X G_{3,X} - 20 X G_{4,\phi_0 X} + 4 X G_{5, \phi_0 \phi_0}+ 3 G_{\text{Tele},I_{2}} + 6 G_{\text{Tele},\phi_0 T_\text{vec}} - 4 G_{\text{Tele},\phi_0 T}) 
    \nonumber \\ & \quad + H^2 \dot{\phi_0} (6 G_{4,X}-6 G_{5,\phi_0}+12 X G_{4,XX} - 8 X G_{5,\phi_0 X}) + H^3 (6 X G_{5,X}+4 X^2 G_{5,XX}) + \dot{\phi_0} (G_{2,X}
    \nonumber \\  & \quad  - 2 G_{3,\phi_0} + 2 G_{4,\phi_0 \phi_0} + G_{\text{Tele},X} - G_{\text{Tele},\phi_0 I_{2}}) \Big] \,, \\
    \mathcal{A}_{5} &= -a \Big[ -2 G_{4,\phi_0} + G_{\text{Tele},I_{2}} + 2 X (G_{3,X} - 2 G_{4,\phi_0 X} + G_{\text{Tele},X I_{2}}) + H \dot{\phi_0} ( 4 G_{4,X} - 4 G_{5,\phi_0} + 3 G_{\text{Tele},I_{2}I_{2}}  
    \nonumber \\ & \quad
    - 6 G_{\text{Tele},X T_\text{vec}} + 4 G_{\text{Tele},X T} + 4 X (2 G_{4,XX} - G_{5,\phi_0 X})) + H^2 ( - 18 G_{\text{Tele},T_\text{vec} I_{2}} + 12 G_{\text{Tele},T I_{2}} + 6 X G_{5,X} 
    \nonumber \\ & \quad 
    + 4 X^2 G_{5,XX})\Big]\,, \\
    \mathcal{A}_{6} &= a \Big[
    \dot{\phi_0} (2 G_{4,\phi_0} - G_{\text{Tele},I_2} - 2X (G_{3,X} - 2G_{4,\phi_0 X}) ) + 2H (2 G_{4} + 3 G_{\text{Tele},T_\text{vec}} - 2 G_{\text{Tele},T} 
    \nonumber \\ & \quad
    + 2X (-4 G_{4,X} + 3 G_{5,\phi_0}) + 4 X^2 (-2 G_{4,XX} + G_{5,\phi_0 X})) - 2 X \dot{\phi_0} H^2 (5 G_{5,X} + 2 X G_{5,XX}) \Big]\,, \\
    \mathcal{A}_{7} &= \frac{2}{9}a \Big[
    9 G_{\text{Tele},T_\text{vec}} + 2 X (-2 G_{\text{Tele},J_{8}} - 5 G_{\text{Tele},J_{5}} + 3 G_{\text{Tele},J_{3}} + 2 X G_{\text{Tele},J_{6}}) \Big]\,, \\ 
    \mathcal{A}_{8} &= -a \Big[
    2 X H^3 (3G_{5,X} + 2 X G_{5,XX}) + H^2 \dot{\phi_0} \big[
    6 G_{4,X} - 6 G_{5,\phi_0} - 2 G_{\text{Tele},J_{3}} + 6 \dot{H} (3 G_{\text{Tele},T_{\text{vec}} J_{3}} - 2 G_{\text{Tele},T J_{3}}) 
    \nonumber \\ & \quad  + 4 X (3 G_{4,XX} - 2 G_{5,\phi_0 X}) - 3 \ddot{\phi_0} G_{\text{Tele},I_2 J_3}\big] + \dot{\phi_0} \big[
    G_{2,X} - 2 G_{3,\phi_0} + 2 G_{4,\phi_0 \phi_0} + G_{\text{Tele},X} - G_{\text{Tele},\phi_0 I_2} 
    \nonumber \\ & \quad - \dot{H} (G_{\text{Tele},J_{3}} + 3 G_{\text{Tele},I_2 I_2}) - \ddot{\phi_0} G_{\text{Tele},X I_2}
    \big]  + H \big[
    -2 G_{4,\phi_0} + G_{\text{Tele},I_2} + 18 \dot{H} G_{\text{Tele},T_{\text{vec}} I_2} 
    \nonumber \\ & \quad
    - 12 \dot{H} G_{\text{Tele},T I_2} + 2 X (3 G_{3,X} - 10 G_{4,\phi_0 X} + 2 G_{5,\phi_0 \phi_0} - G_{\text{Tele},\phi_0 J_{3}} - 3 \dot{H} G_{\text{Tele},I_2 J_3}) - \ddot{\phi_0} (G_{\text{Tele},J_3} 
    \nonumber \\ & \quad
    + 3 G_{\text{Tele},I_2 I_2} + 2 X G_{\text{Tele},X J_3}) \big] \Big] \,, \\
    \mathcal{A}_{9} &= a \Big[
    -2 G_{4,\phi_0} + 2 X [G_{3,X} + H^2 (3 G_{5,X} + 2 X G_{5,XX}) - 2 G_{4,\phi_0 X}] + G_{\text{Tele},I_2} + H \dot{\phi_0} \big[4 ( G_{4,X} + 2 X G_{4,XX} 
    \nonumber \\ & \quad
    -G_{5,\phi_0} - X G_{5,\phi_0 X}) + G_{\text{Tele},J_3} \big]
    \Big]\,, \\ 
    \mathcal{A}_{10} &= - 2 a \Big[ H(-6 G_{\text{Tele},T_{\text{vec}}} + 4 G_{\text{Tele},T}) + \dot{\phi_0} G_{\text{Tele},I_{2}} \Big]\,, \\
    \mathcal{A}_{11} &= \frac{2}{9}a \Big[ 18 ( G_{\text{Tele},T_{\text{vec}}} - G_{\text{Tele},T}) + X (4 G_{\text{Tele},J_{8}} + 10 G_{\text{Tele},J_{5}} + 3 G_{\text{Tele},J_{3}} - 4 X G_{\text{Tele},J_{6}} ) \Big]\,, \\ 
    \mathcal{A}_{12} &= 4 a \Big[ - G_{4} + X (2 G_{4,X} - G_{5,\phi_0} + H \dot{\phi_0} G_{5,X}) \Big] \,, \\
    \mathcal{A}_{13} &= \frac{1}{9} a^2 \Big[2 X (-2 G_{\text{Tele},J_8} + 2 X G_{\text{Tele},J_{6}} - 5 G_{\text{Tele},J_{5}} + 3 G_{\text{Tele},J_3}) + 9 G_{\text{Tele},T_{\text{vec}}} \Big]\,, \\
    \mathcal{A}_{14} &= \frac{2}{9} \Big[ - 18 (G_{4} - G_{\text{Tele},T} + G_{\text{Tele},T_\text{vec}})  + X (36 G_{4,X} - 3 G_{\text{Tele},J_{3}} - 2 (9 G_{5,\phi_0} + 2 G_{\text{Tele},J_{8}} + 5 G_{\text{Tele},J_{5}}))  \nonumber \\ & \qquad  + 18 \dot{\phi_0} X H G_{5,X} + 4 X^2 G_{\text{Tele},J_{6}} \Big] \,, \\
    \mathcal{A}_{15} &= - 3 \Big[ 12 H^3 \left(9 G_{\text{Tele},T_\text{vec}T_\text{vec}} + 4 (-3G_{\text{Tele},T T_\text{vec}} + G_{\text{Tele},TT}) \right) + \dot{\phi_0} (-2 G_{4,\phi_0} + G_{\text{Tele},I_{2}} 
    \nonumber \\ & \quad
    + 2X (G_{3,X} - 2 G_{4,\phi_0 X} + G_{\text{Tele},X I_{2}}) ) + H^2 \dot{\phi_0} (-54 G_{\text{Tele},T_\text{vec} I_2} + 36 G_{\text{Tele},T I_{2}} + 10 X G_{5,X} + 4 X^2 G_{5,XX}) 
    \nonumber \\ & \quad
    + 2H (-2 G_{4} - 3 G_{\text{Tele},T_\text{vec}} + 2 G_{\text{Tele},T} +2X (4G_{4,X} + 2 G_{\text{Tele},X T} - 3 (G_{5,\phi_0} - G_{\text{Tele},I_{2}I_{2}} + G_{\text{Tele},X T_\text{vec}})) 
    \nonumber \\ & \quad
    + 4 X^2 (2 G_{4,XX} - G_{5,\phi_0 X})) \Big]\,, \\
    \mathcal{A}_{16} &= G_{2,\phi_0} + G_{\text{Tele},\phi_0} + 6 H^2 \big[G_{4,\phi_0} + 3 G_{\text{Tele},\phi_0 T_{\text{vec}}} - 2 G_{\text{Tele},\phi T} + X (-4 G_{4,\phi_0 X} - 4 X G_{4,\phi_0 X X} + 3 G_{5,\phi_0 \phi_0} 
    \nonumber \\ & \quad
    + 2 X G_{5,\phi_0 \phi_0 X}) \big] - 2 X (G_{2,\phi_0 X} - G_{3, \phi_0 \phi_0} + G_{\text{Tele},\phi_0 X}) - 2 X H^3 \dot{\phi_0} (5 G_{5,\phi_0 X} + 2 X G_{5,\phi_0 X X}) 
    \nonumber \\ & \quad
    - 6 H \dot{\phi_0} (X G_{3,\phi_0 X} - G_{4,\phi_0 \phi_0} - 2 X G_{4,\phi_0 \phi_0 X} + G_{\text{Tele},\phi_0  I_2}) \,,\\
    \mathcal{A}_{17} &= - 2 G_{4,\phi_0} + 2 X (G_{3,X} - 2 G_{4,\phi_0 X} + H^2 (3 G_{5,X} + 2 X G_{5,XX})) + G_{\text{Tele},I_2} + H \dot{\phi_0} (4 G_{4,X} + 8 X G_{4,XX} 
    \nonumber \\ & \quad 
    - 4 (G_{5,\phi_0} + X G_{5,\phi_0 X}) + G_{\text{Tele},J_{3}})\,, \\
    \mathcal{A}_{18} &= -2 H^3 \big[ 15 X G_{5,X} + 4 X^2 (5 G_{5,XX} + X G_{5,XXX}) - 27 G_{\text{Tele},T_{\text{vec}} I_2} + 18 G_{\text{Tele},T I_2} \big] - 3 H \big[-2 G_{4,\phi_0} 
    \nonumber \\ & \quad
    + G_{\text{Tele},I_2} + 2 X (3 G_{3,X} + 2 X G_{3,XX} - 8 G_{4,\phi_0 X} - 4 X G_{4,\phi_0 XX} + 3 G_{\text{Tele},X I_2}) \big] - 6 H^2 \dot{\phi_0} \big[3 G_{4,X} 
    \nonumber \\ & \quad 
    + X ( 12 G_{4,XX} + 4 X G_{4,XXX} - 7 G_{5,\phi_0 X} - 2 X G_{5,\phi_0 XX} ) - 3( G_{5,\phi_0} - G_{\text{Tele},I_2 I_2} + G_{\text{Tele},X T_{\text{vec}}} )  
    \nonumber \\ & \quad
    + 2 G_{\text{Tele},X T} \big] - \dot{\phi_0} \big[ G_{2,X} - 2 G_{3,\phi_0} + G_{\text{Tele},X} + 2 X (G_{2,XX} - G_{3,\phi_0 X} + G_{\text{Tele},XX} ) \big]\,, \\
    \mathcal{A}_{19} &= \frac{1}{9} \big[ 2 X (- 2 G_{\text{Tele},J_{8}} + 2 X G_{\text{Tele},J_{6}} - 5 G_{\text{Tele},J_5} + 3 G_{\text{Tele},J_3}) + 9 G_{\text{Tele},T_{\text{vec}}} \big]\,,  \\
    \mathcal{A}_{20} &= 18 H^4 ( 9 G_{\text{Tele},T_\text{vec} T_\text{vec}} - 12 G_{\text{Tele},T T_\text{vec}} + 4 G_{\text{Tele}, T T} ) + 3 H^2 \big[ -2 G_4 - 3 G_{\text{Tele},T_\text{vec}} + 2 G_{\text{Tele},T}
    \nonumber \\ &  \quad
    + 2 X ( 7 G_{4,X} + X (16 G_{4,XX} + 4 X G_{4,XXX} - 9 G_{5,\phi_0 X} - 2 X G_{5,\phi_0 X X}) - 6 (G_{5,\phi_0} - G_{\text{Tele},I_2 I_2} + G_{\text{Tele},X T_\text{vec}}) 
    \nonumber \\ & \quad
    + 4 G_{\text{Tele},X T}) \big] + X \big[ G_{2,X} - 2 G_{3,\phi_0} + G_{\text{Tele},X} + 2 X ( G_{2,XX} - G_{3,\phi_0 X} + G_{\text{Tele},XX} ) \big] + 3 H \dot{\phi_0} \big[ -2 G_{4,\phi_0} 
    \nonumber \\ & \quad
    + G_{\text{Tele},I_2} + 2 X ( 2 G_{3,X} + X G_{3,XX} - 5 G_{4,\phi_0 X} - 2 X G_{4,\phi_0 X X} + 2 G_{\text{Tele},X I_2} ) \big] + 2 H^3 \dot{\phi_0} \big[ 15 X G_{5,X}  
    \nonumber \\ & \quad
    +13 X^2 G_{5,XX} + 2 X^3 G_{5,XXX} - 54 G_{\text{Tele},T_\text{vec} I_2} + 36 G_{\text{Tele},T I_2} \big]\,, \\
    \mathcal{A}_{21} &= \frac{2}{9} \big[ 9 ( G_4 + 2 G_{\text{Tele},T_\text{vec}} - G_{\text{Tele},T} ) + X ( -9 G_{5,\phi_0} - 2 G_{\text{Tele},J_8} + 2 X G_{\text{Tele},J_6} - 5 G_{\text{Tele},J_5} - 6 G_{\text{Tele},J_3} \nonumber \\ & \quad 
    - 9 \ddot{\phi_0} G_{5,X}) \big]\,, \\
    \mathcal{A}_{22} &= - 6 G_{4} + 12 X G_{4,X} - 6 X G_{5,\phi_0} + 9 X G_{\text{Tele},I_2 I_2} - 9 G_{\text{Tele},T_\text{vec}} + 6 G_{\text{Tele},T} + 18 H^2 [9 G_{\text{Tele},T_\text{vec} T_\text{vec}} 
    \nonumber \\ & \quad
    -  12 G_{\text{Tele},T T_\text{vec}} + 4 G_{\text{Tele},T T}] + 6 H \dot{\phi_0} [X G_{5,X} - 9 G_{\text{Tele},T_\text{vec} I_2} + 6 G_{\text{Tele},T I_2}]\,, \\
    \mathcal{A}_{23} &= -4 X H^2 G_{5,X} + 4 G_{4,\phi_0} - 2 G_{\text{Tele},I_2} + 4 X ( -2 G_{4,\phi_0 X} + G_{5,\phi_0 \phi_0} - \dot{H} G_{5,X} ) + 4 \ddot{\phi_0} (- G_{4,X} - 2 X G_{4,XX}
    \nonumber \\ & \quad 
    + G_{5,\phi_0} + X G_{5,\phi_0 X}) + H \dot{\phi_0} (-4 G_{4,X} + 4 G_{5,\phi_0} - 4 X G_{5,\phi_0 X} + G_{\text{Tele},J_{3}} - 4 \ddot{\phi_0}(G_{5,X} + X G_{5,XX}) ) \,, \\
    \mathcal{A}_{24} &= -3 \big[ 2 X H^3 (3 G_{5,X} + 2 X G_{5,XX}) + H (6 X G_{3,X} -2 G_{4,\phi_0} + 4 X (-5 G_{4,\phi_0 X} + G_{5,\phi_0 \phi_0}) + 3 G_{\text{Tele}, I_2} 
    \nonumber \\ & \quad
    + 6 G_{\text{Tele},\phi_0 T_\text{vec}} - 4 G_{\text{Tele},\phi_0 T} ) + 2 H^2 \dot{\phi_0} (3 G_{4,X} + 6 X G_{4,XX} - 3 G_{5,\phi_0} - 4 X G_{5,\phi_0 X} ) + \dot{\phi_0} (G_{2,X} -2 G_{3,\phi_0} 
    \nonumber \\ & \quad 
    + 2 G_{4,\phi_0 \phi_0} + G_{\text{Tele},X} - G_{\text{Tele},\phi_0 I_2}) \big]\,, \\
    \mathcal{A}_{25} &= 3 \big[ 2 X G_{3,X} - 2 G_{4,\phi_0} - 4 X G_{4,\phi_0 X} + G_{\text{Tele},I_2} + 2 H^2 (3 X G_{5,X} + 2 X^2 G_{5,XX} - 9 G_{\text{Tele},T_\text{vec} I_2} + 6 G_{\text{Tele},T I_2} ) 
    \nonumber \\ & \quad 
    + 2 X G_{\text{Tele},X I_2} + H \dot{\phi_0} (4 G_{4,X} + 8 X G_{4,XX} - 4 G_{5,\phi_0} - 4 X G_{5,\phi_0 X} + 3 G_{\text{Tele},I_2 I_2} - 6 G_{\text{Tele},X T_\text{vec}} 
    \nonumber \\ & \quad
    + 4 G_{\text{Tele},X T}) \big]\,, \\ 
    \mathcal{A}_{26} &= -\frac{1}{2} \left(G_{2,X} + G_{\text{Tele},X} \right) + G_{3,\phi_0}  - (2 \dot{H} + 3 H^2) ( G_{4,X} - G_{5,\phi_0})  - 2 H^3\dot{\phi_0} (G_{5,X}+X G_{5,XX}) (H^2 + \dot{H}) 
    \nonumber \\ & \quad 
    - (\ddot{\phi_0} + 2 H \dot{\phi_{0}}) (G_{3,X}-3G_{4,\phi_0 X}) + H^2 (5 X (-2 G_{4,XX} + G_{5,\phi_0 X}) - 2X^2 (G_{5,\phi_0 XX} + \ddot{\phi_0} G_{5,XXX})) 
    \nonumber \\ & \quad 
    + 2 H \dot{\phi_0} ( - \dot{H} G_{5,X} -3 \ddot{\phi_0} G_{4,XX} + 2 \ddot{\phi_0} G_{5,\phi_0 X} + X(-2 G_{4,\phi_0 XX} + G_{5,\phi_0 \phi_0 X} - \dot{H} G_{5, XX} + \ddot{\phi_0} ( G_{5,\phi_0 XX}
    \nonumber \\ & \quad 
    -2 G_{4,XXX}) )) - X(G_{3,\phi_0 X} - 2 G_{4,\phi_0 \phi_0 X} + 2 \dot{H} (2 G_{4,XX} - G_{5,\phi_0 X}) + \ddot{\phi_0} (G_{3,XX} - 2 G_{4,\phi_0 XX})) \big]\,, \\ 
    \mathcal{A}_{27} &= \frac{1}{2} \big[-6 H^4 X (3 G_{5,\phi_0 X} + 2 X G_{5,\phi_0 XX}) + G_{2,\phi_0 \phi_0} + G_{\text{Tele},\phi_0 \phi_0} + 3 \dot{H} (2 G_{4,\phi_0 \phi_0} - G_{\text{Tele},\phi_0 I_{2}}) - 3 H \dot{\phi_0} \big( G_{2,\phi_0 X} 
    \nonumber \\ & \quad 
    - 2 G_{3,\phi_0 \phi_0} + 2 X G_{3,\phi_0 \phi_0 X} - 4 X G_{4,\phi_0 \phi_0 \phi_0 X} + G_{\text{Tele},\phi_0 X} + G_{\text{Tele},\phi_0 \phi_0 I_{2}} + \dot{H} (4 G_{4,\phi_0 X} + 8 X G_{4,\phi_0 XX} 
    \nonumber \\ & \quad 
    - 4 G_{5,\phi_0 \phi_0} - 4 X G_{5,\phi_0 \phi_0 X} + 3 G_{\text{Tele},\phi_0 I_2 I_2} - 6 G_{\text{Tele},\phi_0 X T_{\text{vec}}} + 4 G_{\text{Tele},\phi_0 X T})\big) - \ddot{\phi_0}(G_{2,\phi_0 X} - 2 G_{3,\phi_0 \phi_0}  
    \nonumber \\ & \quad 
    + G_{\text{Tele},\phi_0 X}) - 6 H \dot{\phi_0} \ddot{\phi_0} (G_{3,\phi_0 X} - 3 G_{4,\phi_0 \phi_0 X} + X (G_{3,\phi_0 XX} - 2 G_{4,\phi_0 \phi_0 X X}) + G_{\text{Tele},\phi_0 X I_{2}} ) 
    \nonumber \\ & \quad 
    + 3 H^2 (4 G_{4,\phi_0 \phi_0} + 2 X (-3 G_{3,\phi_0 X} + 6 G_{4,\phi_0 \phi_0 X} + G_{5,\phi_0 \phi_0 \phi_0} + 2 X (-2 G_{4,\phi_0 \phi_0 X X} + G_{5,\phi_0 \phi_0 \phi_0 X}))  
    \nonumber \\ & \quad 
    - 3 G_{\text{Tele},\phi_0 I_{2}} - 2 \dot{H} (3 X G_{5,\phi_0 X} + 2 X^2 G_{5,\phi_0 XX} - 9 G_{\text{Tele},\phi_0 T_{\text{vec}} I_{2}} + 6 G_{\text{Tele},\phi_0 T I_{2}}) - 2 G_{4,\phi_0 X} \ddot{\phi_0}  
    \nonumber \\ & \quad 
    + (2 G_{5,\phi_0 \phi_0} + 2 X (-8 G_{4,\phi_0 XX} + 5 G_{5,\phi_0 \phi_0 X} + 2 X (-2 G_{4,\phi_0 XXX} + G_{5,\phi_0 \phi_0 XX})) - 3 G_{\text{Tele},\phi_0 I_2 I_2} ) \ddot{\phi_0} ) 
    \nonumber \\ & \quad 
    + 2 X (- G_{2,\phi_0 \phi_0 X} + G_{3,\phi_0 \phi_0 \phi_0} - G_{\text{Tele},\phi_0 \phi_0 X} - 3 \dot{H} (G_{3,\phi_0 X} - 2 G_{4,\phi_0 \phi_0 X} + G_{\text{Tele},\phi_0 X I_{2}}) - \ddot{\phi_0} (G_{2,\phi_0 XX} 
    \nonumber \\ & \quad 
    - G_{3,\phi_0 \phi_0 X} + G_{\text{Tele},\phi_0 XX})) - 2 H^3 \dot{\phi_0} (9 G_{4,\phi_0 X} - 9 G_{5,\phi_0 \phi_0} + 3 \ddot{\phi_0} G_{5,\phi_0 X} + X (18 G_{4,\phi_0 XX} - 7 G_{5,\phi_0 \phi_0 X} 
    \nonumber \\ & \quad 
    + 2 X G_{5,\phi_0 \phi_0 XX} + (7 G_{5,\phi_0 XX} + 2 X G_{5,\phi_0 XXX}) \ddot{\phi_0})) \big]\,, \\
    \mathcal{A}_{28} &= \frac{1}{2} \left( G_{2,X} +G_{\text{Tele},X} \right) - G_{3,\phi_0} + X( G_{2,XX} - G_{3,\phi_0 X} + G_{\text{Tele},XX}) + 3 H \dot{\phi_0} ( G_{3,X} -3 G_{4,\phi_0 X} 
    \nonumber \\ & \quad
    +X( G_{3,XX}-2 G_{4,\phi_0 X}) + G_{\text{Tele},X I_2}) +  H^3 \dot{\phi_0} (3 G_{5,X} + 7 X G_{5,XX} +2 X^2 G_{5,XXX})
    \nonumber \\ & \quad 
    + 3H^2 ( G_{4,X} - G_{5,\phi_0} + X (8 G_{4,XX}-5 G_{5, \phi_0 X}) + 2 X^2(2 G_{4,XXX}-G_{5,\phi_0 XX}) + \frac{3}{2} G_{\text{Tele},I_2 I_2}) \Big]\,.
\end{align}
\end{subequations}

Furthermore, the coefficients of the gauge invariant variables for the scalar quadratic action \eqref{eq:scalar_gauge_invariant} are the following
\begin{subequations}
\allowdisplaybreaks
\begin{align} 
    \tilde{\mathcal{A}}_{1} &=  X \big(G_{2,X} - 2 G_{3, \phi_0} + X G_{\text{Tele},X}\big) + 2 X^2 \big(G_{2,XX} - G_{3,\phi_0 X} + G_{\text{Tele},XX}\big) - 9 H^2 G_{\text{Tele},T_{\text{vec}}} +6 X H \dot{\phi_0} \big(2 G_{3,X} 
    \nonumber \\ & \quad 
    + X G_{3,XX} - 5 G_{4,\phi_0 X} - 2 X G_{4,\phi_0 XX} + 2 G_{\text{Tele},XI_{2}} + 5 H^2 G_{5,X}\big) + 6 H^2 \big(-G_{4} + G_{\text{Tele},T} + 7 X G_{4,X} 
    \nonumber \\ & \quad 
    + 4 X G_{\text{Tele},XT} + 6 X (-G_{5,\phi_0} + G_{\text{Tele},I_2 I_2} -G_{\text{Tele},XT_{\text{vec}}})\big) + 6 X^2 H^2 \big(4(4G_{4,XX} + X G_{4,XXX}) - 9 G_{5,\phi_0 X}\big)
    \nonumber \\ & \quad 
     +2 H^3 X^2 \big(13 \dot{\phi_0} G_{5,XX} + 2 X \dot{\phi_0} G_{5,XXX}\big) + 3 H \dot{\phi_0} \big(G_{\text{Tele},I_{2}} - 2 G_{4,\phi} - 6 H^2 (6G_{\text{Tele},T_{\text{vec}} I_2} - 4 G_{\text{Tele},T I_{2}})\big)
    \nonumber \\ & \quad 
    + 18 H^4 \big(9 G_{\text{Tele},T_{\text{vec}} T_{\text{vec}}} - 12 G_{\text{Tele},T T_{\text{vec}}} + 4 G_{\text{Tele},TT}\big) - 12 X^3 H^2 G_{5,\phi_0 XX} \,, \\
    \tilde{\mathcal{A}}_{2} &= \frac{1}{9} \left(9 G_{\text{Tele},T_{\text{vec}}} + 6 X G_{\text{Tele},J_{3}} - 10X G_{\text{Tele},J_{5}} + 4X (X G_{\text{Tele},J_{6}} - G_{\text{Tele},J_{8}}) \right) \,, \\
    \tilde{\mathcal{A}}_{3} &= 2 \big(G_{4} - X(\ddot{\phi_0} G_{5,X} + G_{5,\phi_0}) + 2 G_{\text{Tele},T_{\text{vec}}} - G_{\text{Tele},T}\big) + \frac{X}{9} \big( 4(X G_{\text{Tele},J_{6}} - G_{\text{Tele},J_{8}} - 3 G_{\text{Tele},J_{3}})
    \nonumber \\ & \quad 
    - 10 X G_{\text{Tele},J_{5}} \big) \,, \\
    \tilde{\mathcal{A}}_{4} &= 6 \big(X(2 G_{4,X} - G_{5,\phi_0}) -G_{4} + G_{\text{Tele},T}\big) + 6 H \dot{\phi_0} \big(X G_{5,X} + 6 G_{\text{Tele},T I_{2}} - 9 G_{\text{Tele},T_{\text{vec}} I_{2}}\big)
    \nonumber \\ & \quad 
    + 9 \big(X G_{\text{Tele},I_{2} I_{2}} - G_{\text{Tele},T_{\text{vec}}} + 2 H^2 (9 G_{\text{Tele},T_{\text{vec}} T_{\text{vec}}} - 12 G_{\text{Tele},T T_{\text{vec}}} + 4 G_{\text{Tele},TT})\big) \,, \\
    \tilde{\mathcal{A}}_{5} &= \frac{1}{2} \big( G_{2,\phi_0 \phi_0} + G_{\text{Tele},\phi_0 \phi_0} - \ddot{\phi_0}\left(G_{2,\phi_0 X}+ G_{\text{Tele},\phi_0 X} \right)\big) + 3 \dot{H} \left(G_{4,\phi_0 \phi_0} -\frac{1}{2} G_{\text{Tele},\phi_0 I_2} \right) + \ddot{\phi_0} G_{3,\phi_0 \phi_0}
    \nonumber \\  & \quad  
    + X \Big(-G_{2,\phi_0 \phi_0 X} -G_{\text{Tele},\phi_0 \phi_0 X} + G_{3,\phi_0 \phi_0 \phi_0} - 3\dot{H}\left(G_{3,\phi_0 X} -2 G_{4,\phi_0 \phi_0 X} + G_{\text{Tele},\phi_0 X I_2} \right) - \ddot{\phi_0} \big(G_{2,\phi_0 X X} 
    \nonumber \\ & \quad 
    - G_{3,\phi_0 \phi_0 X}  +G_{\text{Tele},\phi_0 XX} \big)\Big) + H \dot{\phi_0} \Big[-\frac{3}{2} \big( G_{2,\phi_0 X} - 2 G_{3,\phi_0 \phi_0} + G_{\text{Tele},\phi_0 X}+ G_{\text{Tele},\phi_0 \phi_0 I_2} \big) + \dot{H} \big(-6G_{4,\phi_0 X} 
    \nonumber \\ & \quad 
    +6G_{5,\phi_0 \phi_0} -\frac{9}{2}G_{\text{Tele},\phi_0 I_2 I_2} - 6 G_{\text{Tele},\phi_0 X T} + 9G_{\text{Tele},\phi_0 X T_{\text{vec}}}\big)-3 \ddot{\phi_0}\big(G_{3,\phi_0 X} - 3 G_{4,\phi_0 \phi_0 X} + G_{\text{Tele},\phi_0 X I_2} \big) 
    \nonumber \\ & \quad 
    +3 X \Big(-G_{3,\phi_0 \phi_0 X} + 2 G_{4,\phi_0 \phi_0 \phi_0 X} + \dot{H} \big(-4G_{4,\phi_0 XX}+2G_{5,\phi_0 \phi_0 X} \big) +\ddot{\phi_0} \big(-G_{3,\phi_0 XX}+2G_{4,\phi_0 \phi_0 XX} \big) \Big) \Big] 
    \nonumber \\ & \quad 
    + 3 H^2 \Big[2 G_{4,\phi_0 \phi_0}-\frac{3}{2} ( G_{\text{Tele},\phi_0 I_2} + \ddot{\phi_0} G_{\text{Tele},\phi_0 I_2 I_2})+ \dot{H}\big(-6G_{\text{Tele},\phi_0 T I_2}+9G_{\text{Tele},\phi_0 T_{\text{vec}} I_2} \big) + \ddot{\phi_0}\big(-G_{4,\phi_0 X} 
    \nonumber \\ & \quad 
    + G_{5,\phi_0 \phi_0} \big) +X \big(-3G_{3,\phi_0 X} + 6 G_{4,\phi_0 \phi_0 X} + G_{5,\phi_0 \phi_0 \phi_0} -3 \dot{H}G_{5,\phi_0 X} - 8 \ddot{\phi_0}G_{4,\phi_0 XX} + 5 \ddot{\phi_0} G_{5,\phi_0 \phi_0 X} \big)  
    \nonumber \\ & \quad 
    + 2 X^2 \big(-2G_{4,\phi_0 \phi_0 XX} + G_{5,\phi_0 \phi_0 \phi_0 X} - \dot{H}G_{5,\phi_0 XX} - 2 \ddot{\phi_0}G_{4,\phi_0 XXX}+ \ddot{\phi_0} G_{5,\phi_0 \phi_0 XX} \big) \Big] - H^3 \dot{\phi_0} \Big[3 (3 G_{4,\phi_0 X} 
    \nonumber \\ & \quad 
    -3 G_{5,\phi_0 \phi_0} + \ddot{\phi_0}G_{5,\phi_0 X}) + X \big(18G_{4,\phi_0 XX}-7 (G_{5,\phi_0 \phi_0 X} -\ddot{\phi_0}G_{5,\phi_0 XX}) \big) + 2 X^2 \big(G_{5,\phi_0 \phi_0 XX}  
    \nonumber \\ & \quad 
    + \ddot{\phi_0}G_{5,\phi_0 XXX} \big) \Big] - 3 X H^4\big(3 G_{5,\phi_0 X} + 2 X G_{5,\phi_0 XX} \big) \,, \\
    \tilde{\mathcal{A}}_{6} &= \frac{1}{2}\Big[G_{2,X} - 2 G_{3,\phi_0} + 2 X (G_{\text{Tele},XX} + G_{2,XX} - G_{3,\phi_0 X}) + 6 H\dot{\phi_0} (G_{3,X} + H^2 G_{5,X} - 3 G_{4,\phi_0 X} - 2 X G_{4,\phi_0 XX}
    \nonumber \\ & \quad 
    + G_{\text{Tele},X I_{2}}) + 2 X H \dot{\phi_0} (2 G_{3,XX} + 7 H^2 G_{5,XX}) + G_{\text{Tele},X} + 6 H^2 \big(G_{4,X} + 4 X (2 G_{4,XX} + X G_{4,XXX})
    \nonumber \\ & \quad 
    - G_{5,\phi_0} - 5 X G_{5,\phi_0 X} - 2 X^2 G_{5,\phi_0 XX} \big) + 4 X^2 H^3 \dot{\phi_0} G_{5,XXX} + 9 H^2 G_{\text{Tele},I_{2} I_{2}}\Big] \,, \\
    \tilde{\mathcal{A}}_{7} &= - \frac{1}{2} \big( G_{2,X} + G_{\text{Tele},X}\big) + G_{3,\phi_0} + \ddot{\phi_0} (2 X G_{4,\phi_0 XX} + 3 G_{4,\phi_0 X} - H^2 G_{5,X} -G_{3,X} - X G_{3,XX}) + 5 X H^2 (G_{5,\phi_0 X} 
    \nonumber \\ & \quad 
    -\ddot{\phi_0} G_{5,XX} - 2 G_{4,XX}) + 2 X H \dot{\phi_0}(-\dot{H} G_{5,XX} - 2 \ddot{\phi_0} G_{4,XXX} - 2 G_{4,\phi_0 XX} + \ddot{\phi} G_{5,\phi_0 XX} + G_{5,\phi_0 \phi_0 X})
    \nonumber \\ & \quad
    - 2 X H^2 (H \dot{\phi_0} G_{5,XX} + X \ddot{\phi_0} G_{5,XXX} + X G_{5,\phi_0 XX}) + 3 H^2 (G_{5,\phi_0} - G_{4,X}) + 2 \dot{H} (G_{5,\phi} - G_{4,X})
    \nonumber \\ & \quad 
    + 2 X \dot{H} (G_{5,\phi_0 X} - 2 G_{4,XX}) + 2 H \dot{\phi_0}(-G_{3,X} - (H^2 + \dot{H}) G_{5,X} - 3 \ddot{\phi_0} G_{4,XX} + 3 G_{4,\phi_0 X} + 2 \ddot{\phi_0} G_{5,\phi X})
    \nonumber \\ & \quad 
    + X (2G_{4,\phi_0 \phi_0 X} - G_{3,\phi_0 X}) \,, \\ 
    \tilde{\mathcal{A}}_{8} &= \frac{1}{3} a^2 \Big[
    X (4 G_{\text{Tele},J_{8}} + G_{\text{Tele},J_{5}} + 3 G_{\text{Tele},I_{2}I_{2}}) - 3 G_{\text{Tele},T_{\text{vec}}} + 6 H \big( H (9 G_{\text{Tele},T_{\text{vec}}T_{\text{vec}}} -12 G_{\text{Tele},T T_{\text{vec}}} 
    \nonumber \\ & \quad 
    + 4 G_{\text{Tele},TT}) + \dot{\phi_0} (-3 G_{\text{Tele},T_{\text{vec}}I_{2}} + 2 G_{\text{Tele},T I_{2}}) \big)
    \Big]\,, \\
    \tilde{\mathcal{A}}_{9} &= 3 \big[
    -12 H^3 [9 G_{\text{Tele},T_{\text{vec}}T_{\text{vec}}} + 4 (-3 G_{\text{Tele},T T_{\text{vec}}} + G_{\text{Tele},TT})] + 2 H [ 2 G_{4} + 4 X^2 (-2 G_{4,XX} + G_{5,\phi_0 X}) 
    \nonumber \\ & \quad 
    + 3 G_{\text{Tele},T_{\text{vec}}} - 2 G_{\text{Tele},T} + 2 X ( -4 G_{4,X} + 3 G_{5,\phi_0} - 3 G_{\text{Tele},I_{2}I_{2}} + 3 G_{\text{Tele},X T_{\text{vec}}} - 2 G_{\text{Tele},XT} ) ] 
    \nonumber \\ & \quad 
    - 2 H^2 \dot{\phi_0} (5 X G_{5,X} + 2X^2 G_{5,XX} - 27 G_{\text{Tele},T_{\text{vec}}I_{2}} + 18 G_{\text{Tele},T I_{2}}) - \dot{\phi_0} [-2 G_{4,\phi_0} + G_{\text{Tele},I_{2}} + 2 X (G_{3,X} 
    \nonumber \\ & \quad 
    - 2 G_{4,\phi_0 X} + G_{\text{Tele},X I_{2}})]
    \big]\,, \\
    \tilde{\mathcal{A}}_{10} &= \frac{2}{9} \big[
    18 (G_{4} + G_{\text{Tele},T_{\text{vec}}} - G_{\text{Tele},T}) + X (-36 G_{4,X} + 18 G_{5,\phi_0} + 4 G_{\text{Tele},J_{8}} - 4 X G_{\text{Tele},J_{6}} + 10 G_{\text{Tele},J_{5}} 
    \nonumber \\ & \quad 
    + 3 G_{\text{Tele},J_{3}} - 18 H \dot{\phi_0} G_{5,X})
    \big]\,, \\
    \tilde{\mathcal{A}}_{11} &= G_{2,\phi_0} + G_{\text{Tele},\phi_0} + 6H^2 [ G_{4,\phi_0} + X (-4 G_{4,\phi_0 X} + 3 G_{5,\phi_0 \phi_0} -4 X G_{4,\phi_0 XX} + 2 X G_{5,\phi_0 \phi_0 X}) + 3 G_{\text{Tele},\phi_0 T_{\text{vec}}} 
    \nonumber \\ & \quad 
    -2 G_{\text{Tele},\phi_0 T} ] - 2 X (G_{2,\phi_0 X} - G_{3,\phi_0 \phi_0} + G_{\text{Tele},\phi_0 X}) - 2 H^3 X \dot{\phi_0} (5 G_{5,\phi_0 X} + 2 X G_{5,\phi_0 XX}) - 6 H \dot{\phi_0} [-G_{4,\phi_0 \phi_0} 
    \nonumber \\ & \quad 
    + X (G_{3,\phi_0 X} - 2 G_{4,\phi_0 \phi_0 X}) + G_{\text{Tele},\phi_0 I_{2}}]\,, \\ 
    \tilde{\mathcal{A}}_{12} &= - \big[
    H^{3} (30 X G_{5,X} + 40 X^2 G_{5,XX} + 8 X^3 G_{5,XXX} - 54 G_{\text{Tele},T_{\text{vec}}I_{2}} + 36 G_{\text{Tele},T I_{2}}) + 3 H [-2 G_{4,\phi_0} + G_{\text{Tele},I_{2}} 
    \nonumber \\ & \quad 
    + 2 X (3 G_{3,X} - 8 G_{4,\phi_0 X} + 2 X (G_{3,XX}-2G_{4,\phi_0 XX}) +3 G_{\text{Tele},X I_{2}})] + 6 H^2 \dot{\phi_0} [ 3 G_{4,X} + X (12 G_{4,XX} 
    \nonumber \\ & \quad 
    + 4 X G_{4,XXX} - 7 G_{5,\phi_0 X} - 2 X G_{5,\phi_0 XX}) - 3 ( G_{5,\phi_0} - G_{\text{Tele},I_2 I_2} + G_{\text{Tele},X T_{\text{vec}}} ) + 2 G_{\text{Tele},X T}] + \dot{\phi_0} [ G_{2,X} 
    \nonumber \\ & \quad 
    - 2 G_{3,\phi_0} + G_{\text{Tele},X} + 2 X (G_{2,XX} - G_{3,\phi_0 X} + G_{\text{Tele},XX}) ]
    \big]\,, \\
    \tilde{\mathcal{A}}_{13} &= 
    2 G_{4,\phi_0} - 2 X \big(G_{3,X} + H^2 (3G_{5,X} + 2 X G_{5,XX}) - 2 G_{4,\phi_0 X}\big) - G_{\text{Tele},I_{2}} - H \dot{\phi_0} \big( 4 (G_{4,X} - G_{5,\phi_0} 
    \nonumber \\ & \quad 
    + X(2G_{4,XX} - G_{5,\phi_0 X})) + G_{\text{Tele},J_{3}} \big) \,,\\
    \tilde{\mathcal{A}}_{14} &=  \frac{\tilde{\mathcal{A}}_{9} a}{3} \,, \\
    \tilde{\mathcal{A}}_{15} &= 3\dot{\phi_0}\Big(2G_{3,\phi_0}-G_{2,X}-G_{\text{Tele},X}-2G_{4,\phi_0 \phi_0}+G_{\text{Tele},\phi_0 I_{2}}\Big)+6H^2\dot{\phi_0}\Big(3(G_{5,\phi_0}-G_{4,X})+2X(2G_{5,\phi_0 X}
    \nonumber \\ & \quad 
    -3G_{4,XX})\Big)+3H\Big(2G_{4,\phi_0}+4G_{\text{Tele},\phi_0 T}-6G_{\text{Tele},\phi_0 T_{\text{vec}}}-3G_{\text{Tele},I_{2}}\Big)+6XH\Big(2(5G_{4,\phi_0 X}-G_{5,\phi_0 \phi_0})\\  \nonumber&-3(G_{3,X}+H^2G_{5,X})\Big)-12X^2H^3G_{5,XX}\,, \\
    \tilde{\mathcal{A}}_{16}   &=  3\big(2XG_{3,X}-2G_{4,\phi_0}+G_{\text{Tele},I_{2}}\big)+6H\dot{\phi_0}\big(2G_{4,X}-2G_{5,\phi_0}-3G_{\text{Tele},X T_{\text{vec}}}+2G_{\text{Tele},XT}\big)
    \nonumber \\ & \quad 
    +6X\big(G_{\text{Tele},XI_{2}}-2G_{4,\phi_0 X}\big) + 9 H \dot{\phi_0} G_{\text{Tele},I_{2}I_{2}}+12XH\dot{\phi_0}\big(2G_{4,XX}-G_{5,\phi_0 X}\big)
    \nonumber \\ & \quad +6H^2\Big(6G_{\text{Tele},TI_{2}}-9G_{\text{Tele},T_{\text{vec}}I_{2}}+X(3G_{5,X}+2XG_{5,XX})\Big) \,, \\
    \tilde{\mathcal{A}}_{17} &= 4H\dot{\phi_0}\big(G_{5,\phi_0}-G_{4,X}-\ddot{\phi_0}G_{5,X}-X(\ddot{\phi_0}G_{5,XX}+G_{5,\phi_0 X})\big)+4\ddot{\phi_0}\big(XG_{5,\phi_0 X}+G_{5,\phi_0}-G_{4,X}-2XG_{4,XX}\big)
    \nonumber \\ & \quad 
    + 4 G_{4,\phi_0} - 2 G_{\text{Tele},I_{2}} - 4 X \dot{H} G_{5,X} + 4 X \big(G_{5,\phi_0 \phi_0} - 2 G_{4,\phi_0 X}-H^2 G_{5,X}\big) + H \dot{\phi_0} G_{\text{Tele},J_{3}} \,, \\
    \tilde{\mathcal{A}}_{18} &= -a \Big[4 \big(G_{4} - G_{\text{Tele},T} + X (G_{5,\phi_0} - 2 G_{4,X})\big) + 4 H \dot{\phi_0} \big(-X G_{5,X} + 9 G_{\text{Tele},T_{\text{vec}}I_{2}} - 6 G_{\text{Tele},TI_{2}}\big)
    \nonumber \\ & \quad 
    + 6 \big(G_{\text{Tele},T_{\text{vec}}} - X G_{\text{Tele},I_{2}I_{2}}\big) + 12 H^2 \big(12 G_{\text{Tele},TT_{\text{vec}}} - 9 G_{\text{Tele},T_{\text{vec}}T_{\text{vec}}} - 4 G_{\text{Tele},TT}\big)\Big] \,, \\
    \tilde{\mathcal{A}}_{19} &= -a \Big[\dot{\phi_0} \big(G_{2,X} - 2 G_{3,\phi_0} + 2 G_{4,\phi_0 \phi_0} + G_{\text{Tele},X} - G_{\text{Tele},\phi_0 I_{2}}\big) + H \big(3 G_{\text{Tele},I_2} - 2 G_{4,\phi_0} + 6 G_{\text{Tele},\phi_0 T_{\text{vec}}}
    \nonumber \\ & \quad 
    -4 G_{\text{Tele},\phi_0 T}\big) + 6 H^2 \dot{\phi_0} \big(G_{4,X} -G_{5,\phi_0}\big) + 2 X H \big(3 G_{3,X} + 6 H \dot{\phi_0} G_{4,XX} - 10 G_{4,\phi_0 X} + 2 G_{5,\phi_0 \phi_0}\big)
    \nonumber \\ & \quad 
    +2XH^2\big(3HG_{5,X}+2XHG_{5,XX}-4\dot{\phi_0}G_{5,\phi_0 X}\big)\Big] \,, \\
    \tilde{\mathcal{A}}_{20} &= -a \Big[2 X \big(2 G_{4,\phi_0 X} - G_{3,X} - G_{\text{Tele},XI_{2}}\big) -G_{\text{Tele},I_{2}} + 2G_{4,\phi_0} + H \dot{\phi_0} \big(4(G_{5,\phi_0} - G_{4,X} - 2 X G_{4,XX}
    \nonumber \\ & \quad 
    + X G_{5,\phi_0 X} - G_{\text{Tele},XT}) + 3(2 G_{\text{Tele},XT_{\text{vec}}} - G_{\text{Tele},I_{2}I_{2}})\big) + H^2 \big(6(3 G_{\text{Tele},T_{\text{vec}}I_{2}} - 2 G_{\text{Tele},TI_{2}} - X G_{5,X})
    \nonumber \\ & \quad 
    -4 X^2 G_{5,XX}\big)\Big]\,.
\end{align}
\end{subequations}

For the quadratic vector action \eqref{eq:vector_action} we have the following coefficients
\begin{subequations}
\allowdisplaybreaks
\begin{align}
    \mathcal{C}_{1} &= \frac{1}{18} \left[ 
    2 G_{\text{Tele},T_{\text{ax}}} 
    + 3 [3 (G_{4} - G_{\text{Tele},T}) + X (-6 G_{4,X} + 3G_{5,\phi} + 2 G_{\text{Tele},J_{10}} + 6 G_{\text{Tele},J_{8}} + 3 G_{\text{Tele},J_{5}} - 3 H \dot{\phi} G_{5,X})] \right]\,, \\
    \mathcal{C}_{2} &= G_{\text{Tele},T_{\text{vec}}} + \frac{X}{18} \left[ -2 G_{\text{Tele},J_{8}}
    - 5 G_{\text{Tele},J_{5}} - 6 G_{\text{Tele},J_{3}} + 2 X G_{\text{Tele},J_{6}}\right]\,, \\
    \mathcal{C}_{3} &= 
    2 (G_{4} - G_{\text{Tele},T}) + X [-4 G_{4,X} + 2 G_{5,\phi} + 4 G_{\text{Tele},J_{8}} + G_{\text{Tele},J_{5}} - 2 H \dot{\phi} G_{5,X}]\,, \\
    \mathcal{C}_{4} &= \frac{1}{9} [3 X (4G_{\text{Tele},J_{10}} + 3 G_{\text{Tele},J_{5}}) + 4 G_{\text{Tele},T_{\text{ax}}}]\,, \\
    \mathcal{C}_{5} &= G_{\text{Tele},T_{\text{vec}}} + \frac{2 X}{9} [-2 G_{\text{Tele},J_{8}}
    - 5 G_{\text{Tele},J_{5}} + 3 G_{\text{Tele},J_{3}} + 2 X G_{\text{Tele},J_{6}}]\,, \\
     \mathcal{C}_{6} &= \frac{1}{18}[ 2 G_{\text{Tele},T_{\text{ax}}} +
    9 (G_{4}-G_{\text{Tele},T}) -3 X (6 G_{4,X} - 3 G_{5,\phi} + 4 G_{\text{Tele},J_{10}} - 6 G_{\text{Tele},J_{5}} + 3 H \dot{\phi} G_{5,X})
    ]\,, \\
    \mathcal{C}_{7} &= -\frac{1}{9} [9 (G_{4} - G_{\text{Tele},T}) - 2 G_{\text{Tele},T_{\text{ax}}} + 3 X (- 6 G_{4,X} + 3 G_{5,\phi} - 3  H \dot{\phi} G_{5,X} + G_{\text{Tele},J_{10}} + 3 G_{\text{Tele},J_{5}})]\,, \\
    \mathcal{C}_{8} &=  \dot{\phi} G_{\text{Tele},I_{2}} + 2 H \left(- 3 G_{\text{Tele},T_{\text{vec}}} + 2 G_{\text{Tele},T} \right)\,, \\
    \mathcal{C}_{9} &= 2 \left( G_{4} + X \left(-2 G_{4,X} + G_{5,\phi} - H \dot{\phi} G_{5,X}\right) \right) \,, \\
    \mathcal{C}_{10} &= \frac{1}{9}\left( 18 \left(- G_{\text{Tele},T_{\text{vec}}} + G_{\text{Tele},T}\right) + X \left(-4 G_{\text{Tele},J_{8}} + 4 X G_{\text{Tele},J_{6}} - 10 G_{\text{Tele},J_{5}} - 3 G_{\text{Tele},J_{3}}\right) \right) \,, \\
    \mathcal{C}_{11} &= - \frac{2}{9} \left( 3 X \left(G_{\text{Tele},J_{10}} + 3 G_{\text{Tele},J_{5}}\right) - 2 G_{\text{Tele},T_{\text{ax}}} - 9 G_{\text{Tele},T} \right)\,.
\end{align}
\end{subequations}
While the coefficients $\tilde{\mathcal{C}}_1 - \tilde{\mathcal{C}}_{9}$ for the gauge invariant quadratic vector action \eqref{eq:gauge_invariant_vectors} are
\begin{subequations}
\allowdisplaybreaks
\begin{align}
    \tilde{\mathcal{C}}_1 &= \frac{1}{9} \left(9 G_{\text{Tele}, T_{\text{vec}}} + X (6 G_{\text{Tele}, J_3} - 10 G_{\text{Tele},J_5} + 4 X G_{\text{Tele}, J_6} - 4 G_{\text{Tele}, J_8})\right) \,, \\
    \tilde{\mathcal{C}}_2 &= \frac{1}{9} \left(9 (G_4 - G_{\text{Tele}, T}) + 2 G_{\text{Tele}, T_{\text{ax}}} + X (9 G_{5,\phi} - 18 G_{4,X} - 12 G_{\text{Tele}, J_{10}} + 18 G_{\text{Tele}, J_{5}} - 9 H \dot{\phi} G_{5,X}) \right) \,, \\
    \tilde{\mathcal{C}}_{3} &= - 2 X \dot{H} \dot{\phi} G_{\text{Tele}, I_2 J_{10}} - 6 X \dot{H} \dot{\phi} G_{\text{Tele}, I_2 J_5} + \frac{4}{3} \dot{H} \dot{\phi} G_{\text{Tele},T_{\text{ax}} I_2} + 6 \dot{H} \dot{\phi} G_{\text{Tele},T I_2} - 8 X H \dot{H} G_{\text{Tele},T J_{10}}
    \nonumber \\ & \quad 
    + 12 X H \dot{H} G_{\text{Tele},T_{\text{vec}} J_{10}} - 24 X  H \dot{H} G_{\text{Tele},T J_5} + 36 X H \dot{H} G_{\text{Tele}, T_{\text{vec}} J_5} + \frac{16}{3} H \dot{H} G_{\text{Tele},T T_{\text{ax}}} - 8 H \dot{H} G_{\text{Tele},T_{\text{ax}} T_{\text{vec}}}
    \nonumber \\ & \quad 
    - 36 H \dot{H} G_{\text{Tele},T T_{\text{vec}}} + 24 H \dot{H} G_{\text{Tele},T T} - 2 X H \ddot{\phi} G_{\text{Tele}, I_2 J_{10}} - 6 X H \ddot{\phi} G_{\text{Tele},I_2 J_5} + \frac{4}{3} H \ddot{\phi} G_{\text{Tele},T_{\text{ax}},I_2}
    \nonumber \\ & \quad 
    + 6 H \ddot{\phi} G_{\text{Tele},T I_2} - \frac{4}{3} X H G_{\text{Tele}, J_{10}} - 4 X H G_{\text{Tele}, J_5} +\frac{8}{9} H G_{\text{Tele},T_{\text{ax}}}-6 H G_{\text{Tele},T_{\text{vec}}} + \dot{\phi} G_{\text{Tele}, I_2}
    \nonumber \\ & \quad 
    - \frac{2}{3} X \dot{\phi} G_{\text{Tele},\phi J_{10}} - \frac{2}{3} X \dot{\phi} \ddot{\phi} G_{\text{Tele},X J_{10}} - \frac{2}{3} \dot{\phi} \ddot{\phi} G_{\text{Tele}, J_{10}} - 2 X \dot{\phi} G_{\text{Tele},\phi J_5} - 2 X \dot{\phi} \ddot{\phi} G_{\text{Tele}, X J_5} - 2 \dot{\phi} \ddot{\phi} G_{\text{Tele}, J_5}
    \nonumber \\ & \quad 
    + \frac{4}{9} \dot{\phi} \ddot{\phi} G_{\text{Tele},X T_{\text{ax}}} + \frac{4}{9} \dot{\phi} G_{\text{Tele},\phi T_{\text{ax}}} + 2 \dot{\phi } \ddot{\phi} G_{\text{Tele},X T} + 2 \dot{\phi} G_{\text{Tele},\phi T} + 8 H G_{\text{Tele},T} \,, \\
    \tilde{\mathcal{C}}_4 &= -\frac{1}{9} \left(9(G_4-G_{\text{Tele}, T}) - 2 G_{\text{Tele}, T_{\text{ax}}} + X (9 G_{5,\phi} - 18 G_{4,X} + 3 G_{\text{Tele}, J_{10}} + 9 G_{\text{Tele}, J_{5}} - 9 H \dot{\phi} G_{5,X}) \right)\,, \\
    \tilde{\mathcal{C}}_5 &= \frac{1}{9} \left(18 G_{\text{Tele}, T_{\text{vec}}} + 4 G_{\text{Tele}, T_{\text{ax}}} + X (4 G_{\text{Tele}, J_{8}} - 4 X G_{\text{Tele}, J_{6}} - 8 G_{\text{Tele}, J_{5}} + 3 G_{\text{Tele}, J_{3}} - 6 G_{\text{Tele}, J_{10}}) \right)\,, \\
    \tilde{\mathcal{C}}_6 &= \frac{1}{9} \left(4 G_{\text{Tele}, T_{\text{ax}}} + X (9 G_{\text{Tele}, J_{5}} + 12 G_{\text{Tele}, J_{10}}) \right) \,, \\
    \tilde{\mathcal{C}}_7 &=\tilde{\mathcal{C}}_6 \,, \\
    \tilde{\mathcal{C}}_8 &= \frac{1}{18} \left(18 G_{\text{Tele}, T_{\text{vec}}} - X(6 G_{\text{Tele}, J_{3}} + 5 G_{\text{Tele}, J_{5}} + 2 G_{\text{Tele}, J_{8}} - 2 X G_{\text{Tele}, J_{6}}) \right) \,, \\
    \tilde{\mathcal{C}}_9 &= \frac{1}{18} \left(9 (G_4-G_{\text{Tele}, T}) + 2 G_{\text{Tele},  T_{\text{ax}}} + X(9 G_{5,\phi} - 18 G_{4,X} + 6 G_{\text{Tele}, J_{10}} + 9 G_{\text{Tele}, J_{5}} + 18 G_{\text{Tele}, J_{8}} - 9 H \dot{\phi} G_{5,X}) \right)\,.
\end{align}
\end{subequations}

\section{Gauge Choices}\label{appendix_gauge_choices}

Section~\ref{sec:cosmological_perturbations} provides a system of equations for an SVT decomposition without imposing any gauge choices. Furthermore, it was shown that the perturbation of the BDLS action can be expressed in terms of gauge invariant quantities given by Eq.~\eqref{eq:gauge_invariant_quantities}. Alternatively, through the gauge transformation in Eq.~\eqref{eq:gauge_invariant_calculations}, the gauge invariant quantities $\{\sigma, u_{i}, h_{ij}\}$ are identified but non-gauge invariant variables dependent on each other can be lumped together to form the following groups: $\{\delta\phi, \Phi, \psi, \beta\}$, $\{B, E\}$, $\{v_{i}, w_{i}, V_{i}\}$. A gauge choice is done by setting one from each group to vanish. Here, we list some of the commonly used gauge choices within the scalar and vector sectors.

All gauge choices can be obtained through the original action perturbations given by Eqs~(\ref{eq:scalar-quadratic-gauge-dependent-action}, \ref{eq:pseudoscalar_action}, \ref{eq:vector_action}, \ref{eq:ten_pert_action}). These calculations cannot be done through the minimized gauge invariant action explored in Sec.~\ref{sec:stability}, as certain gauge choices may result in undefined regions with a null denominator. We present a simplified version of these results obtained through integration by parts and collecting like terms.

\subsection{Scalars}

In the scalar sector, we explore the flat, unitary, Newtonian and synchronous gauge. It is known that in the synchronous gauge, residual gauge freedom remains due to spatial coordinate transformations that are time-independent. These correspond to arbitrary spatial translations or mode-dependent redefinitions of spatial coordinates. To fully fix the gauge, additional conditions are required, such as aligning the perturbation center of mass with a specific fluid or setting one metric potential to zero. Properly addressing this residual freedom is crucial to avoid spurious results and ensure consistent interpretation of physical observables in theoretical and observational cosmology \cite{Ma:1995ey,Malik:2008im,1992ApJ...395...34B,Bruni:1996im,1984PThPS..78....1K}.

\subsubsection{Flat Gauge}
The \textit{flat gauge} is characterized by $\psi=0$ and $E=0$, for which Eq.~\eqref{eq:scalar-quadratic-gauge-dependent-action} simplifies to
\begin{align}\label{flat_gauge_scalar}
     \mathcal{S}_{\rm S}^{\text{Flat}}=\int \dd t \, \dd^3 x a^3 \Bigg[& \dot{\delta \phi}\Bigg(F_1 \Phi-F_2 \frac{\nabla ^2 \beta}{a}+F_3 \frac{\nabla^2 B}{a}\Bigg)+\delta \phi\Bigg(F_4 \Phi-F_5 \frac{\nabla ^2 \beta}{a}+F_6 \frac{\nabla^2 B}{a}-F_2 \frac{\nabla ^2 \Phi}{a^2}\Bigg)-F_7 \frac{\Phi \nabla^2 \beta}{a}
     \\ \nonumber & +F_8 \frac{\Phi \nabla^2 B}{a}+F_{9}\Phi^2 +F_{10}\delta \phi^2+ F_{11}\frac{\left(\nabla \delta \phi\right)^2}{a^2}+ F_{12}\frac{\left(\nabla^2  B \right)^2}{a^2} +F_{13} \Bigg(\nabla \dot{\beta}- \frac{\nabla \Phi}{a} \Bigg)^2 + F_{14} \dot{\delta\phi}^2 \Bigg] 
    \end{align}
    where
    \begin{subequations}
    \allowdisplaybreaks
    \begin{align}
        \nonumber
        F_{1}=& \dot{\phi_0} \big(X \left(-2 G_{\text{Tele},XX}-2 G_{2,XX}+2 G_{3,\phi_0 X}\right)-G_{\text{Tele},X}-G_{2,X}+2 G_{3,\phi_0 }\Big)+H \Big(X \big(-18 G_{\text{Tele},XI_2}-18 G_{3,X}
        \\ \nonumber &+48 G_{4,\phi_0 X}\big)-3 G_{\text{Tele},I_2}+X^2 \left(24 G_{4,\phi_0 XX}-12 G_{3,XX}\right)+6 G_{4,\phi_0 }\big)+H^2 \dot{\phi_0} \big(-18 G_{\text{Tele},I_2 I_2}-12 G_{\text{Tele},XT}
        \\ \nonumber &+18 G_{\text{Tele},XT_{\textrm{vec}}} +X^2 \left(12 G_{5,\phi_0 XX}-24 G_{4,XXX}\right)+X \left(42 G_{5,\phi_0 X}-72 G_{4,XX}\right)-18 G_{4,X}+18 G_{5,\phi_0 }\big)
        \\&+H^3 \big(-36 G_{\text{Tele},TI_2}+54 G_{\text{Tele},T_{\textrm{vec}}I_2}-8 X^3 G_{5,XXX}-40 X^2 G_{5,XX}-30 X G_{5,X}\big)\,,
        \\ \nonumber
        F_{2}=& 2 G_{4,\phi_0 }-G_{\text{Tele},I_2}-X \left(2 G_{3,X}-4 G_{4,\phi_0 X}\right)-H \dot{\phi_0} \Big(G_{\text{Tele},J_3}+X \left(8 G_{4,XX}-4 G_{5,\phi_0 X}\right)+4 G_{4,X}-4 G_{5,\phi_0}\Big)
        \\ & -H^2 \left(4 X^2 G_{5,XX}+6 X G_{5,X}\right)\,,
        \\ 
        F_{3}=& F_2 + H \dot{\phi_0} \left(-3 G_{\text{Tele},I_2 I_2}+G_{\text{Tele},J_3}-4 G_{\text{Tele},X T}+6 G_{\text{Tele},X T_{\text{vec}}}\right)-6 H^2 \left(2 G_{\text{Tele},T I_2}-3 G_{\text{Tele},T_{\text{vec}} I_2}\right)
        \nonumber \\ & -2 X G_{\text{Tele},X I_2} \,,
        \\ \nonumber
        F_{4}=& -2 X G_{\text{Tele},\phi_0 X}+G_{\text{Tele},\phi_0 }-2 X G_{2,\phi_0 X}+G_{2,\phi_0 }+2 X G_{3,\phi_0 \phi_0 } + H \dot{\phi_0} X\big(-6 G_{\text{Tele},\phi_0 I_2}-6 X G_{3,\phi_0 X}
        \\ \nonumber &+12 X G_{4,\phi_0 \phi_0 X} +6 G_{4,\phi_0 \phi_0 }\big)+H^2 X\big(-12 G_{\text{Tele},\phi T}+18 G_{\text{Tele},\phi_0 T_{\text{vec}}}-24 X^2 G_{4,\phi_0 XX}-24 X G_{4,\phi_0 X}
        \\ &+6 G_{4,\phi_0 } + 12 X^2 G_{5,\phi_0 \phi_0 X} +18 X G_{5,\phi_0 \phi_0}\big) +H^3 \dot{\phi_0} X\left(-4 X^2 G_{5,\phi_0 XX}-10 X G_{5,\phi_0 X}\right)\,,
        \\ \nonumber
        F_{5}=& F_6 -H \big(\dot{H} \left( 6 X G_{\text{Tele},I_2 J_3}+12 G_{\text{Tele},T I_2}-18 G_{\text{Tele},T_{\text{vec}} I_2}\right)+\ddot{\phi_0} \left(3 G_{\text{Tele},I_2 I_2}+2 X G_{\text{Tele},X J_3}+ G_{\text{Tele},J_3} \right)
        \\ \nonumber & +2 G_{\text{Tele},I_2}+2 X G_{\text{Tele},\phi_0 J_3}-4 G_{\text{Tele},\phi_0 T}+6 G_{\text{Tele},\phi_0 T_{\text{vec}}}\big)-H^2 \dot{\phi_0} \big(3 \big(4 \dot{H} G_{\text{Tele},T J_3}-6 \dot{H} G_{\text{Tele},T_{\text{vec}} J_3}
        \\ & +\ddot{\phi_0} G_{\text{Tele},I_2 J_3}\big) +2 G_{\text{Tele},J_3}\big)-\dot{\phi_0} \left(3 \dot{H} G_{\text{Tele},I_2 I_2}+\dot{H} G_{\text{Tele},J_3}+\ddot{\phi_0} G_{\text{Tele},X I_2}\right) \,,
        \\ \nonumber
        F_{6}=& \dot{\phi_0} \big(-G_{\text{Tele},\phi_0 I_2}+G_{\text{Tele},X}+G_{2,X}-2 G_{3,\phi_0}+2 G_{4,\phi_0 \phi_0}\big)+H \big(3 G_{\text{Tele},I_2}-4 G_{\text{Tele},\phi_0 T}+6 G_{\text{Tele},\phi_0 T_{\text{vec}}}
        \\ \nonumber &+X \left(6 G_{3,X}-20 G_{4,\phi_0 X}+4 G_{5,\phi_0 \phi_0 }\right)-2 G_{4,\phi_0}\big)+H^2 \dot{\phi_0} \big(X \left(12 G_{4,XX}-8 G_{5,\phi_0 X}\right)+6 G_{4,X}-6 G_{5,\phi_0}\big)
        \\ & +H^3 \left(4 X^2 G_{5,XX}+6 X G_{5,X}\right) \,,
        \\ \nonumber
        F_{7}=& \dot{\phi_0} \big(-G_{\text{Tele},I_2}+X \left(4 G_{4,\phi_0 X}-2 G_{3,X}\right)+2 G_{4,\phi_0}\big)+H \big(-4 G_{\text{Tele},T}+6 G_{\text{Tele},T_{\text{vec}}}+X^2 \left(8 G_{5,\phi_0 X}-16 G_{4,XX}\right)
        \\& +X \left(12 G_{5,\phi_0}-16 G_{4,X}\right)+4 G_{4}\big)+H^2 \dot{\phi_0} \left(-4 X^2 G_{5,XX}-10 X G_{5,X}\right) \,,
        \\ \nonumber
        F_{8}=& F_7 + X H \big(12 G_{\text{Tele},I_2 I_2}+8 G_{\text{Tele},X T}-12 G_{\text{Tele},X T_{\text{vec}}}\big)+H^2 \dot{\phi_0} \left(36 G_{\text{Tele},T I_2}-54 G_{\text{Tele},T_{\text{vec}} I_2}\right)
        \\ &+H^3 \big(-144 G_{\text{Tele},T T_{\text{vec}}}+48 G_{\text{Tele},TT}+108 G_{\text{Tele},T_{\text{vec}}T_{\text{vec}}}\big)+2 X \dot{\phi_0} G_{\text{Tele},XI_2} \,,
        \\ \nonumber
        F_{9}=&X \left(G_{\text{Tele},X}+G_{2,X}-2 G_{3,\phi_0}\right)+ X^2 \left(2 G_{\text{Tele},XX}+2 G_{2,XX}-2 G_{3,\phi_0 X}\right)+H \dot{\phi_0} \Big(X \left(12 G_{\text{Tele},XI_2}+12 G_{3,X} \right.
        \\ \nonumber & \left. -30 G_{4,\phi_0 X}\right) +3 G_{\text{Tele},I_2}+X^2 \left(6 G_{3,XX}-12 G_{4,\phi_0  XX}\right)-6 G_{4,\phi_0}\Big)+H^2 \Big(X \left(36 G_{\text{Tele},I_2 I_2}+24 G_{\text{Tele},XT} \right.
        \\ \nonumber & \left. -36 G_{\text{Tele},XT_{\text{vec}}}+42 G_{4,X} -36 G_{5,\phi_0}\right)+6 G_{\text{Tele},T}-9 G_{\text{Tele},T_{\text{vec}}}+X^3 \left(24 G_{4,XXX}-12 G_{5,\phi_0 XX}\right)
        \\ \nonumber &+X^2 \left(96 G_{4,XX}-54 G_{5, \phi_0 X}\right)-6 G_{4}\Big) +H^3 \dot{\phi_0} \left(72 G_{\text{Tele},TI_2}-108 G_{\text{Tele},T_{\text{vec}}I_2}+4 X^3 G_{5,XXX}+26 X^2 G_{5,XX} \right.
        \\ & \left. +30 X G_{5,X}\right) + H^4 \left(-216 G_{\text{Tele},TT_{\text{vec}}}+72 G_{\text{Tele},TT}+162 G_{\text{Tele},T_{\text{vec}}T_{\text{vec}}}\right)\,, 
        \\ \nonumber
        F_{10}=& \frac{1}{2}\big(-3 \dot{H} G_{\text{Tele},\phi_0 I_2}- \ddot{\phi_0} G_{\text{Tele},\phi_0 X}+ G_{\text{Tele},\phi_0 \phi_0}- G_{2,\phi_0 X} \ddot{\phi_0}+G_{2,\phi_0 \phi_0 }+G_{3,\phi_0 \phi_0 } \ddot{\phi}+6 \dot{H} G_{4,\phi_0 \phi_0 } \big)
        \\ \nonumber & +X \big(-3 \dot{H} G_{\text{Tele},\phi_0 XI_2}-\ddot{\phi_0} G_{\text{Tele},\phi_0 XX}-G_{\text{Tele},\phi_0 \phi_0 X}+G_{3, \phi_0 \phi_0 \phi_0}-G_{2,\phi_0 XX} \ddot{\phi_0}-G_{2,\phi_0 \phi_0 X}-3 G_{3,\phi_0 X} \dot{H}
        \\ \nonumber &+G_{3,\phi_0 \phi_0 X} \ddot{\phi_0 }+6 G_{4,\phi_0 \phi_0 X} \dot{H}\big)+H \dot{\phi_0} \big(-\frac{9}{2} \dot{H} G_{\text{Tele},\phi_0 I_2 I_2}-6 \dot{H} G_{\text{Tele},\phi_0 X T}+9 \dot{H} G_{\text{Tele},\phi_0 XT_{\text{vec}}}
        \\ \nonumber &-3 \ddot{\phi_0} G_{\text{Tele},\phi_0 X I_2} -\frac{3}{2} G_{\text{Tele},\phi_0 \phi_0 I_2}  -\frac{3}{2} G_{\text{Tele},\phi_0 X}-\frac{3 G_{2, \phi_0 X}}{2} +X \big(-3 G_{3,\phi_0 XX} \ddot{\phi_0}-3 G_{3,\phi_0 \phi_0 X}
        \\ \nonumber&+6 G_{4,\phi_0 \phi_0 XX} \ddot{\phi_0} - 12 G_{4,\phi_0 XX} \dot{H} + 6 G_{4,\phi_0 \phi_0 \phi_0 X}+6 G_{5,\phi_0 \phi_0 X} \dot{H}\big)-3 G_{3,\phi_0 X} \ddot{\phi_0}+3 G_{3,\phi_0 \phi_0 } -6 G_{4,\phi_0 X} \dot{H}
        \\ \nonumber &+9 G_{4,\phi_0 \phi_0 X} \ddot{\phi_0}+6 G_{5,\phi_0 \phi_0 } \dot{H}\big) +H^2 \big(-18 \dot{H} G_{\text{Tele},\phi_0 T I_2}+27 \dot{H} G_{\text{Tele},\phi_0 T_{\text{vec}}I_2}-\frac{9}{2} \ddot{\phi_0} G_{\text{Tele},\phi_0 I_2 I_2}-\frac{9}{2} G_{\text{Tele},\phi_0 I_2}
        \\ \nonumber &+X \big(-9 G_{3,\phi_0 X}-24 G_{4,\phi_0 XX} \ddot{\phi_0} + 18 G_{4,\phi_0 \phi_0 X}-9 G_{5,\phi_0 X} \dot{H}+15 G_{5,\phi_0 \phi_0 X} \ddot{\phi_0}+3 G_{5,\phi_0 \phi_0 \phi_0 }\big)
        \\ \nonumber &+X^2 \big(-12 G_{4,\phi_0 XXX} \ddot{\phi_0 }-12 G_{4,\phi_0 \phi_0 XX}-6 G_{5,\phi_0 XX} \dot{H}+6 G_{5,\phi_0 \phi_0 XX} \ddot{\phi_0}+6 G_{5,\phi_0 \phi_0 \phi_0 X}\big)-3 G_{4,\phi_0 X} \ddot{\phi_0}
        \\ \nonumber &+6 G_{4,\phi_0 \phi_0 }+3 G_{5,\phi_0 \phi_0 } \ddot{\phi_0}\big)+H^3 \dot{H} \dot{\phi_0} \big(X \big(-18 G_{4,\phi_0 XX}-7 G_{5,\phi_0 XX} \ddot{\phi_0} +7 G_{5,\phi_0 \phi_0 X}\big)-9 G_{4,\phi_0 X}
        \\ &+X^2 \left(-2 G_{5,\phi_0 XXX} \ddot{\phi_0}-2 G_{5,\phi_0 \phi_0 XX}\right) -3 G_{5,\phi_0 X} \ddot{\phi_0}+9 G_{5,\phi_0 \phi_0 }\big) +H^4 \left(-6 X^2 G_{5,\phi_0 XX}-9 X G_{5,\phi_0 X}\right)
        \\ \nonumber
        F_{11}=& -\frac{1}{2} \big(G_{\text{Tele},X} + G_{2,X}\big) - G_{3,X} \ddot{\phi_0}+G_{3,\phi_0}+X \left(-G_{3,XX} \ddot{\phi_0 }-G_{3,\phi_0 X}-4 G_{4,XX} \dot{H}+2 G_{4,\phi_0 XX} \ddot{\phi_0}\right.
        \\ \nonumber & \left. +2 G_{4,\phi_0 \phi_0 } + 2 G_{5,\phi_0 X} \dot{H}\right) +H \dot{\phi_0} \big(-2 G_{3,X}+X \left(-4 G_{4,XXX} \ddot{\phi_0}-4 G_{4,\phi_0 XX}-2 G_{5,XX} \dot{H}+2 G_{5,\phi_0 XX} \ddot{\phi_0}\right.
        \\ \nonumber & \left. +2 G_{5,\phi_0 \phi_0 X}\right) -6 G_{4,XX} \ddot{\phi_0} + 6 G_{4,\phi_0 X}-2 G_{5,X} \dot{H}+4 G_{5,\phi_0 X} \ddot{\phi_0}\big) -2 G_{4,X} \dot{H}+3 G_{4,\phi_0 X} \ddot{\phi_0}+2 G_{5,\phi_0} \dot{H} 
        \\ \nonumber& +H^2 \left(X \left(-10 G_{4,XX}-5 G_{5,XX} \ddot{\phi_0}+5 G_{5,\phi_0 X}\right)-3 G_{4,X} +X^2 \left(-2 G_{5,XXX} \ddot{\phi_0}-2 G_{5,\phi_0 XX}\right)-G_{5,X} \ddot{\phi_0} \right.
        \\ & \left. +3 G_{5,\phi_0}\right) +H^3 \left(-2 X G_{5,XX}-2 G_{5,X}\right) \dot{\phi_0}\,, 
        \\ \nonumber
        F_{12}=& \frac{X}{3} \left(3G_{\text{Tele},I_2 I_2}+G_{\text{Tele},J_5}+4 G_{\text{Tele},J_8}\right)-G_{\text{Tele},T_{\text{vec}}} +H \dot{\phi_0} \left(4 G_{\text{Tele},T I_2}-6 G_{\text{Tele},T_{\text{vec}}I_2}\right)
        \\ &+H^2 \left(-24 G_{\text{Tele},TT_{\text{vec}}}+8 G_{\text{Tele},TT}+18 G_{\text{Tele},T_{\text{vec}}T_{\text{vec}}}\right)\,,
        \\
        F_{13}=& \frac{1}{9}\Big(9G_{\text{Tele},T_{\text{vec}}}+X \left(6 G_{\text{Tele},J_3}-10 G_{\text{Tele},J_5}-4 G_{\text{Tele},J_8}\right)+4 X^2 G_{\text{Tele},J_6}\Big)\,, \\
        F_{14} =& \frac{1}{2} \Big[ G_{2,X} + 2 \left(X(G_{2,XX} - G_{3,\phi X}) - G_{3,\phi}\right) + 6H^2 \left( G_{4,X} - G_{5,\phi} + X (G_{4,XX} - 5 G_{5,\phi X}) +2 X^2 (2 G_{4,XXX} \right.
        \nonumber \\ & \left. -G_{5,\phi XX}) + \tfrac{3}{2} G_{\text{Tele},I_{2}I_{2}}\right) + G_{\text{Tele},X} + 2 X G_{\text{Tele},XX} + 2 H^3 \dot{\phi} \left(3 G_{5,X} + X (7 G_{5,XX} + 2 X G_{5,XXX})\right) 
        \nonumber \\ & 
        + 6 H \dot{\phi} \left(G_{3,X} - 3 G_{4,\phi X} + G_{\text{Tele},X I_{2}} + X (G_{3,XX} - 2 G_{4,\phi XX})\right)\Big]
    \end{align}
    \end{subequations}
    where $F_{12}$ and $F_{13}$ are purely teleparallel coefficients. Moreover, difference between coefficients of $\beta$ and $B$ quantities stem from the teleparallel contribution of the BDLS action. For this case, only $\delta\phi$ and $\beta$ appear to be temporally dynamical with $\Phi$ and $B$ being the auxiliary modes.

    Following a minimization through obtaining field equations for auxiliary modes $\Phi$ and $B$, and diagionalization of the kinetic matrix, the final action can be given by
    \begin{align}
        \mathcal{S}_{\text{S}}^{\text{Flat}} = \int \dd t \, \frac{\dd k^3}{(2\pi)^{\frac{3}{2}}} \Bigg[& \left(\tilde{F}_{1} -\frac{1}{2} \dot{\tilde{F}}_{2}\right)\Psi_{1}^2 + \left(\tilde{F}_{3} -\frac{1}{2} \dot{\tilde{F}}_{4}\right)\Psi_{2}^2 + \left(\tilde{F}_{5}-\dot{\tilde{F}}_{7}\right) \Psi_{1} \Psi_{2} + \left(\tilde{F}_{6}-\tilde{F}_{7}\right) \dot{\Psi}_{1} \Psi_{2} \nonumber \\ & + \tilde{F}_{8} \dot{\Psi}_{1}^2 + \tilde{F}_{9} \dot{\Psi}_{2}^2 \Bigg]\,,
    \end{align}
    resulting in two propagating DoFs where coefficients $\tilde{F}_{i}$ are presented in Ref.~\cite{Caruana2024}. The number of DoFs for subclasses given in Table~\ref{tab:comparison_literature} is remains the same in the flat gauge.


\subsubsection{Unitary Gauge}
Next, we consider the \textit{unitary gauge} where the assumptions of $\delta \phi=0$ and $E=0$ are implemented in Eq.~\eqref{eq:scalar-quadratic-gauge-dependent-action}, resulting in
\begin{align}\label{unitary_scalar} \nonumber
     \mathcal{S}_{\rm S}^{\text{Unitary}}=\int \text{d}t \, \text{d}^3 x \, a^3 \Bigg[& U_1\Bigg(2 \dot{\psi} \frac{\nabla^2 B}{a}-3\dot{\psi}^2\Bigg)+U_2\frac{(\nabla\psi)^2}{a^2}+U_3 \Phi^2 +2 U_4 \Phi \Bigg(3\dot{\psi}- \frac{\nabla ^2 B}{a} \Bigg) +U_5 \Phi \frac{\nabla ^2 \beta}{a}   \\ & +  U_6 \Bigg( \dot{\psi} \frac{\nabla^2 \beta}{a}+\Phi \frac{\nabla^2 \psi}{a^2}\Bigg)+U_7\Bigg(\frac{(\nabla\Phi)^2}{a^2} + \left(\nabla \dot{\beta} \right)^2 + 2 \frac{\Phi \nabla^2\dot{\beta}}{a}\Bigg) + U_8 \frac{\left(\nabla^2 B \right)^2}{a^2}  + U_{9}\psi \nabla^2 \beta\Bigg]\,,
\end{align}
where
\begin{subequations}
\allowdisplaybreaks
\begin{align}
    \nonumber
        U_1 =& 2G_{4}-G_{\text{Tele},T}+3G_{\text{Tele},T_{\text{vec}}} +X\left(-4G_{4,X}+2G_{5,\phi_0}-3G_{\text{Tele},I_2 I_2}\right)+H \dot{\phi_0}\left( -2XG_{5,X}-12 G_{\text{Tele},T I_2} \right. 
        \\ & \left. +18G_{\text{Tele},T_{\text{vec}} I_2}\right) +H^2\left(-24 G_{\text{Tele},TT}+72G_{\text{Tele},TT_{\text{vec}}}-54G_{\text{Tele}T_{\text{vec}} T_{\text{vec}}}\right) \,,
        \\
        U_2 =& 2G_4 -2G_{\text{Tele}, T}+4G_{\text{Tele}, T_{\text{vec}}}+\frac{4}{9}X^2G_{\text{Tele}, J_6}-\frac{X}{9} \big( 2G_{5, \phi_0}+2G_{5,X} \ddot{\phi_0}+12G_{\text{Tele}, J_3}+10G_{\text{Tele}, J_5}+4G_{\text{Tele}, J_8} \big) \,,
        \\ \nonumber
        U_3 =& X \left(G_{\text{Tele},X}+G_{2,X}-2 G_{3,\phi_0 }\right)+ X^2 \left(2 G_{\text{Tele},XX}+2 G_{2, XX}-2 G_{3, \phi_0 X}\right)
        \\ \nonumber&+H \dot{\phi_0} \left(X \left(12 G_{\text{Tele},X,I_2}+12 G_{3,X}-30 G_{4,\phi_0 X}\right)+3 G_{\text{Tele},I_2}+X^2 \left(6 G_{3,XX}-12 G_{4,\phi_0 XX}\right)-6 G_{4,\phi_0 }\right)
        \\ \nonumber &+H^2 \left(X \left(36 G_{\text{Tele},I_2 I_2}+24 G_{\text{Tele},X T}-36 G_{\text{Tele},X T_{\text{vec}}}+42 G_{4,X}-36 G_{5,\phi_0}\right)+6 G_{\text{Tele},T}-9 G_{\text{Tele},T_{\text{vec}}} \right.
        \\ \nonumber &\left.+X^3 \left(24 G_{4,XXX}-12 G_{5,\phi_0 XX}\right)+X^2 \left(96 G_{4,XX}-54 G_{5,\phi_0 X}\right)-6 G_{4}\right)+H^3 \dot{\phi_0} \left(72 G_{\text{Tele},T I_2}-108 G_{\text{Tele},T_{\text{vec}} I_2} \right.
        \\&\left.+4 X^3 G_{5,XXX}+26 X^2 G_{5,XX}+30 X G_{5,X}\right)+18 H^4 \left(-12 G_{\text{Tele},T T_{\text{vec}}}+4 G_{\text{Tele},TT}+9 G_{\text{Tele},T_{\text{vec}}T_{\text{vec}}}\right) \,, 
        \\ \nonumber
        U_4 =& \dot{\phi_0} \left(X \left(-G_{\text{Tele},XI_2}-G_{3,X}+2 G_{4,\phi_0 X}\right)-\frac{1}{2}G_{\text{Tele},I_2}+G_{4,\phi_0}\right)+H \left(X \left(-6 G_{\text{Tele},I_2 I_2}-4 G_{\text{Tele},X T}+6 G_{\text{Tele},X T_{\text{vec}}} \right.\right.
        \\ \nonumber& \left. \left. -8 G_{4,X}+6 G_{5,\phi_0}\right)-2 G_{\text{Tele},T}+3 G_{\text{Tele},T_{\text{vec}}}+X^2 \left(4 G_{5,\phi_0 X}-8 G_{4,XX}\right)+2 G_{4}\right)+H^2 \dot{\phi_0} \left(-18 G_{\text{Tele},T I_2} \right.
        \\&\left.+27 G_{\text{Tele},T_{\text{vec}}I_2}-2 X^2 G_{5,XX}-5 X G_{5,X}\right)+H^3 \left(72 G_{\text{Tele},TT_{\text{vec}}}-24 G_{\text{Tele},TT}-54 G_{\text{Tele},T_{\text{vec}}T_{\text{vec}}}\right) \,, 
        \\ \nonumber 
        U_{5} =& 2 H \big[ 2 G_{4} + 2 X (-4 G_{4,X} + 3 G_{5,\phi} + 2 X (-2 G_{4,XX} + G_{5,\phi X})) + 3 G_{\text{Tele},T_{\text{vec}}} - 2 G_{\text{Tele},T} - H X \dot{\phi} (5 G_{5,X} 
        \nonumber \\ & 
        + 2 X G_{5,XX}) \big] - \dot{\phi} (-2 G_{4,\phi} + 2 X (G_{3,X} - 2 G_{4,\phi X}) + G_{\text{Tele},I_{2}}) \,,
        \\ \nonumber
        U_6 =& 4\left( G_{\text{Tele},T}- G_{\text{Tele},T_{\text{vec}}}- G_{4}+X G_{5_X} H \dot{\phi_0}\right)  -\frac{1}{9}X \left(6 G_{\text{Tele},J_3}+20 G_{\text{Tele},J_5}+8 G_{\text{Tele},J_8}-72 G_{4,X}+36 G_{5,\phi_0 }\right)
        \\ &+\frac{8}{9} X^2 G_{\text{Tele},J_6}\,, \\
        U_7 =& G_{\text{Tele},T_{\text{vec}}} +\frac{X}{9} \left(6 G_{\text{Tele},J_3}-10 G_{\text{Tele},J_5}-4 G_{\text{Tele},J_8}\right)+\frac{4}{9} X^2 G_{\text{Tele},J_6} \,,
        \\ \nonumber
        U_8 =& -G_{\text{Tele},T_{\text{vec}}}+\frac{X}{3} \left(3 G_{\text{Tele},I_2 I_2}+G_{\text{Tele},J_5}+4 G_{\text{Tele},J_8}\right)+H \dot{\phi_0} \left(4 G_{\text{Tele},TI_2}-6 G_{\text{Tele},T_{\text{vec}}I_2}\right)
        \\ &+H^2 \left(-24 G_{\text{Tele},T T_{\text{vec}}}+8 G_{\text{Tele},TT}+18 G_{\text{Tele},T_{\text{vec}}T_{tvec}}\right)\,,
        \\  \nonumber
        U_9=&\frac{1}{9}\Big[\dot{\phi_0} \big(-18 G_{\text{Tele},I_2}+X \left(\ddot{\phi_0}\left(-6 G_{\text{Tele},XJ_3}-20 G_{\text{Tele},XJ_5}+16 G_{\text{Tele},J_6}-8 G_{\text{Tele},XJ_8}\right)-6 G_{\text{Tele},\phi_0 J_3} \right.
        \\  \nonumber & \left.-20 G_{\text{Tele},\phi_0 J_5} - 8 G_{\text{Tele},\phi_0 J_8}\right)
        +\ddot{\phi_0} \left(-6 G_{\text{Tele},J_3}-20 G_{\text{Tele},J_5}-8 G_{\text{Tele},J_8}+36 G_{\text{Tele},XT}-36 G_{\text{Tele},XT_{\text{vec}}}\right)
        \\ \nonumber&+X^2 \left(8 \ddot{\phi_0} G_{\text{Tele},XJ_6}+8 G_{\text{Tele},\phi_0 J_6}\right) + 36 G_{\text{Tele},\phi_0 T}
        -36 G_{\text{Tele},\phi_0 T_{\text{vec}}}\big)+ H \big(\dot{H} \big(X \left(-72 G_{\text{Tele},TJ_3} \right.
        \\ \nonumber& \left. +108 G_{\text{Tele},T_{\text{vec}}J_3}-240 G_{\text{Tele},T J_5}+360 G_{\text{Tele},T_{\text{vec}}J_5} -96 G_{\text{Tele},TJ_8}+144 G_{\text{Tele},T_{\text{vec}}J_8}\right)+X^2 \left(96 G_{\text{Tele},TJ_6} \right.
        \\ \nonumber & \left.- 144 G_{\text{Tele},T_{\text{vec}}J_6}\right)-1080 G_{\text{Tele},TT_{\text{vec}}}+432 G_{\text{Tele},TT}+648 G_{\text{Tele},T_{\text{vec}}T_{\text{vec}}}\Big) +X \left(\ddot{\phi_0} \left(-18 G_{\text{Tele},I_2 J_3} \right.\right.
        \\ \nonumber & \left.\left. -60 G_{\text{Tele},I_2 J_5}-24 G_{\text{Tele},I_2J_8}\right)-12 G_{\text{Tele},J_3}-40 G_{\text{Tele},J_5}-16 G_{\text{Tele},J_8}\right)+36 G_{\text{Tele},T_{\text{vec}}}
        \\ \nonumber &+X^2 \left(24 \ddot{\phi_0} G_{\text{Tele},I_2J_6}+16 G_{\text{Tele},J_6}\right)+\ddot{\phi_0} \left(108 G_{\text{Tele},T I_2}-108 G_{\text{Tele},T_{\text{vec}}I_2}\right)\big) +\dot{H} \dot{\phi_0} \big(X \left(-18 G_{\text{Tele},I_2J_3} \right.
        \\ & \left. -60 G_{\text{Tele},I_2 J_5}-24 G_{\text{Tele},I_2J_8}\right)+24 X^2 G_{\text{Tele},I_2J_6}+108 G_{\text{Tele},T I_2}-108 G_{\text{Tele},T_{\text{vec}}I_2}\big) \Big]\,.
    \end{align}
    \end{subequations}
for which $\{U_{7}, U_{8}, U_{9}\}$ are coefficients arising from the purely teleparallel sector of the gravitational theory. The unitary gauge has been favoured in scalar-tensor theories, so the standard Horndeski result given in Ref.~\cite{Kobayashi:2019hrl} can be obtained by taking the limit $G_{\text{Tele} }\rightarrow 0$ limit. First and foremost, the coefficients $\{U_{7}, U_{8}, U_{9}\}$ completely disappear, eliminating in particularly the contribution proportional to $k^4$ (upon applying Fourier transformation). Next, in metric-based theories, there is no distinction between the off-diagional terms since the metric is said to be symmetric. The relationship between tetrad and metric given by Eq.~\eqref{eq:metr_trans} indicates that $\beta$ and $B$ can be combined to have a single off-diagonal variable $\mathcal{B} = -B + \beta$. Provided these modifications are implemented, the Horndeski gravity result obtained in Ref.~\cite{Kobayashi:2019hrl} is attained. For the general BDLS action, the final action is of the form
\begin{align}
    \mathcal{S}_{\text{S}}^{\text{Unitary}} = \int \dd t \, \frac{\dd^3k}{(2\pi)^{\frac{3}{2}}} \Bigg[&  
    \left(\tilde{U}_{1} -\frac{1}{2} \dot{\tilde{U}}_{2}\right)\Psi_{1}^2 + \left(\tilde{U}_{3} -\frac{1}{2} \dot{\tilde{U}}_{4}\right)\Psi_{2}^2 + \left(\tilde{U}_{5}-\dot{\tilde{U}}_{7}\right) \Psi_{1} \Psi_{2} + \left(\tilde{U}_{6}-\tilde{U}_{7}\right) \dot{\Psi}_{1} \Psi_{2} \nonumber \\ & + \tilde{U}_{8} \dot{\Psi}_{1}^2 + \tilde{U}_{9} \dot{\Psi}_{2}^2
    \Bigg]\,,
\end{align}
where the procedure and details of coefficients $\tilde{U}_{i}$ for $i \in [1,9]$ have been included in Ref.~\cite{Caruana2024}. In-line with the gauge-invariant and flat gauge, only two dynamical modes propagate following a diagonalization of the kinetic term.



\subsubsection{Newtonian Gauge}

The \textit{Newtonian / longitudinal gauge} has been commonly used in teleparallel gravity, where $E=0$ and $\beta = B$. In Ref.~\cite{Izumi:2012qj}, the gauge choice was further restricted by imposing $\beta = 0$, leading to potentially overfixing the gauge. The action transforms to
\begin{align}
    \mathcal{S}_{\text{S}}^{\text{Newtonian}} = \int \text{d}t \, \text{d}^3x\, a^3 \Bigg[&
    \mathcal{A}_{1} \left(\frac{\nabla^{2} \beta}{a^2}\right)^2 + \frac{ \beta }{a^2}\left((\mathcal{A}_{2}+\mathcal{A}_{6}) \nabla^2 \Phi + (\mathcal{A}_{3}+\mathcal{A}_{12}) \nabla^2 \dot{\psi} + (\mathcal{A}_{4}+\mathcal{A}_{8}) \nabla^2 \delta\phi  \right.
    \nonumber \\ & \left. + (\mathcal{A}_{5}+\mathcal{A}_{9}) \nabla^2 \dot{\delta\phi}\right) + \mathcal{A}_{10} \frac{\beta}{a^2}  \nabla^2 \psi  + \frac{\dot{\beta}}{a^2} \left(\mathcal{A}_{7} \nabla^2 \Phi + \mathcal{A}_{11} \nabla^{2}\psi \right) + \mathcal{A}_{13} \left(\frac{\nabla \dot{\beta}}{a}\right)^2 
    \nonumber \\ & + \Phi \left(\mathcal{A}_{14} \frac{\nabla^2 \psi}{a^2} + \mathcal{A}_{15} \dot{\psi} + \mathcal{A}_{16} \delta\phi + \mathcal{A}_{17} \frac{\nabla^2 \delta\phi}{a^2} + \mathcal{A}_{18} \dot{\delta\phi}\right) + \mathcal{A}_{19} \left( \frac{\nabla \Phi}{a}\right)^2 + \mathcal{A}_{20}  \Phi^2 
    \nonumber \\ & 
    + \mathcal{A}_{21} \left( \frac{\nabla \psi}{a}\right)^2 + \mathcal{A}_{22} \dot{\psi}^2 + \mathcal{A}_{23} \frac{\nabla \psi \nabla \delta\phi}{a^2} + \mathcal{A}_{24} \delta\phi \dot{\psi} + \mathcal{A}_{25} \dot{\delta\phi} \dot{\psi} + \mathcal{A}_{26} \left( \frac{\nabla \delta\phi}{a}\right)^2 
    \nonumber \\ & + \mathcal{A}_{27} \delta\phi^2 + \mathcal{A}_{28} \dot{\delta\phi}^2
    \Bigg] \,.
\end{align}
with very minor modifications to the original action~\eqref{eq:scalar-quadratic-gauge-dependent-action}. Here, $\{\delta\phi, \beta, \psi\}$ are all dynamical modes and only $\Phi$ is non-dynamical. Varying with the latter mode and diagionalizing the kinetic matrix by relabelling the dynamical modes in terms of new mode $\{\Psi_{1},\Psi_{3},\Psi_{3}\}$, the action becomes of the form
\begin{align}
    \mathcal{S}_{\text{S}}^{\text{Newtonian}} = \int \dd t \, \frac{\dd k^3}{(2\pi)^{\frac{3}{2}}} \bigg[&
    \tilde{N}_{1} \Psi_{1}^2 + \tilde{N}_{2} \Psi_{2}^2 + \tilde{N}_{3} \Psi_{3}^2 +  \Psi_{1} \left( \tilde{N}_{4} \Psi_{2} + \tilde{N}_{6} \Psi_{3} + \tilde{N}_{7} \dot{\Psi}_{3} + \tilde{N}_{9}  \dot{\Psi}_{2}\right) + \Psi_{2} \left( \tilde{N}_{5}  \Psi_{3}  + \tilde{N}_{8} \dot{\Psi}_{3} \right) 
    \nonumber \\ & 
    + \tilde{N}_{10} \dot{\Psi}_{1}^2 + \tilde{N}_{11} \dot{\Psi}_{2}^2 + \tilde{N}_{12} \dot{\Psi}_{3}^2
    \bigg]\,,
\end{align}
where coefficients $\tilde{N}_{i}$ for $i \in [1,12]$ are detailed in Ref.~\cite{Caruana2024}. The longitudinal mode suggests an additional dynamical scalar mode for the general BDLS model, but further investigation would be required as discrepancy in the number of DoFs does not occur for widely used subclasses of BDLS. For the particular case of $\beta = 0$, the number reduced to two.

\subsubsection{Synchronous Gauge}

The \textit{synchronous gauge} is obtained by setting $\Phi=0$ and $\beta = B$ and very popular in numerical studies while it is the one mostly used in Boltzmann solvers like CLASS and CAMB. The action for the gauge choice reads
\begin{align}
    \mathcal{S}_{\text{S}}^{\text{Synchronous}} = \int \text{d}t \,\text{d}^3 x a^3 \Bigg[&
    \mathcal{A}_{1} \left(\frac{\nabla^{2}(\beta - a \dot{E})}{a^2}\right)^2 + \frac{\left( \beta - a \dot{E}\right)}{a^2}\left( \mathcal{A}_{3} \nabla^2 \dot{\psi} + \mathcal{A}_{4} \nabla^2 \delta\phi + \mathcal{A}_{5} \nabla^2 \dot{\delta\phi}\right) + \frac{\beta}{a^2} \left(  \mathcal{A}_{8} \nabla^2 \delta\phi + \mathcal{A}_{9} \nabla^2 \dot{\delta\phi}  \right.
    \nonumber \\ & \left. + \mathcal{A}_{10} \nabla^2 \psi + \mathcal{A}_{12} \nabla^2 \dot{\psi} \right) + \mathcal{A}_{11}\frac{\dot{\beta}}{a^2}  \nabla^{2}\psi  + \mathcal{A}_{13} \left(\frac{\nabla \dot{\beta}}{a}\right)^2 + \mathcal{A}_{21} \left( \frac{\nabla \psi}{a}\right)^2 + \mathcal{A}_{22} \dot{\psi}^2 + \mathcal{A}_{23} \frac{\nabla \psi \nabla \delta\phi}{a^2}
    \nonumber \\ &  + \mathcal{A}_{24} \delta\phi \dot{\psi} + \mathcal{A}_{25} \dot{\delta\phi} \dot{\psi} + \mathcal{A}_{26} \left( \frac{\nabla \delta\phi}{a}\right)^2 + \mathcal{A}_{27} \delta\phi^2 + \mathcal{A}_{28} \dot{\delta\phi}^2
    \Bigg] \,,
\end{align}
showing very minor modifications from the action given by Eq.~\eqref{eq:scalar-quadratic-gauge-dependent-action}. Here, all modes $\{\delta\phi, \beta, \psi, E\}$ are dynamical, as $\{\Phi, B\}$ are typically the auxiliary modes in the system but have been eliminated through the gauge choice. Since mode $E$ appears to have at least a first order derivative, a relabeling it as $\mathcal{E} = \dot{E}$ allows treating $\mathcal{E}$ as an non-dynamical mode. Thus, the final action is of the form
\begin{align}
    \mathcal{S}_{\text{S}}^{\text{Synchronous}} = \int \text{d}t \frac{\text{d}^3 k}{(2\pi)^{\frac{3}{2}}} \Bigg[&
    \tilde{S}_{1} \Psi_{1}^2 + \tilde{S}_{3} \Psi_{3}^2 + \Psi_{1} \left( \tilde{S}_{4}  \Psi_{2} + \tilde{S}_{5} \dot{\Psi}_{2} + \tilde{S}_{6} \Psi_{3} + \tilde{S}_{7} \dot{\Psi}_{3}\right) + \Psi_{3} \left( \tilde{S}_{8} \Psi_{2}  + \tilde{S}_{9} \dot{\Psi}_{2} \right) 
    \nonumber \\ & 
    + \tilde{S}_{10} \dot{\Psi}_{1}^2 + \tilde{S}_{11} \dot{\Psi}_{2}^2 + \tilde{S}_{12} \dot{\Psi}_{3}^2
    \Bigg] \,,
\end{align}
with coefficient details included in the supplementary document of Ref.~\cite{Caruana2024}. Similar to the longitudinal gauge choice, the substitution of $\beta = B$ yields an extra scalar DoF from the rest of the gauge choices. While further setting $\beta = 0$, the two dynamical modes are obtained but may lead to an overfixed gauge.

\subsection{Vectors}

The vector sector contains a gauge invariant mode $u_{i}$ and the remaining modes $\{v_{i}, w_{i}, V_{i}\}$ can be expressed as a linear combination of each other. The gauge choice $V_{i} = 0$ is favoured as it eliminates contributions related to the Levi-Civita tensor associated with the pseudovector. Regardless, all possible gauge choices are presented here.

In the case of $v_{i}=0$, the action transforms to
\begin{align}
    \mathcal{S}_{\text{V}}^{v_{i} = 0} = \int \text{d}t\, \text{d}^3x\, a^{3} \bigg[&
    \frac{\mathcal{C}_{2}}{a^2}\left( (\nabla^2 \textbf{w})^2 + (\nabla \textbf{V})^2+2 \,  (\nabla\times\textbf{V})(\nabla^{2} \textbf{w}) \right)  + \mathcal{C}_{4} \, \dot{\textbf{V}} \left( \dot{\textbf{V}} + \frac{1}{a} \nabla\times \textbf{v} \right)
    \nonumber \\ & 
    + \mathcal{C}_{3} \, (\nabla\dot{\textbf{w}})^2 + \mathcal{C}_{5} \,\dot{\textbf{u}}^2 + \mathcal{C}_{6} \frac{(\nabla \textbf{u})^2}{a^2} + \frac{\mathcal{C}_{8}}{a} \Big( (\nabla \textbf{u})(\nabla \textbf{w}) + \, \textbf{V}(\nabla\times \textbf{u}) \Big) + \mathcal{C}_{9}  \frac{(\nabla \textbf{u}) \, (\nabla \dot{\textbf{w}})}{a}
    \nonumber \\ & 
    + \frac{\mathcal{C}_{10}}{a}  \Big((\nabla \dot{\textbf{u}})(\nabla \textbf{w}) + \, \textbf{V}(\nabla\times \dot{\textbf{u}})\Big)  - \frac{1}{a} \textbf{V}\left( \mathcal{C}_{11} \, (\nabla\times \dot{\textbf{u}}) + \frac{1}{a^2} \frac{\text{d}}{\text{d}t} \left( a^2  \mathcal{C}_{11}\right)  (\nabla\times \textbf{u}) \right)
\bigg]\,,
\label{eq:vetor_vi=0}
\end{align}
 such that modes $\{u_{i}, w_{i}, V_{i}\}$ are dynamical modes. In the other gauge choices, it was noted that $v_{i}$ is the only mode that behaves as an auxiliary mode.

The gauge choice of $w_{i} = 0$, followed by substitution of the equation of the auxiliary mode $v_{i}$ results in 
\begin{align}
    \mathcal{S}_{\text{V}}^{w_{i} = 0} = \int \text{d}t\, \text{d}^3x\, a^{3} \bigg[&
    \mathcal{C}_{2} \frac{(\nabla \textbf{V})^2}{a^2} + \mathcal{C}_{4}\left(1 + \frac{\mathcal{C}_{4}}{4\mathcal{C}_{1}}\right) \, \dot{\textbf{V}}^2 + \left(\mathcal{C}_{6} - \frac{\mathcal{C}_{7}^2}{4\mathcal{C}_{1}}\right) \frac{(\nabla \textbf{u})^2}{a^2} + \mathcal{C}_{5} \, (\dot{\textbf{u}})^2 
    \nonumber \\ &
    + \frac{1}{a} \left( \mathcal{C}_{8} - \frac{1}{a^2} \frac{\text{d}}{\text{d}t} \left( a^2 \mathcal{C}_{10}\right)\right) \textbf{V} \, (\nabla\times \textbf{u})  + \frac{1}{a} \left(\mathcal{C}_{11} - \mathcal{C}_{10} - \frac{\mathcal{C}_{4} \mathcal{C}_{7}}{2 \mathcal{C}_{1}}\right) \dot{\textbf{V}} \, (\nabla\times \textbf{u}) \bigg)
\bigg]\,,
\end{align}
for a total of two dynamical modes $\{u_{i}, V_{i}\}$ provided $\mathcal{C}_{1} \neq 0$, noting a difference in the number of dynamical modes when compared to Eq.~\eqref{eq:vetor_vi=0}.

For the gauge choice $V_{i} = 0$,
\begin{align}
      \mathcal{S}_{\text{V}}^{V_{i} = 0} = \int \text{d}t\, \text{d}^3x\, a^{3} \bigg[&
     \mathcal{C}_{2} \frac{(\Delta \textbf{w})^2 }{a^2} + \mathcal{C}_{3}\left( 1 - \frac{\mathcal{C}_{3}}{4 \mathcal{C}_{1} }\right) \, (\nabla\dot{\textbf{w}})^2 + \left(\mathcal{C}_{6} - \frac{\mathcal{C}_{7}^2}{4\mathcal{C}_{1}}\right)\, (\dot{\textbf{u}})^2+ \mathcal{C}_{7} \frac{(\nabla \textbf{u})^2}{a^2} 
     \nonumber \\ & 
     + \left(\mathcal{C}_{8}  - \frac{1}{a^2}\frac{\text{d}}{\text{d}t} \left(a^2 \mathcal{C}_{10}\right)\right) \frac{(\nabla \textbf{u}) \, (\nabla \textbf{w})}{a} + \left( \mathcal{C}_{9} - \mathcal{C}_{10} + \frac{\mathcal{C}_{3}\mathcal{C}_{7}}{2 \mathcal{C}_{1}}\right)  \frac{(\nabla \textbf{u}) \, (\nabla \dot{\textbf{w}})}{a}
 \bigg]\,,
\end{align}
where now the two dynamical modes are $\{u_{i}, w_{i}\}$ provided $\mathcal{C}_{1} \neq 0$. The latter two results, along with the gauge invariant analysis, suggests that the gauge choice of $v_{i} = 0$ requires further analysis within the BDLS framework.

\bibliographystyle{plain}
\bibliography{references}

\end{document}